\documentclass{aa}  
\usepackage[colorlinks=true, linkcolor=blue, citecolor=blue]{hyperref}
\usepackage{graphicx}
\usepackage{txfonts}
\usepackage{xcolor}
\usepackage{amsmath}	    
\usepackage{amssymb}	    
\usepackage{siunitx}        
\usepackage{nicefrac}       
\usepackage{mathtools}      
\usepackage{xcolor}         
\usepackage{textcomp}       
\graphicspath{{figures/}}
\usepackage{graphicx}	    
\usepackage{stfloats}  		
\usepackage{booktabs}
\usepackage{calc}           
\newcommand{\orcid}[1]{\protect\href{https://orcid.org/#1}{\protect\includegraphics[width=8pt]{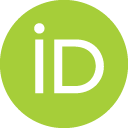}}}
\usepackage{siunitx}
\DeclareSIUnit\dex{dex}
\DeclareSIUnit\year{yr}
\DeclareSIUnit\fwhm{FWHM}
\DeclareSIUnit\ppt{ppt}
\DeclareSIUnit\ppm{ppm}
\DeclareSIUnit\ppmh{ppm\,h^{-\nicefrac{1}{2}}}
\DeclareSIUnit\arcsec{arcsec}
\DeclareSIUnit\pixel{pixel}
\DeclareSIUnit\electron{{e^-}}
\DeclareSIUnit\proton{{p^+}}
\DeclareSIUnit\electronvolt{{eV}}
\DeclareSIUnit\adu{ADU}
\DeclareSIUnit\dn{DN}
\DeclareSIUnit\hit{hits}
\DeclareSIUnit\events{events}
\DeclareSIUnit\photon{photons}
\DeclareSIUnit\volt{V}
\DeclareSIUnit\magnitude{mag}
\DeclareSIUnit\au{AU}
\DeclareSIUnit\pc{pc}
\DeclareSIUnit\exposure{exposure}
\DeclareSIUnit\Rsun{R_{\odot}}
\DeclareSIUnit\Msun{M_{\odot}}
\DeclareSIUnit\smbhb{SMBHB}
\DeclareSIUnit\agn{AGN}
\DeclareSIUnit\deg{deg}
\newcommand{\paren}[1]{\left(#1\right)} 
\newcommand{\parenf}[1]{\left[#1\right]}

\newcommand{\sole}[1]{{#1}_{\odot}}

\newcommand{\tx}[2]{{#1}_{\text{#2}}} 			   	
\newcommand{\pmx}[3]{${#1}_{#2}^{#3}$} 	
\newcommand{\plato}{\textit{Plato}}
\newcommand{\kepler}{\textit{Kepler}}
\newcommand{\platosim}{\texttt{PlatoSim}}

\newcommand{\Pb}{{\mathcal{P}}}

\defcitealias{smith2018kepler}{S18}
\defcitealias{hu2020spikey}{H20}

\makeatletter
\newcommand{\Hline}{%
    \noalign {\ifnum 0=`}\fi \hrule height 1pt
    \futurelet \reserved@a \@xhline
}
\newcolumntype{\Vline}{@{\hskip\tabcolsep\vrule width 1pt\hskip\tabcolsep}}
\makeatother

\begin{document} 

\title{Exploring the faint limit of the ESA \textit{Plato} space mission:}
\subtitle{A case study of extragalactic research with the self-lensing \\ supermassive black hole binary candidate, Spikey}
\author{
N.~Jannsen\inst{1,2}\orcid{0000-0003-4670-9616} \and
P.~Huijse\inst{2}\orcid{0000-0003-3541-1697} \and
K.~Park\inst{3}\orcid{0009-0007-4908-1314} \and
J.~Davelaar\inst{4,5}\orcid{0000-0002-2685-2434} \and
Z.~Haiman\inst{3,6,7}\orcid{0000-0003-3633-5403} \and
D.~J.~D'Orazio\inst{8,9,10}\orcid{0000-0002-1271-6247} \and
C.~Aerts\inst{2,11,12}\orcid{0000-0003-1822-7126}
}
\institute{
Isaac Newton Group of Telescopes, Apartado 321, E-38700 Santa Cruz de La Palma, Canary Islands, Spain \\
\email{nicholasj@ing.iac.es}) \and
Institute of Astronomy, KU Leuven, Celestijnenlaan 200D, 3001 Leuven, Belgium  \and
Institute of Science and Technology Austria (ISTA), Am Campus 1, 3400 Klosterneuburg, Austria \and
Department of Astrophysical Sciences, Peyton Hall, Princeton University, Princeton, NJ 08544, USA \and
NASA Hubble Fellowship Program, Einstein Fellow \and
Department of Astronomy, Columbia University, New York, NY 10027, USA \and
Department of Physics, Columbia University, New York, NY 10027, USA \and
Space Telescope Science Institute, 3700 San Martin Drive, Baltimore, MD 21218, USA \and 
Department of Physics and Astronomy, Johns Hopkins University, 3400 North Charles Street, Baltimore, MD 21218, USA \and
Niels Bohr International Academy, Niels Bohr Institute, Blegdamsvej 17, DK-2100 Copenhagen, Denmark \and
Department of Astrophysics, IMAPP, Radboud University Nijmegen, PO Box 9010, 6500 GL Nijmegen, The Netherlands \and
Max Planck Institute for Astronomy, Koenigstuhl 17, 69117 Heidelberg, Germany
}
\date{\today}
 
\abstract
{The search for supermassive black hole (SMBH) binaries has, in recent years, seen the dawn of exploration with several hundred candidates claimed from photometric and spectroscopic surveys monitoring active galactic nuclei (AGNs). While only a select number persist to date, the advent of upcoming high-precision wide-field photometric missions motivates continuing the pursuit of confirming SMBH binaries (SMBHBs) in the optical.}
{We explore the possibility of using the \textit{Plato} space mission of ESA to detect photometric signatures of SMBH binarity. Since many accretion-related signatures are not uniquely related to binaries, nor easily disentangled, the distinct photometric signature, strict occurrence in phase, and duration of a gravitational self-lensing flare (SLF) can strongly reduce such signal confusion. Motivated by the \textit{Kepler} observation of Spikey, the best known SLF candidate to date, this work aims to benchmark the scientific outcome if \textit{Plato} were to observe Spikey-like objects via its Guest Observer programme.}
{Starting from the \textit{Gaia} database, we assembled a catalogue of 12,226 bright ($G < 19$) high-probability quasars (or QSOs) for the two pointing fields of \textit{Plato}'s nominal mission. This \textit{Plato} quasar catalogue will be pivotal for future follow-up observations of larger photometric searches such as the Vera Rubin LSST survey. We used the \textit{Plato} camera simulator, \texttt{PlatoSim}, to realistically explore the noise budget in \textit{Plato}'s faint limit, while generating mock light curves to benchmark \plato{}'s ability to recover signatures of SMBH binarity. These signatures were secured using nested sampling for robust probabilistic modelling.}
{We show that, although not originally designed for this purpose, \textit{Plato} is capable of detecting Spikey-like SMBHB candidates through their relativistic photometric signatures using Bayesian inference and evidence. \textit{Plato} will, in particular, be able to confirm or rule out Spikey and Spikey-like objects with a limiting magnitude of $G\leq18$.}
{With a minimum \SI{2}{\year} baseline per pointing field, we show that \textit{Plato} could not only play an essential role in future SMBHB research, but it may be an integral part of the observational fleet of continuous high-precision facilities monitoring SMBHB candidates in the near future.}
\keywords{quasars: supermassive black holes -- methods: numerical -- methods: statistical -- techniques: photometric -- catalogs} 
 
\maketitle

\begin{figure*}[t!]
\center
\includegraphics[width=1.75\columnwidth]{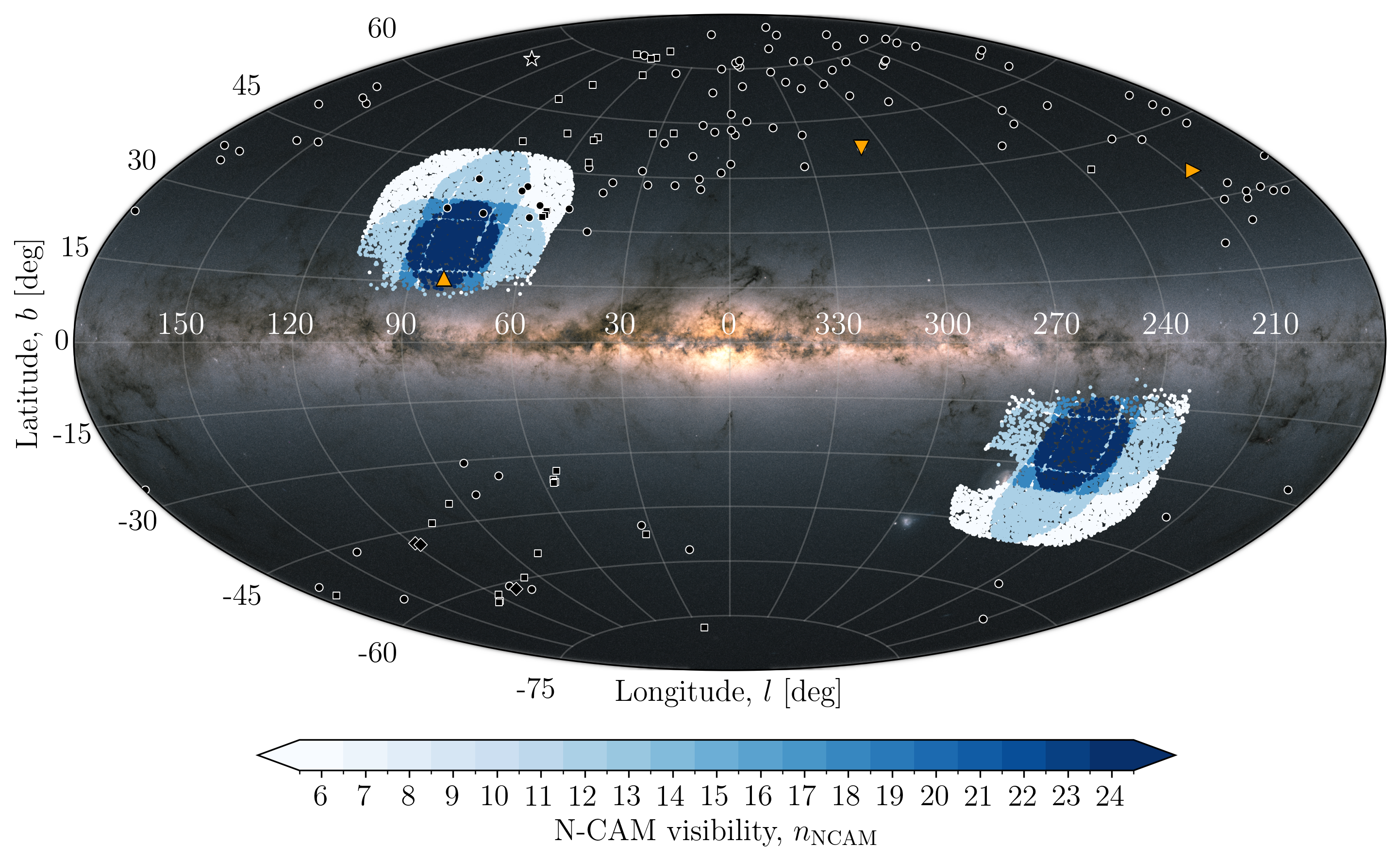}
\caption[]
{\textit{Plato} catalogue of high-probability quasar candidates plotted in an aitoff sky projection in galactic coordinates. The lower right and upper left data collection correspond to 12,226 quasar candidates from the LOPS2 and LOPN1, colour coded after N-CAM visibility, $\tx{n}{CAM}$. The different markers are SMBHB candidates from the literature: \citet[][circles: 111]{graham2015systematic}, \citet[][squares: 33]{charisi2016populsation}, \citet[][star: 1]{liu2018did, liu2019supermassive}, \citet[][upward-pointing triangle: 1]{hu2020spikey}, and \citet[][diamonds: 3]{chen2024searching}. Meanwhile, the orange triangles (upward-, downward-, and rightward-pointing) are three actively debated SMBHB candidates in the literature to date (OJ 287, PG 1302-102, and Spikey). The background image shows the brightness measurements of the $\sim$1.7 billion sources of the \textit{Gaia} DR2 release \citep{gaia2018dr2}.}
\label{fig:aitoff_projection}
\end{figure*}

\section{Introduction}\label{sec:introduction}

The field of supermassive black holes (SMBHs) has undergone a renaissance thanks to numerous discoveries of such objects in recent years \citep[e.g.][]{eht2019m87, eht2022sgr}. These detections built on decades of research, revealing that SMBHs cause the most energetic phenomena in our Universe and play an essential role in the evolution of galaxies throughout cosmic time \citep{sargent1978dynamical, magorrian1998demography, kormendy2013coevolution}. Since galaxies are built over cosmic time through a succession of mergers, it is also expected that they often host compact SMBH binaries \citep[SMBHBs;][]{begelman1980massive}. However, the demography and properties of SMBHBs are yet to be deduced and understood. While the existence of SMBH pairs have mainly been detected at kiloparsec (kpc) separation \citep[e.g.][observed as dual AGNs]{komossa2003discovery, comerford2015merger}, the observational evidence for sub-pc, gravitationally bound SMBHBs is limited and inconclusive. Finding and expanding the arsenal of SMBHB candidates is, among other things, vital for understanding the co-evolution with their host galaxies. 

One way to find SMBHBs is from pulsar timing array (PTA) facilities \citep[e.g.][]{hobbs2010international, kramer2013european, agazie2024nangrav}, which are sensitive to nanohertz gravitational wave (GW) emission sent out during the inevitable inspiraling phase of short-period SMBH pairs. A population of merging SMBHBs is predicted to be the dominant source for the GW background, for which evidence has been steadily emerging \citep{agazie2025nanograv}. However, to understand how SMBHBs evolve and merge, multi-messenger surveys \citep[such as][]{sesana2012multimessenger, charisi2022multimessenger} including electromagnetic observations are needed to build a population from which demographic studies can be conducted. Time-domain surveys in particular can monitor the periodic variability in the light curves of active galactic nuclei (AGNs), which in turn can disclose binarity, either by accretion rate modulation due to orbital motion \citep[e.g.][]{macfadyen2008eccentric, dorazio2013accretion, westernacher2022multiband, dorazio2023observational} or by relativistic effects such as Doppler boosting and gravitational self-lensing \citep[e.g.][]{dorazio2015relativistic,haiman2017electromagnetic, dorazio2018periodic, hu2020spikey, davelaar2022self, dorazio2023observational, krauth2024self, porter2025parameter}. 

Although apparent periodicity in AGN light curves is a promising indicator of SMBHBs, with more than hundreds of cyclic signals proposed as candidates \citep[][see Fig.~\ref{fig:aitoff_projection}]{graham2015systematic, charisi2016populsation, liu2018did, chen2024searching}, AGNs without a binary component also exhibit stochastic variability across a broad range of timescales due to turbulent accretion \citep{macleod2010modeling} or due to disc hot spots \citep{vos2022polarimetric}. As a result, fake apparent periodicity can often be mimicked by stochastic noise \citep{vaughan2016false, witt2022quasars, zhu2020supermassive, elbadry2026active, lin2026lomb, huijse2026periodic}. At present, this has rendered the majority of time-domain candidates uncertain. Three highly debated candidates that have the longest baselines are the following (triangles of Fig.~\ref{fig:aitoff_projection}): OJ 287 \citep{valtonen2008massive, komossa2021project}, PG 1302-102 \citep{graham2015systematic, xin2020testing}, and Spikey \citep{hu2020spikey, kun2020spikey, sorabella2022chandra}. 

While Doppler boosting (from purely circular orbits) alone serves as a poor diagnostic in time-domain searches, the wavelength dependence of the boosting amplitude allows a unique validation of binarity \citep[as in the case of PG 1302-102;][]{dorazio2015relativistic}. However, the distinct photometric signature, strict occurrence in phase, and duration of gravitational self-lensing flare (SLF) taken together offer less confusion than Doppler boosting alone \citep{kelly2021gravitational}. Similar to special-relativistic Doppler boosting, SLFs are a robust general-relativistic prediction that is, at most, only weakly wavelength dependent if at all \citep[][see also Appendix~\ref{app:model_lensing}]{dorazio2018periodic}. This work is particularly motivated by the \textit{Kepler} observation \citep[][hereafter \citetalias{smith2018kepler}]{smith2018kepler} of the SMBHB candidate `Spikey', whose variability is very well fit by an eccentric, unequal-mass binary exhibiting both a Doppler modulation and a SLF \citep[][hereafter \citetalias{hu2020spikey}]{hu2020spikey}.
  
In the hunt for photometric signatures of SMBHBs, we present here the feasibility of doing so with the upcoming medium-class ESA mission: PLAnetary Transits and Oscillation of stars \citep[\textit{Plato};][]{rauer2025plato}. Over its \SI{4}{\year} nominal mission, \textit{Plato} will observe carefully selected targets during a long-observational phase (LOP) in the South (LOPS2) and in the North (LOPN1), each for a duration of two years \citep{nascimbeni2025plato}. While the payload and mission strategy have been designed to observe the intrinsic and extrinsic variability of bright ($V<\SI{13}{\magnitude}$) solar-type stars, an Open Time Guest Observer (GO) programme will ensure that 8\% of the telemetry budget gets allocated to complementary science targets proposed by the worldwide community \citep{heras2024eas}. To fully exploit the mission, a \textit{Plato} complementary science programme (PLATO-CS) has been designed \citep{tkachenko2024eas}. This paper delivers a study within PLATO-CS activities to promote transient phenomena and extragalactic science as a future hunt for SMBHBs.

The \plato{} payload contains two `fast' cameras (F-CAMs) and 24 `normal' cameras (N-CAMs) operating at a cadence of \SI{2.5}{\second} and \SI{25}{\second}, respectively. On-board processing further allows the N-CAMs to downlink light curves (extracted by a binary aperture mask) at a cadence of \SI{50}{\second} and \SI{600}{\second}, equivalent to respectively two and 24 consecutive exposures. The F-CAMs' main purpose is to secure a stable fine guidance and altitude control of the spacecraft, but they are also ideal to measure bright stars ($V \leq 8$).
The N-CAMs on the other hand are the primary science instruments and are separated into four groups of six cameras, each group tilted symmetrically \SI{9}{\degree} away from the boresight pointing. The result is an increase in effective sky coverage and a characteristic footprint in the sky (the so-called N-CAM visibility flower) of co-pointing cameras, counting $\tx{n}{CAM} \in \{6, 12, 18, 24\}$. Thus, together with the source magnitude, the $\tx{n}{CAM}$ parameter naturally sets the noise budget at mission level, as we use below.

As a result of the innovative multi-telescope payload design, a single \plato{} field covers more than five percent of the entire sky \citep[$\sim$\SI{2149}{\deg\squared}, asymmetrically split into 824, 847, 153, and \SI{325}{\deg\squared} for respectively one, two, three, and four overlapping camera groups;][]{pertenais2021unique}. This field is so large that all the targets to be observed must be decided upon prior to the start of the observations, because full-field imaging is not feasible. While the targets for the \plato{} Core Science programme for LOPS2 have been fixed \citep{montalto2026}, those for PLATO-CS will be decided by ESA based on the response to its calls for GO proposals\footnote{\url{https://www.cosmos.esa.int/web/plato/ao-1}}. Combined with high-cadence and continuous photometric monitoring, \plato{} will allow studying AGN variability over a broad range of time scales. This will be suitable to disentangle photometric features of SMBH(B)s. Thus, as a follow-up observatory, \plato{} has the potential to play an essential role in SMBHB searches of ongoing and future surveys, such as the Legacy Survey of Space and Time \citep[LSST;][]{ivezic2019lsst}. 
 
In this paper, we assemble a \textit{Plato} quasar catalogue for the two planned LOPs during the nominal mission (Sect.~\ref{sec:catalogue}). We present the photometric SMBHB model used in our simulations in Sect.~\ref{sec:model}, detail our end-to-end simulation study in Sect.~\ref{sec:simulations}, and present our results on Bayesian inference of Spikey-like objects in Sect.~\ref{sec:results}. We discuss our results and future prospects in Sect.~\ref{sec:discussion} and offer concluding remarks in Sect.~\ref{sec:conclusions}.
 
\section{A \textit{Plato} catalogue of quasars}\label{sec:catalogue}

We start by assembling a catalogue of bright quasars that may be observed with one or more of \plato{}'s cameras and that fall on CCD silicon. This catalogue was used for the first \plato{} GO proposal (AO-1)$^1$ and will be used for further simulation studies before mission launch.

\subsection{Catalogue definition}\label{sec:catalogue_creation}

We create a catalogue of high-probability AGNs from the two LOPs using the \textit{Plato} camera simulator \citep[\platosim{};][]{jannsen2024platosim}. Figure~\ref{fig:aitoff_projection} shows the resultant sources plotted in a galactic projection. The white-to-darkblue patterns illustrate the N-CAM visibility and the overlaid coloured markers are time-domain candidate SMBHBs from the literature.

We start from the third data release \citep[DR3;][]{gaia2023dr3} of the \textit{Gaia} mission \citep{gaia2016mission} to query candidate AGNs in each of the two LOPs with a \textit{Gaia} magnitude below $G<19$. The reason for using \textit{Gaia} and not a more standardised quasar catalogue is that \textit{Gaia} and \plato{} share a similar passband, and secondly, \textit{Gaia} is the only magnitude limited all-sky survey that provides the necessary astrometric information of nearby sources which are needed to perform our pixel-based simulations. While Appendix~\ref{app:cat} provide a detailed catalogue definition, the main steps are: (i) to apply effective sample purity cuts using \textit{Gaia} photometric and spectroscopic observations, (ii) removing spurious astrometric solutions of \textit{Gaia} using a astrometric reliability diagnostic c.f. \cite{rybizki2022classifier}, (iii) differentiate quasars from other field sources using variability metrics c.f. \cite{butler2011optimal}, and (iv) cross-match our sample with the all-sky \textit{Gaia}-unWISE spectroscopic quasar catalogue (Quaia) of \cite{storeyfisher2024quaia}. Quaia effectively accounts for extinction due to its integrated ground-based spectroscopic observation, hence, further justifying the usage of \textit{Gaia}.

As a result of the cross-matching and the above cuts, we find a total of 12,226 high-probability quasars (5660 in the LOPS2 and 6566 in the LOPN1). Figure~\ref{fig:aitoff_projection} shows that the AGN classification itself is sensitive to extinction, with fewer sources near the galactic plane or in the vicinity of the LMC (which explains the lower count of quasars in the LOPS2). 

\begin{figure}[t!]
\center
\includegraphics[width=\columnwidth]{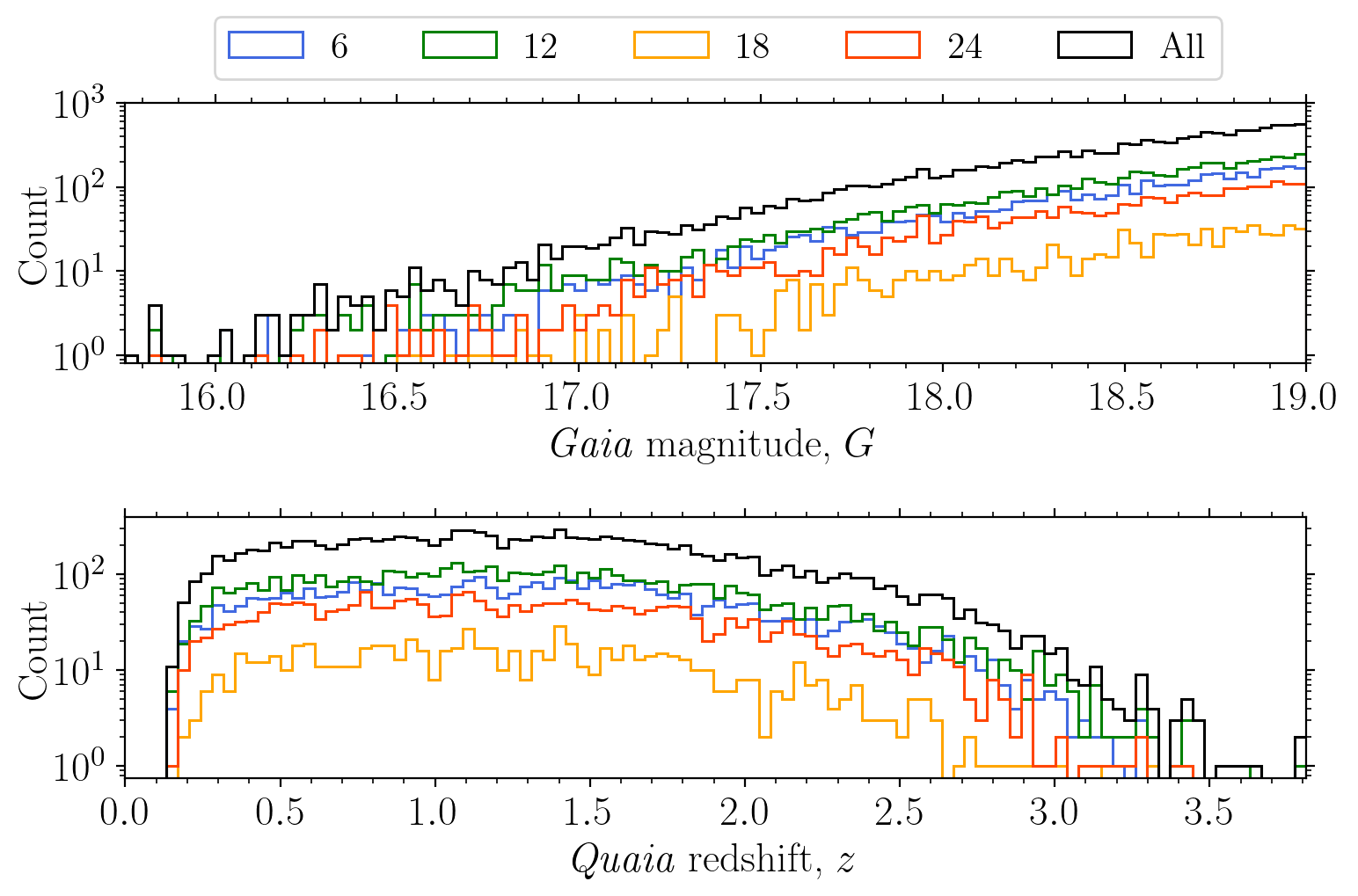}
\caption[]
{Distributions of \textit{Plato} magnitude (top panel) and \textit{Quaia} redshift (bottom panel) for the \textit{Plato} catalogue of quasar candidates. The solid black line represents all quasars, while the blue, green, orange, and red solid lines represent quasars observable with 6, 12, 18, and 24 N-CAMs, respectively. The N-CAM visibility, $\tx{n}{CAM}$, is set by \plato{}'s complex footprint in the sky and controls the noise budget to first order (see Sect.~\ref{sec:catalogue_noise}).} 
\label{fig:histogram_quasars}
\end{figure}

Figure~\ref{fig:histogram_quasars} shows the magnitude distribution (top panel) and the redshift distribution (lower panel) for the total (black lines) and individual (coloured lines) camera visibilities of our final quasar catalogue. The magnitude distribution follows a classical power law (expected for an increased volume of isotropically distributed sources), counterbalanced by the finite time of their formation (the so-called redshift cutoff). The lower panel shows that Quaia's redshift estimates agree well with a main quasar formation epoch of $z\sim 0.3$--3 \citep{shaver1996decrease}, which corresponds to a look-back time of $t_z \sim 3.5$--\SI{11.7}{\giga\year}. 

While different types of AGN (blazars, quasars, LINER, seyfert-I, seyfert-II, BLRG, NLRG, etc.) show a wide range of spectral energy distributions (SEDs), the rest-frame SED of quasars are surprisingly universal since the cosmic noon \citep[i.e. $z \lesssim 3$;][]{cai2024composite}. Unlike a simple blackbody spectrum, quasars are typically brightest at shorter wavelengths. Hence, a blue passband is ideal for observations of nearby quasars, while a redder passband is better suited for quasars at higher redshift. Thus, for the bulk of our quasar candidates, the apparent source intensity will be slightly higher in the \textit{Plato} $\mathcal{P}$-passband compared to \textit{Gaia}'s mean $G$-passband. For example, a typical quasar spectrum transmitted at a redshift of $z = 1$ will be $(\mathcal{P}-G)_0 \approx 0.12$ brighter with \textit{Plato} compared to \textit{Gaia} after extinction correction. A generic quasar spectrum \citep[cf.][]{cai2024composite} and \textit{Gaia}'s colour information is used to convert to $\mathcal{P}$.

\subsection{Noise considerations}\label{sec:catalogue_noise}

Since \textit{Plato} was designed to observe bright sources, we investigate the underlying noise budget of our AGN catalogue in the faint limit. This was done by producing \SI{1}{\day} duration light curves for each non-variable AGN observable with  $\tx{n}{CAM}$ cameras and computing the noise-to-signal ratio (NSR) over \SI{1}{\hour} bins. 

\begin{figure}[t!]
\center
\includegraphics[width=\columnwidth]{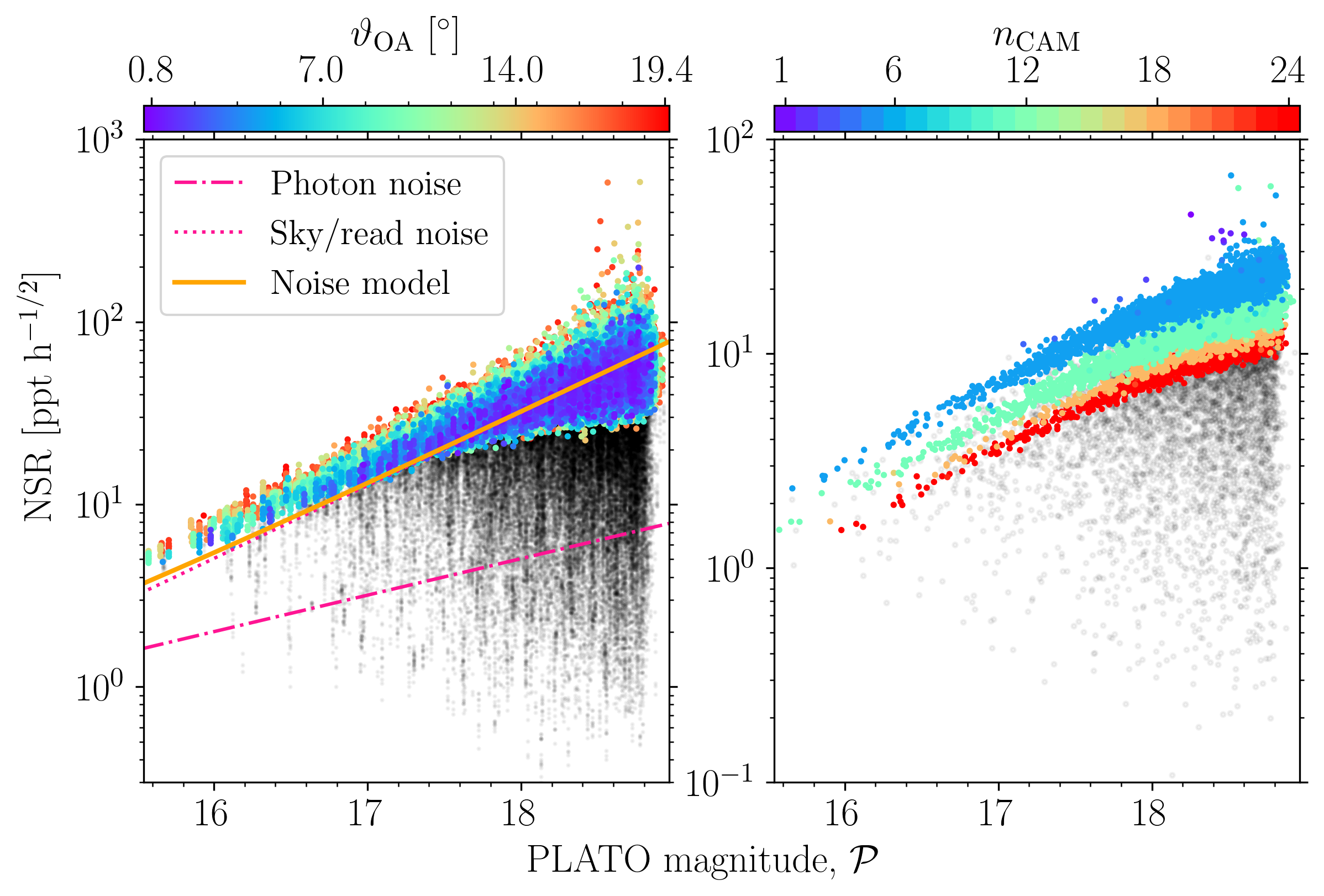}
\caption[]
{N-CAM noise budget of our quasar catalogue for beginning-of-life conditions. Left: camera-level NSR estimates colour-coded after (distorted) gnomonic radial distance from the optical axis, $\vartheta_{\rm OA}$. Shown is a theoretical noise prediction (orange solid line), constituted by a jitter noise limit (negligible here), photon noise limit (pink dashed-dotted line), and the sky/read noise limit (pink dotted line). Right: mission-level NSR estimates colour-coded after camera visibility, $n_{\rm CAM}$. The black points in both panels are quasars with $\text{SPR}>0.01$. This study was used to help shape the guidelines of the first \plato{} GO call: \url{https://www.cosmos.esa.int/web/plato/recipes}.} 
\label{fig:nsr_quasars}
\end{figure}

As shown in Fig.~\ref{fig:nsr_quasars}, we followed the methodology of \cite{jannsen2024platosim} and computed the NSR as a function of magnitude at camera level (left panel) and at mission level (right panel). We highlight that the photometric extraction was done using the on-board photometry module, which computes a stellar pollution ratio \citep[SPR;][]{marchiori2019flight}. The SPR is a key unit metric to estimate how much third-light leaks into the on-board aperture mask. In the faint limit, contamination is increasingly suppressing source variability. In Fig.~\ref{fig:nsr_quasars} we illustrate this by plotting all sources with $\text{SPR}>0.01$ in black markers, clearly pinpointing sources with a large NSR scatter (downwards due to a clear overestimation of the mean flux level).
 
At camera level (left panel of Fig.~\ref{fig:nsr_quasars}), each NSR value is determined from the individual camera light curves. Thus, each source has a unique barycentric position in the 4-CCD focal plane array (FPA) and gnomonic radial distance from the optical axis ($\vartheta_{\rm OA}$, see colour bar). As each \textit{Plato} camera utilises axi-symmetric (refractive) dioptics, a decreasing gradient in transmission efficiency with $\vartheta_{\rm OA}$ enforces an increasing NSR gradient. In the same panel we show the theoretical camera noise model for $\vartheta_{\rm OA}=\SI{0}{\degree}$ (orange solid line), which consists of three noise contributions related to spacecraft jitter (bright end; not visible), sky background and read noise (faint end; pink dotted line), and photon noise (interlacing region; pink dashed-dotted line). As expected, we see that the noise budget of our simulations is dominated by sky background and detector readout noise, while photon noise is negligible except for the brightest sources. 

At mission level (right panel of Fig.~\ref{fig:nsr_quasars}), each NSR value is calculated from a final reduced multi-camera light curve. We therefore represent each data point according to the N-CAM visibility. Beyond $\Pb \gtrsim 17.5$, we observe an increasing deviation of NSR compared to the simple noise model. While source visibility can take any value up to 24 N-CAMs (due to camera and CCD misalignment), the NSR estimate is naturally dominated by the expected visibilities, $\tx{n}{CAM} \in \{6, 12, 18, 24\}$. Just as the mission level NSR estimates suffer from the same underlying noise effects as at camera level, each NSR--visibility curve shows a similar signal suppression beyond $\Pb \gtrsim 17.5$.

While signal suppression in the faint end is typically associated with detector systematics and charge redistribution effects (e.g. charge-transfer inefficiency \citep[][for \plato{}]{mishra2026impact}, gain non-linearity, or thermally induced charge from a temporal and spatial dark current), the observed deviation for $\Pb > 17.5$ is attributed to digitalisation noise. Thus, at beginning-of-life (BOL), \textit{Plato}'s limiting magnitude is around $\Pb \sim 17.5$, depending on the technical details (such as the $\tx{\theta}{OA}$ and intra-pixel position). While it is clear that any NSR estimate below the theoretical noise model is an underestimation of the true NSR, Fig.~\ref{fig:nsr_quasars} shows an indication that $\Pb \sim 18$-19 is \plato{}'s ultimate detection limit. While this limit depends on the signal amplitude, we show in Sect.~\ref{sec:results_observation} that percent-level variability (such as that of Spikey), the detection limit is close to $\Pb \sim 18$ for mission BOL. 

\subsection{Bright quasar candidates}\label{sec:catalogue_bright}

\begin{figure}[h!]
\center
\includegraphics[width=\columnwidth]{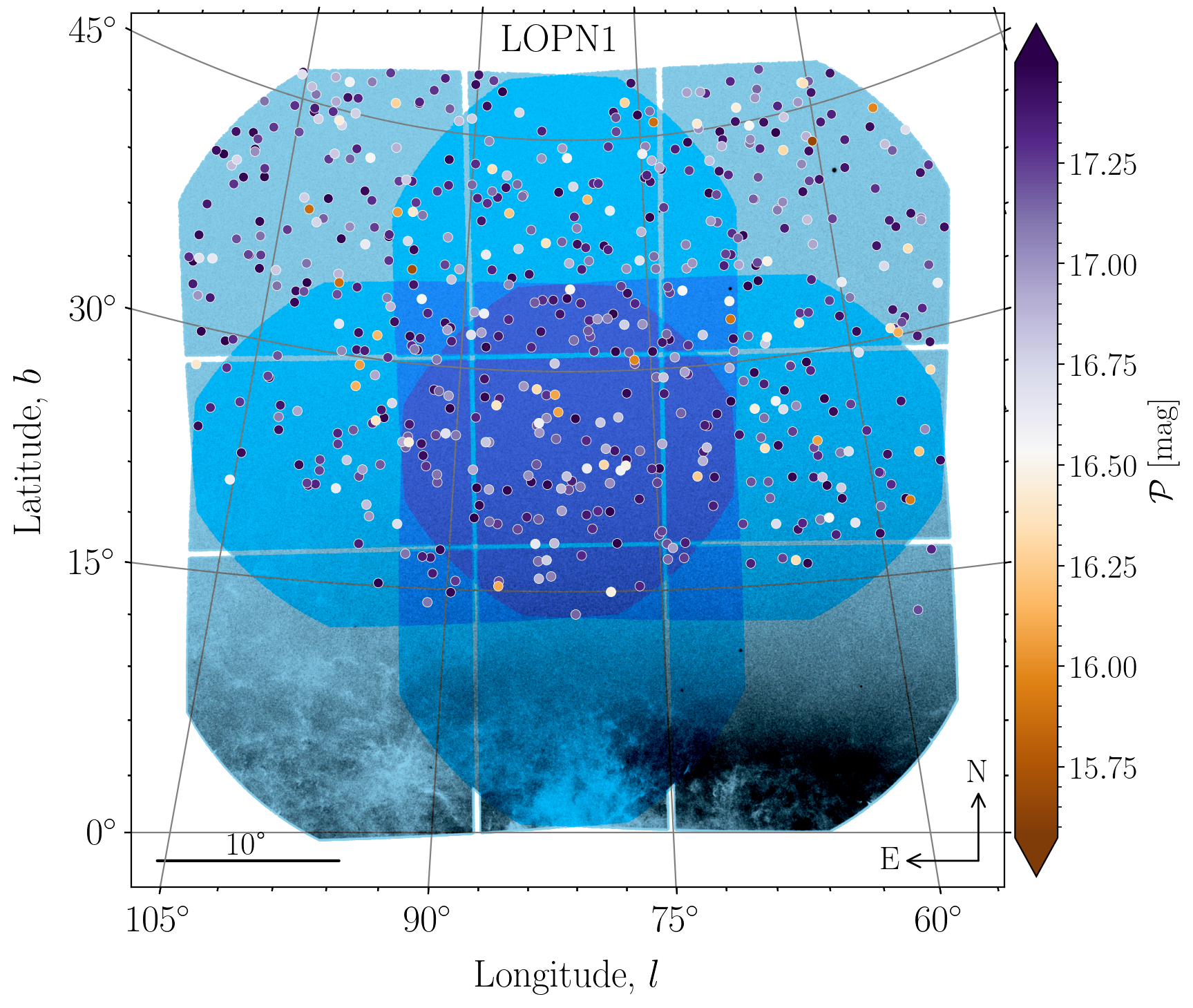}
\includegraphics[width=\columnwidth]{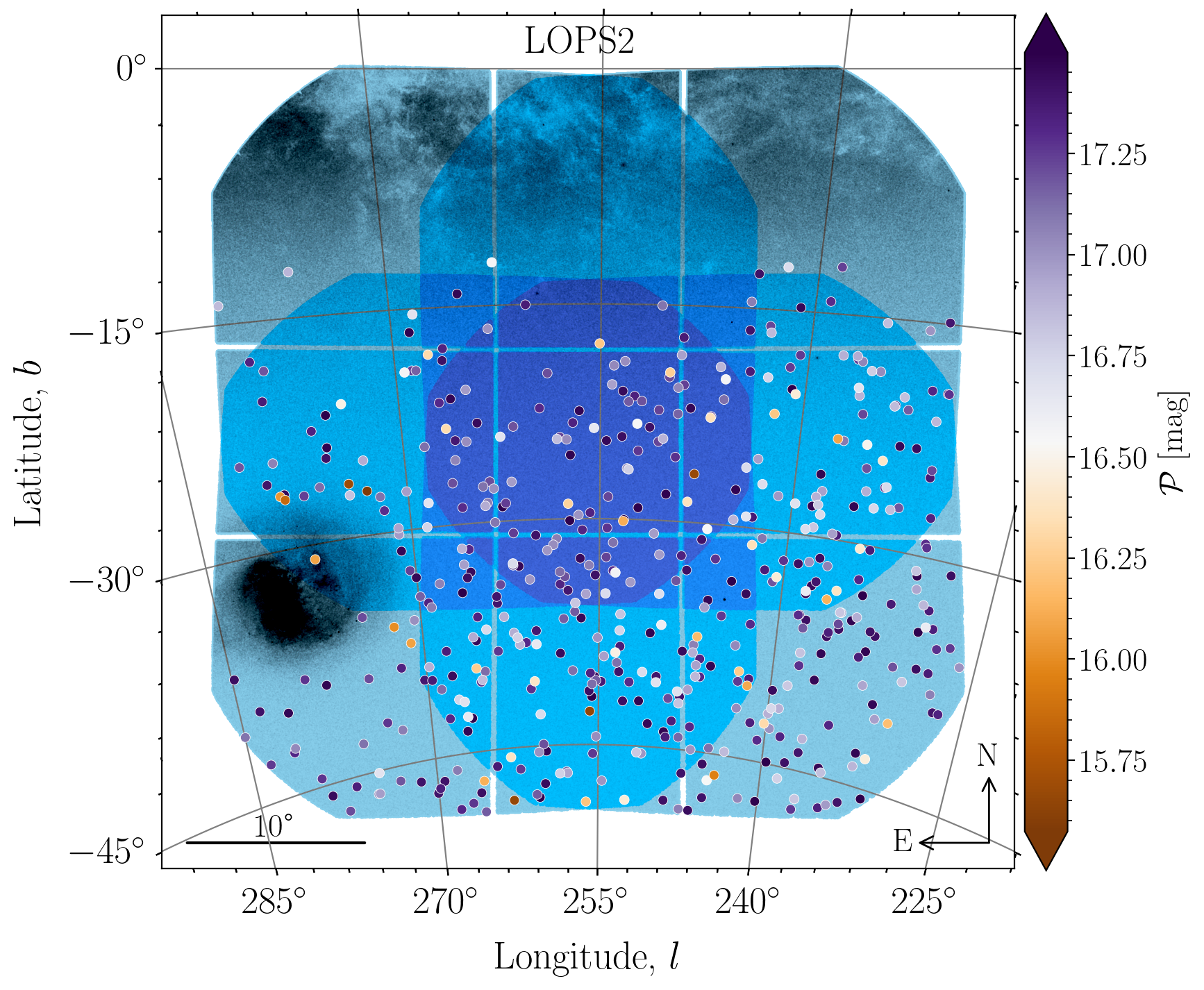}
\caption[]
{Illustration of the two nominal \plato{} fields called LOPN1 (top panel) and LOPS2 (bottom panel) shown in galactic coordinates $(l, b)$. The N-CAM footprint of $\tx{n}{CAM} \in \{6, 12, 18, 24\}$ co-pointing cameras is illustrated with an increasingly darker shade of blue, while the bright subset of high-probability quasars are coloured after magnitude (up to $G<17.5$). The black transparent map highlights dense stellar sky regions such as the location of the Milky Way plane, the Large Magellanic Cloud (within the LOPS2), and globular/open star clusters.}
\label{fig:pointing_fields}
\end{figure}

With the above limiting magnitude in mind, we may consider the number of sources in our catalogue that would be optimal for future follow-up studies. Within the  LOPS2 (LOPN1), our catalogue contains 1452 (1732) sources for $\Pb<18$, 494 (600) sources for $\Pb<17.5$, and 170 (177) sources for $\Pb<17$. Figure~\ref{fig:pointing_fields} shows the sky location of quasars brighter than $\Pb<17.5$ within the LOPN1 (top) and LOPS2 (bottom). We discuss the implications of the number statistic above on quasar-related science investigations in Sect.~\ref{sec:discussion}.

Our bright quasar catalogue is a helpful tool for \plato{} GO proposals. Along with the public \plato{} variability catalogue available in \citet{Kliapets2026}, it can used by the astronomical community to select and refine GO targets.

\section{Photometric model}\label{sec:model}

We simulate the observational signature of SMBHBs with three model components contributing to the mean flux level: (i) intrinsic stochastic quasar variability, $\Delta F_{\rm Quasar}$, (ii) Doppler boosting, $\Delta F_{\rm Doppler}$, and (iii) gravitational self-lensing, $\Delta F_{\rm Lensing}$. The two latter model components are deterministic due to their relativistic nature. Defining the primary black hole as the more massive object, with the mass ratio $q \equiv M_2/M_1 \leq 1$, we have
\begin{align}\label{eq:F_model}
F / F_0 
&= (1 + \Delta F_{\rm Quasar}) \, (1 + \Delta F_{\rm Doppler}) \, (1 + \Delta F_{\rm Lensing}) \nonumber \\
&= 1 + (1-L_{\nu}) \, Q_1 \, D_1^{3-\alpha_{\nu,1}} M_1 + L_{\nu} \, Q_2 \, D_2^{3-\alpha_{\nu,2}} \, M_2 \,.
\end{align}
Here $Q_1$ ($Q_2$) is the AGN accretion variability factor of the primary (secondary), $D_1$ ($D_2$) is the line-of-sight (LOS) Doppler factor of the primary (secondary), and $M_1$ ($M_2$) is the self-lensing magnification factor when the secondary (primary) acts as the lens. Following \citetalias{hu2020spikey}, we have introduced the luminosity ratio $L_{\nu} \equiv L_{\nu,2} / (L_{\nu,1} + L_{\nu,2})$ as a scale factor for the light contribution of each source. 

Equation~\eqref{eq:F_model} assumes that the mini-discs of the central black holes outshine the circum-binary disc from which they are accreting. We elaborate on this assumption in Sect.~\ref{sec:discussion}. To avoid complex hydrodynamical simulations, we adopt the model of \cite{dorazio2018periodic} describing a steady-state, optically thick, geometrically thin accretion disc. In this formalism, the spectral index $\alpha$ describes the spectral slope at the peak emission frequency of such a disc. The spectral indices of the two SMBHs are generally different, however, we assume they are identical to simplify our future model framework. Furthermore, we choose to ignore chromatic effects when considering the luminosity ratio, which reduces its expression to $L \equiv L_2 / (L_1 + L_2)$. However, this is not always a good approximation for SLFs, and we therefore discuss its impact in Sect.~\ref{sec:discussion}. Lastly, we may ignore higher-order terms between the quasar variability, modelled as a damped random walk (DRW), and relativistic effects for small changes in the relative flux (thus modelling $Q_1 + Q_2 = Q$). With these assumptions, the model reduces to  
\begin{equation}\label{eq:QDM}
F / F_0 = 1 + Q + (1-L) \, D_1^{3-\alpha} \, M_1 + L \, D_2^{3-\alpha} \, M_2 \equiv \mathcal{Q\,DM} \,.
\end{equation}
The pure relativistic (Doppler and SLF) model of Eq.~\eqref{eq:QDM} is
\begin{equation}\label{eq:DM}
F / F_0 = 1 + (1-L) \, D_1^{3-\alpha} \, M_1 + L \, D_2^{3-\alpha} \, M_2 \equiv \mathcal{DM} \,,
\end{equation}  
while the AGN variability model is simply
\begin{equation}\label{eq:Q}
F / F_0 = 1 + Q \equiv \mathcal{Q} \,.
\end{equation}  
The models of Eq.~(\ref{eq:QDM}--\ref{eq:Q}), and their short-hand definitions $\{\mathcal{Q\,DM},\,\mathcal{DM},\,\mathcal{Q}\}$, are used for the analysis in Sect.~\ref{sec:results}. We elaborate on each model component (i.e. $\mathcal{Q}$, $\mathcal{D}$, $\mathcal{M}$) in Appendix~\ref{app:model}. 

For all forthcoming simulations, we fix the barycentric LOS velocity to $v_z = 0$, the spectral index of the two mini-discs to $\alpha = 2.09$ (c.f. the mean posterior values of \citetalias{hu2020spikey}), and the longitude of the ascending node to $\Omega=\pi/2$.

\section{End-to-end mock simulations}\label{sec:simulations}

With the quasar catalogue of Sect.~\ref{sec:catalogue} and the SMBHB model template introduced in Sect.~\ref{sec:model}, we now present how all simulated \plato{} light curves analysed in Sect.~\ref{sec:results} were generated and post-processed.

\subsection{Simulated light curves}\label{sec:sims_platosim} 

\begin{figure}[t!]
\center
\includegraphics[width=0.85\columnwidth]{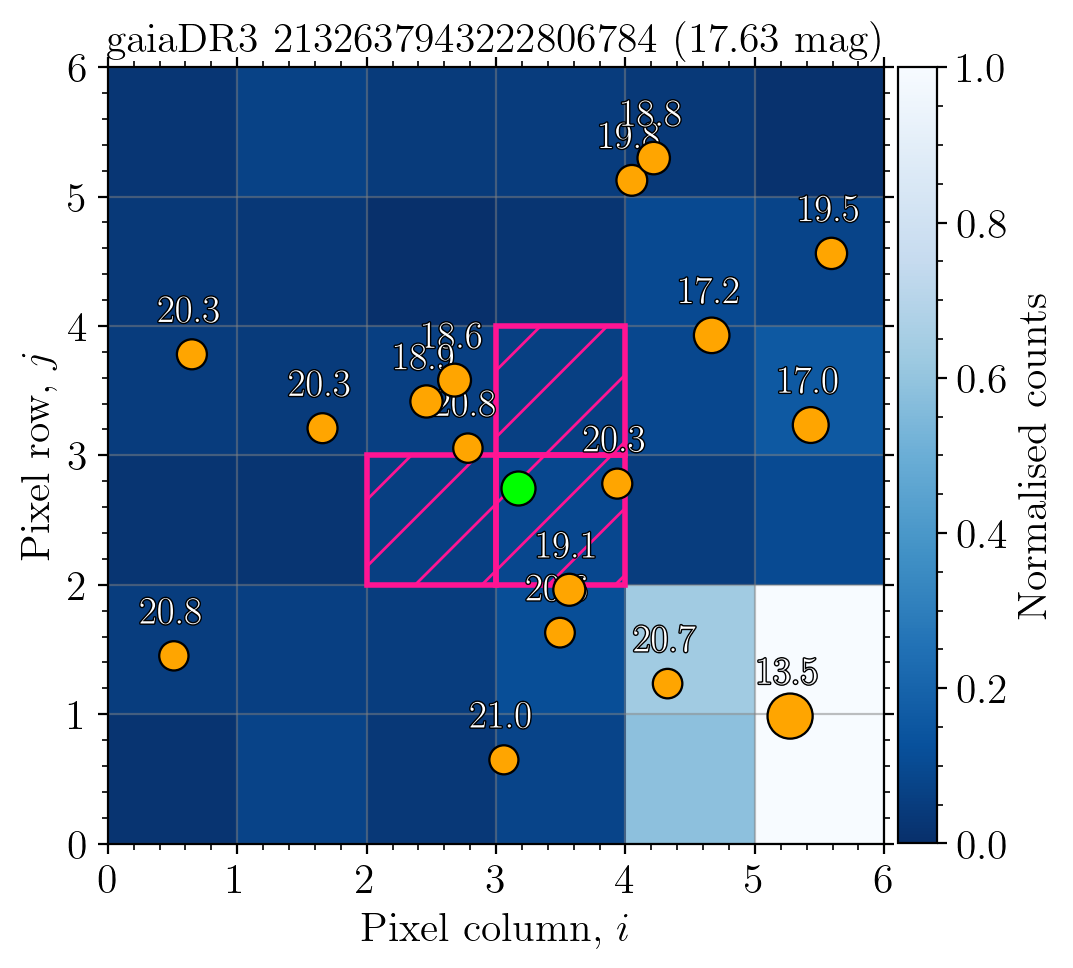}
\caption[]
{\platosim{} simulation of a $6\times6$ imagette for the SMBHB candidate Spikey. The location of Spikey is shown with a green-filled circle. Nearby contaminating sources are shown with orange-filled circles that are scaled linearly in size according to their magnitude (written above each circle) relative to that of the target. The pink-hashed pixels show the on-board aperture mask.} 
\label{fig:imagette}
\end{figure}

We use \platosim{} to produce imagery time series of $6\times6$ pixels, while injecting a $\mathcal{Q\,DM}$ model template of Eq.~\eqref{eq:QDM} into the analytic \plato{} PSF. Figure~\ref{fig:imagette} illustrates the imagette of Spikey seen by a single \plato{} camera taken with an exposure time of \SI{25}{\second}. Even though Spikey is a relatively bright quasar of $\Pb\sim17.6$, it is clear that stellar contamination will be a challenge for \textit{Plato} imagery compared to \textit{Kepler} due to the difference in plate scales as we discuss in the following.

At run-time, the on-board aperture photometry algorithm \citep{marchiori2019flight} is applied to extract the photometry. While the extraction of \textit{Kepler} light curves of quasars benefited from custom aperture masks \citep{smith2016finding}, a similar strategy for \plato{} can only be accomplished if pixel data is requested (which is far more expensive in terms of telemetry). However, since \plato{} has a much larger plate scale compared to \textit{Kepler}%
\footnote{Specifically being $\SI{15}{\arcsec\per\pixel}/\SI{3.98}{\arcsec\per\pixel} = 3.78$ times larger c.f. \cite{borucki2010kepler}.}, %
and since poor (or failing) PSF photometry performance is expected for faint sources \citep{mitchell2026following}, the added benefit of pixel data is small. 

For each target source, we simulate $\tx{n}{CAM}$ raw light curves for a total duration of four years. The latter was done to test the SMBHB detectability in the case that the nominal \SI{2}{\year} LOPs are extended to the full mission duration and that a (potential) \SI{4}{\year} mission extension will take place.%
\footnote{For a 24 N-CAM observation this amounts to 192 (384) individual mission quarter light curves for a \SI{2}{\year} (\SI{4}{\year}) baseline.} %
With a current launch date set for early January 2027, and with an expected commissioning phase of three months, we assume that \textit{Plato} will begin its nominal phase around the beginning of April 2027. With this in mind for all source simulated, we account for time-dependent effects that slowly reduce the optical throughput by setting the start date of the LOPS2 (LOPN1) field to the beginning of April 2027 (2029), equivalent to \SI{0.25}{\year} (\SI{2.25}{\year}) after mission BOL.

\subsection{Reduction pipeline}\label{sec:sims_reduction}

To post-process each light curve, we have implemented a small reduction pipeline that can robustly combine multi-camera observations in the presence of intrinsic AGN variability. To alleviate our pipeline's computational complexity, we only include minor (expected) camera drifts due to thermo-elastic distortion. The pipeline first handles all light curves from each mission quarter separately by averaging data points of the same time-stamp, and then bin the data per \SI{1}{\hour}. Next, all quarter light curves are stitched to each other following \cite{handberg2014automated}, by running a LOWESS regression \citep{cleveland1979robust, cleveland1981lowess} on a \SI{10}{\day} data segment before and after each mission quarter gap and calculating the Theil--Sen median slope \citep{theil1950rank, sen1968estimates} on the smoothed signal. Any residual signal jump is found by evaluating the Theil--Sen slope at the times on either side of the time midpoint between mission quarters and adding the difference to all time stamps after the mission quarter gap. Next, a second-order polynomial is used on the residual mission-level light curve to correct for long-term trends due to the ageing effects of the camera optics and CCDs. The uncertainty on the flux measurements is computed as in \cite{jannsen2025mocka}, accounting for the internal scatter in the time series (affected by previous reduction steps). Lastly, to correct for gain non-linearity and digitalisation, an inverse normalisation using the second-order polynomial trend is applied. A \SI{1}{\day} binned version of the light curve is stored for further analysis.

\subsection{Bayesian model inference}\label{sec:sims_bayesian}

We recover the simulated system parameters (shown in Table~\ref{tab:priors}) using two Bayesian inference techniques: first, with a traditional Markov chain Monte Carlo (MCMC) sampling technique implemented in \texttt{NumPyro} \citep{phan2019composable, bingham2019pyro} and secondly using nested sampling with the package \texttt{JAXNS} \citep{albert2020jaxns} and \texttt{UltraNest} \citep{buchner2021ultranest}. The choice of nested sampling was made to fit an arbitrary model and account for multimodal or non-Gaussian parameter spaces (expected for DRW modelling). After thorough testing, \texttt{JAXNS} was used to produce all future results and plots due to its superiority in robustness and speed. We use 2000 live-points with a minimum sample of 10,000 maximum likelihood evaluations throughout this study. The termination condition is set to when the remaining evidence is $\text{d}\ln z < 10^{-4}$ of the current total evidence, $z$.

For model evaluation we use the maximum likelihood function: $\ln \mathcal{L}(y_n|\{x_i\}, m_n, \sigma_n|)$ for normal distributed data, where $\{x_i\}$ are model parameters, $m_n$ is the model given $\{x_i\}$, $y_n$ is the light curve data, and $\sigma_n$ is the photometric standard error. For modelling, we fix the barycentric velocity to $v_z = 0$ (equivalent to Spikey) as our focus is on model comparison. As shown in Table~\ref{tab:priors}, we use wide uninformed flat priors for all $\mathcal{DM}$ model parameters. To fit the $\mathcal{Q}$ model, a Gaussian process with an exponential covariance kernel is employed (being the definition of the DRW). As each light curve contains prior knowledge of the typical DRW amplitude $\sigma$, we use a log-normal prior centred on the observed standard deviation. While a DRW time scale has been established from samples of AGNs \citep[e.g.][]{vaughan2016false}, we use our prior knowledge from \kepler{} and a log-normal prior centred on \SI{100}{\day} with twice a standard deviation. From testing, the results are barely affected by the value of the prior.  

Lastly, we perform a model comparison by employing our nested sampling on the model (i) $\mathcal{Q}$, (ii) $\mathcal{DM}$, and (iii) $\mathcal{Q\,DM}$. We assess which model is favoured by the data using the Bayesian inference $\ln z$. Following \cite{kass1995bayes}, we use $(\ln z_1 - \ln z_2) > 5$ as criterion to select between models.

\section{Results}\label{sec:results}

In this section we investigate if \plato{} can deliver photometric measurements allowing for the detection of SMBHB signatures. We first revisit the SMBHB candidate `Spikey' observed by the \kepler{} mission in Sect.~\ref{sec:results_spikey}, while \plato{}'s ability to discover potential Spikey-like SMBHB photometric signatures is presented in Sect.~\ref{sec:results_observation}.

\subsection{The case study of Spikey}
\label{sec:results_spikey}

As a proof-of-concept of \textit{Plato}'s ability to observe SMBH binarity, we consider the \textit{Kepler} candidate discovered by \citetalias{hu2020spikey}, called Spikey (KIC\,11606854). With a mean magnitude in the \kepler{} (\plato{}) passband of $\tx{K}{P} = 17.75$ ($\Pb=17.63$), Spikey was one out of 21 AGNs discovered in the \kepler{} long-cadence (\SI{30}{\min}) data by \citetalias{smith2018kepler}. Thanks to \kepler{}'s high photometric precision, \citetalias{smith2018kepler} first recognised the flare-like event but it remained a mystery: the flare duration was too short to be a supernova afterglow and its decaying (exponential) feature was inconsistent with a traditional tidal disruption event within a standard SMBH accretion disc. \citetalias{hu2020spikey} performed probabilistic modelling of Spikey, and found that the \kepler{} light curve is very well fit by a self-lensing, unequal-mass SMBHB system with an eccentric orbit. In the observer's frame (i.e. accounting for the redshift $z=0.918$)\footnote{Since $z=0.918$ (c.f. Table~1; \citetalias{smith2018kepler}) is well aligned and within the errors of $z=0.92(8)$ from Quaia, we adopt this value to be consistent with the study of \citetalias{hu2020spikey}.}, the orbital period is $T=P(1+z)=\SI{2.194}{\year}$. This period did not allow a detection of two SLFs during the $\sim$\SI{2.9}{\year} duration (i.e. $\sim$11.5 mission quarters) of the \textit{Kepler} observation, to fully confirm the SMBHB nature unambiguously. 

Spikey has been followed up by the Chandra X-ray observatory \citep{sorabella2022chandra} and with the very long baseline interferometer (VLBI) in radio \citep{kun2020spikey}. While the VLBI images did show a wiggled jet, consistent with the binary hypothesis, no self-lensing spike was observed in either wavelength regimes. While an increasing timing uncertainty of the expected SLF occurrence can explain this discrepancy, the data quality, duration, and cadence of these two surveys are also dwarfed by the \textit{Kepler} observations. For example, \cite{sorabella2022chandra} showed that the low cadence of the Chandra observation cannot rule out a \SI{10}{\day} flare. However, for the VLBI observation, \cite{kun2020spikey} mentioned that it is uncertain whether the radio signal is being emitted from the jet, and so it may not be subject to self-lensing.

Spikey will be visible in the LOPN1 and observable with all 24 N-CAMs. Hence, it is an ideal target to follow up with \textit{Plato}. Since the first SLF was observed by \textit{Kepler} around 55750 MJD, using the observed orbital period, the next lensing flare within the LOPN1 observational window (i.e. 8th since \textit{Kepler}) is expected to occur around June 21, 2029. With a scheduled launch date of January 1, 2027 and a \SI{90}{\day} commissioning phase, the ephemeris of the SLF (measured from the beginning of the LOPN1) is $t_0=\SI{0.905}{\year}$. Propagating the error of the observed period and accounting for the relativistic orbital precession of $\delta\omega \sim \SI{1.3}{\degree}$ per orbit \citepalias[altering the SLF timing with $\delta t_0 \sim \SI{1.5}{\day}$ per orbit;][]{hu2020spikey}, amounts to a relative uncertainty of the first predicted LOPN1 event to $t_0 \sim 0.9\pm\SI{0.3}{\year}$. Thus, if Spikey is indeed an SMBHB, then \plato{} will securely capture at least one SLF.

\tabcolsep=5pt
\begin{table*}
\caption[]{Model parameters, priors, inferences, and uncertainties from \texttt{JAXNS} nested sampling of Spikey.}
\makebox[\linewidth][c]{%
\tiny
\begin{tabular}{lllc|cc|cc|cc}
\hline\hline
Parameter & Meaning & Prior & Injected & \multicolumn{2}{c}{\textit{Kepler} result} & \multicolumn{2}{c}{\plato{} result} &  \multicolumn{2}{c}{\textit{Kepler} + \plato{}} \\
		  & 		&    	& (c.f. \citetalias{hu2020spikey}) & Case I & Case II & Case I & Case II & Case I & Case II \\
\hline
$\tau$		 [day]				& DRW damping time scale	& $\mathcal{LN}(0, 10^3)$ & 31 		& \pmx{124}{-37}{+79} 			& \pmx{169}{-31}{+62}
																								& \pmx{156}{-57}{+105} 			& \pmx{143}{-35}{+188} 
																								& \pmx{135}{-36}{+63}			& \pmx{211}{-5}{+5} \\
$\sigma$	 [ppt]				& DRW variable scale		& $\mathcal{LN}(0, 10^3)$ & 10   	& \pmx{17}{-3}{+5} 				& \pmx{12}{-2}{+2}
																								& \pmx{22}{-4}{+6}				& \pmx{10}{-1}{+6}
																								& \pmx{18}{-3}{+4}				& \pmx{13.6}{-0.3}{+0.4} \\\cmidrule{5-5}\cmidrule{7-7}\cmidrule{9-9}%
$t_0$ 		 [yr] 				& Time of ephemeris 		& $\mathcal{U}(0, 3)$	 & 1.050	& \pmx{1.042}{-0.001}{+0.001} 	& \pmx{1.07}{-0.02}{+0.01}
																								& \pmx{2.58}{-0.02}{+0.02}		& \pmx{0.89}{-0.03}{+0.03}
																								& \pmx{1.046}{-0.001}{+0.001} 	& \pmx{1.045}{-0.004}{+0.004} \\
$P$ 		 [yr]				& Orbital rest-frame period & $\mathcal{U}(0, 5)$ 	 & 1.144  	& \pmx{1.130}{-0.004}{+0.004} 	& \pmx{3.3}{-1.3}{+0.7}
																								& \pmx{0.88}{-0.01}{+0.01}		& \pmx{0.9}{-0.1}{+0.9}
																								& \pmx{1.14369}{-0.00004}{+0.00004}& \pmx{3.0499}{-0.0001}{+0.0001} \\
$i$ 		 [deg]				& Orbital inclination		& $\mathcal{U}(0, 90)$ 	 & 81.952 	& \pmx{85.1}{-0.2}{+0.2} 		& \pmx{59}{-19}{+16}
																								& \pmx{81.8}{-1.0}{+0.8}		& \pmx{80}{-3}{+2}
																								& \pmx{83.9}{-0.2}{+0.1}		& \pmx{80.5}{-0.6}{+0.4} \\
$e$								& Orbital eccentricity 		& $\mathcal{U}(0, 1)$ 	 & 0.524  	& \pmx{0.418}{-0.006}{+0.006} 	& \pmx{0.92}{-0.10}{+0.02}
																								& \pmx{0.42}{-0.02}{+0.02} 		& \pmx{0.77}{-19}{+18}
																								& \pmx{0.457}{-0.006}{+0.005}	& \pmx{0.767}{-0.008}{+0.010} \\
$\omega$ 	 [deg]				& Argument of periapse 		& $\mathcal{U}(0, 360)$  & 84.626 	& \pmx{80.6}{-0.5}{+0.5} 		& \pmx{119}{-44}{+17}
																								& \pmx{78}{-2}{+2}				& \pmx{85}{-63}{+22}
																								& \pmx{81.5}{-0.4}{+0.5}		& \pmx{77}{-3}{+2} \\
$\log M_1$ 	 [$\log \sole{M}$] & Primary binary mass 		& $\mathcal{U}(5, 11)$	 & 7.4 	  	& \pmx{7.23}{-0.06}{+0.07} 		& \pmx{9.3}{-0.6}{+0.4}
																								& \pmx{7.9}{-0.2}{+0.1}			& \pmx{8.0}{-0.2}{+0.3}
																								& \pmx{7.31}{-0.02}{+0.03} 		& \pmx{7.84}{-0.04}{+0.06} \\
$\log M_2$	 [$\log \sole{M}$] & Secondary binary mass		& $\mathcal{U}(5, 11)$	 & 6.7 	  	& \pmx{6.71}{-0.07}{+0.07} 		& \pmx{7.2}{-1.1}{+0.5}
																								& \pmx{6.0}{-0.3}{+0.3} 		& \pmx{6.9}{-1.3}{+0.7}
																								& \pmx{6.94}{-0.03}{+0.03}		& \pmx{6.78}{-0.06}{+0.07} \\ 
$\alpha$						& Spectral index 			& $\mathcal{U}(-4, 4)$ 	 & 2.09   	& \pmx{-0.3}{-0.4}{+0.2} 		& \pmx{2.6}{-0.5}{+0.1}
																								& \pmx{1.5}{-0.3}{+0.3}			& \pmx{1.8}{-0.6}{+0.4}
																								& \pmx{0.97}{-0.05}{+0.04}		& \pmx{2.31}{-0.08}{+0.08} \\
$L$								& Binary luminosity ratio	& $\mathcal{U}(0, 1)$ 	 & 0.89   	& \pmx{0.44}{-0.06}{+0.06} 		& \pmx{0.26}{-0.07}{+0.14}
																								& \pmx{0.4}{-0.1}{+0.1}			& \pmx{0.6}{-0.1}{+0.1}
																								& \pmx{0.618}{-0.005}{+0.005}	& \pmx{0.52}{-0.01}{+0.01} \\
\hline
\end{tabular}
}
\label{tab:priors}
\tablefoot{All parameters are in the binary rest frame (with $z=0.918$), and a 50, 16, and 84 per cent quantile estimate is shown for each derived parameter. Priors are either uniform ($\mathcal{U}$) or log-normal ($\mathcal{LN}$) distributed. The injected values are from \citetalias{hu2020spikey}, Table~1. Model results are shown for the \SI{1}{\day} binned light curve of \textit{Kepler}, \plato{}, and \textit{Kepler}+\plato{}, where the \SI{2}{\year} duration \plato{} light curve was used. For each dataset, Case I reflects an independent model fit of $\mathcal{Q}$ and $\mathcal{DM}$ (with parameters separated by the horizontal lines), and Case II is a combined $\mathcal{Q\,DM}$ model fit. Note that for \plato{} the injected value is $t_0=\SI{0.905}{\year}$.}
\end{table*}

\begin{figure*}
\center
\includegraphics[width=1.164\columnwidth]{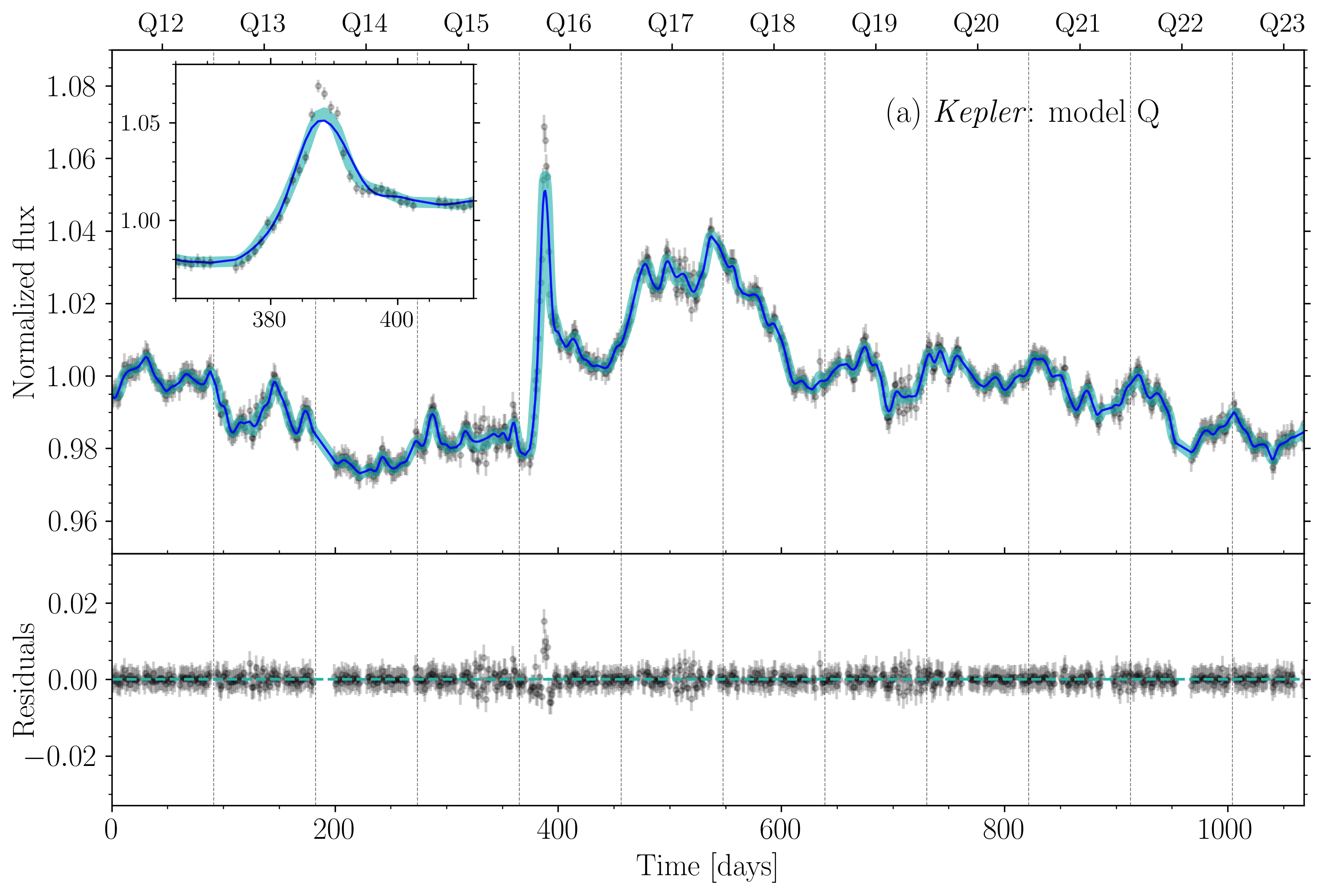}
\includegraphics[width=0.836\columnwidth]{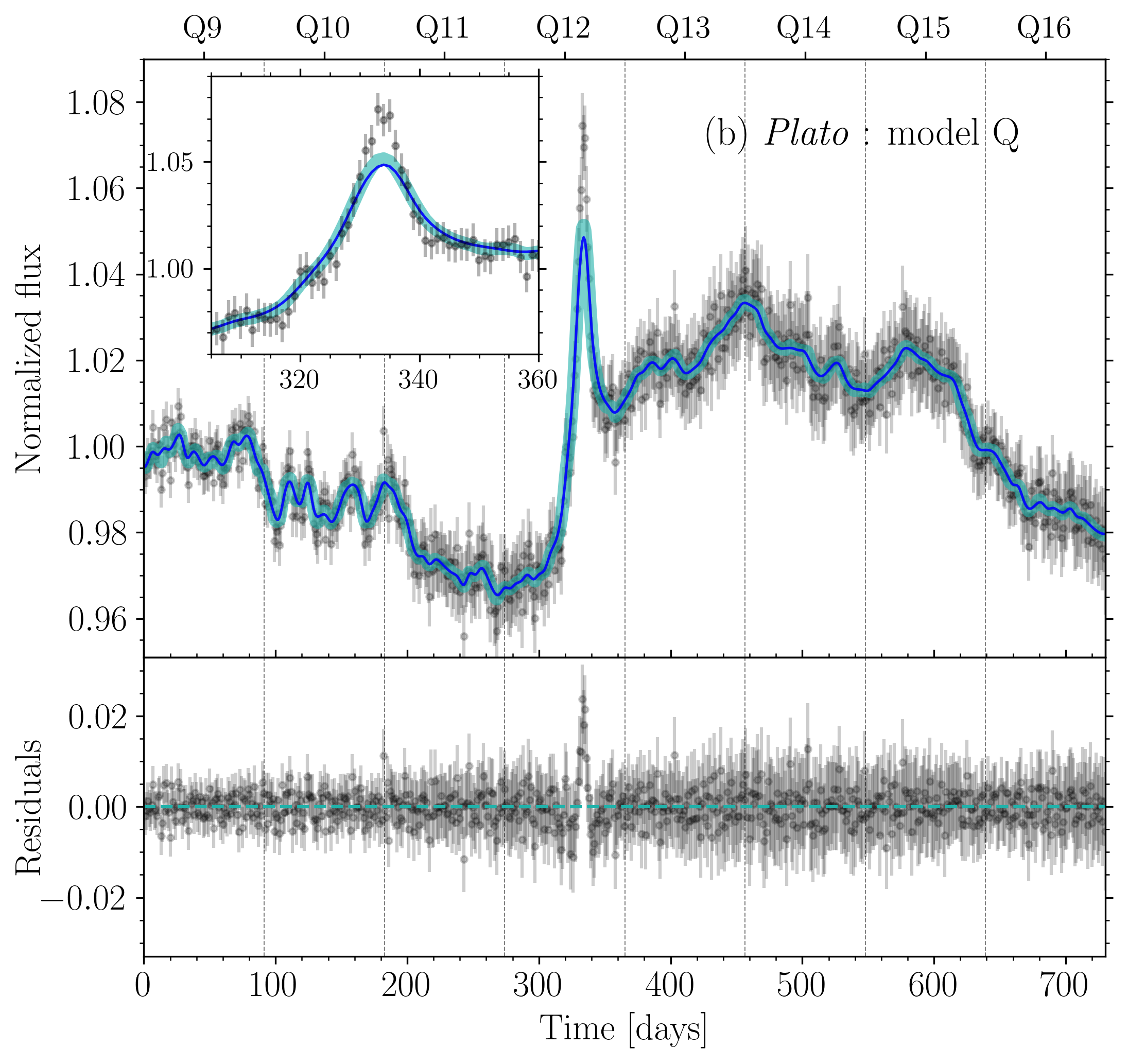}
\includegraphics[width=1.164\columnwidth]{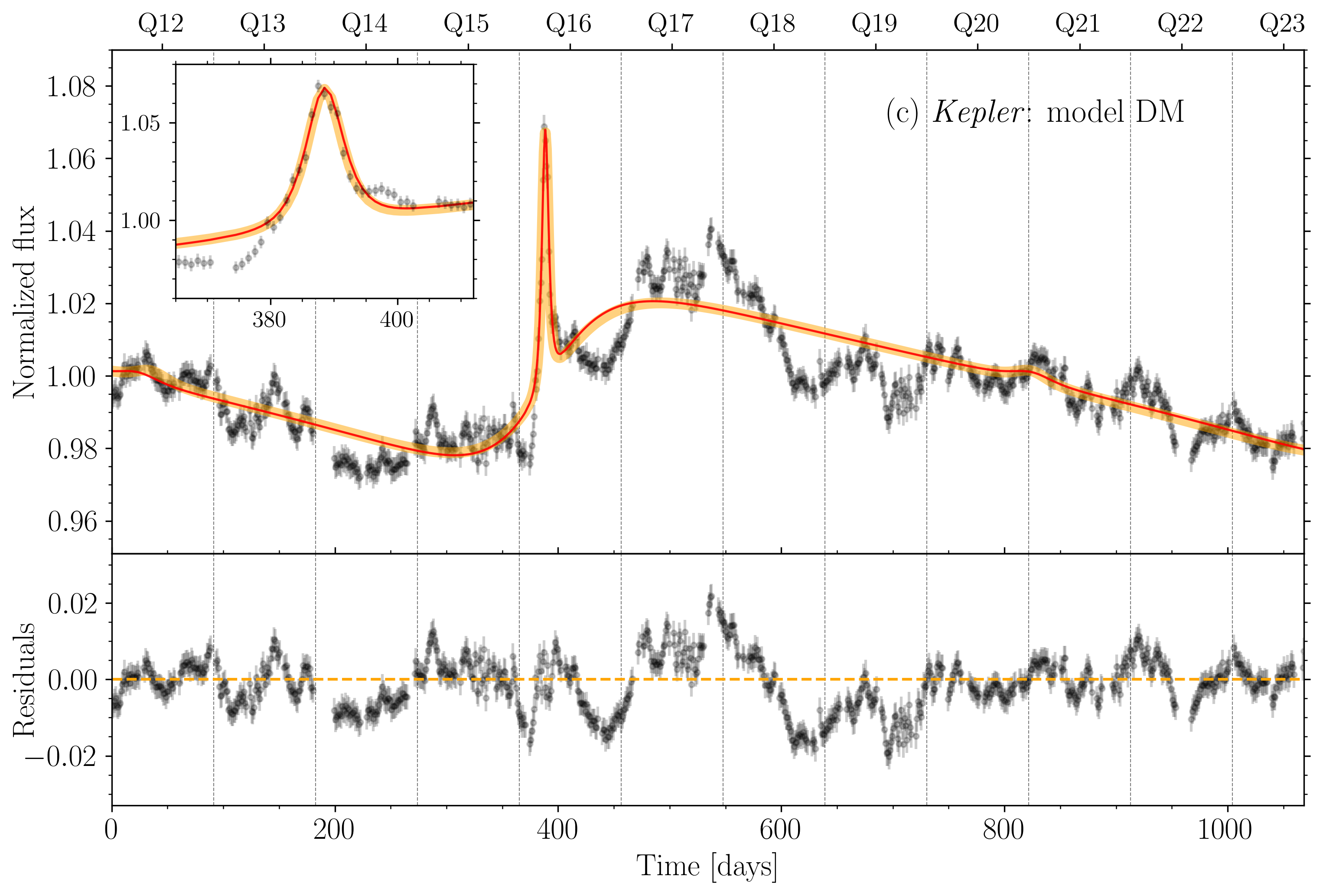}
\includegraphics[width=0.836\columnwidth]{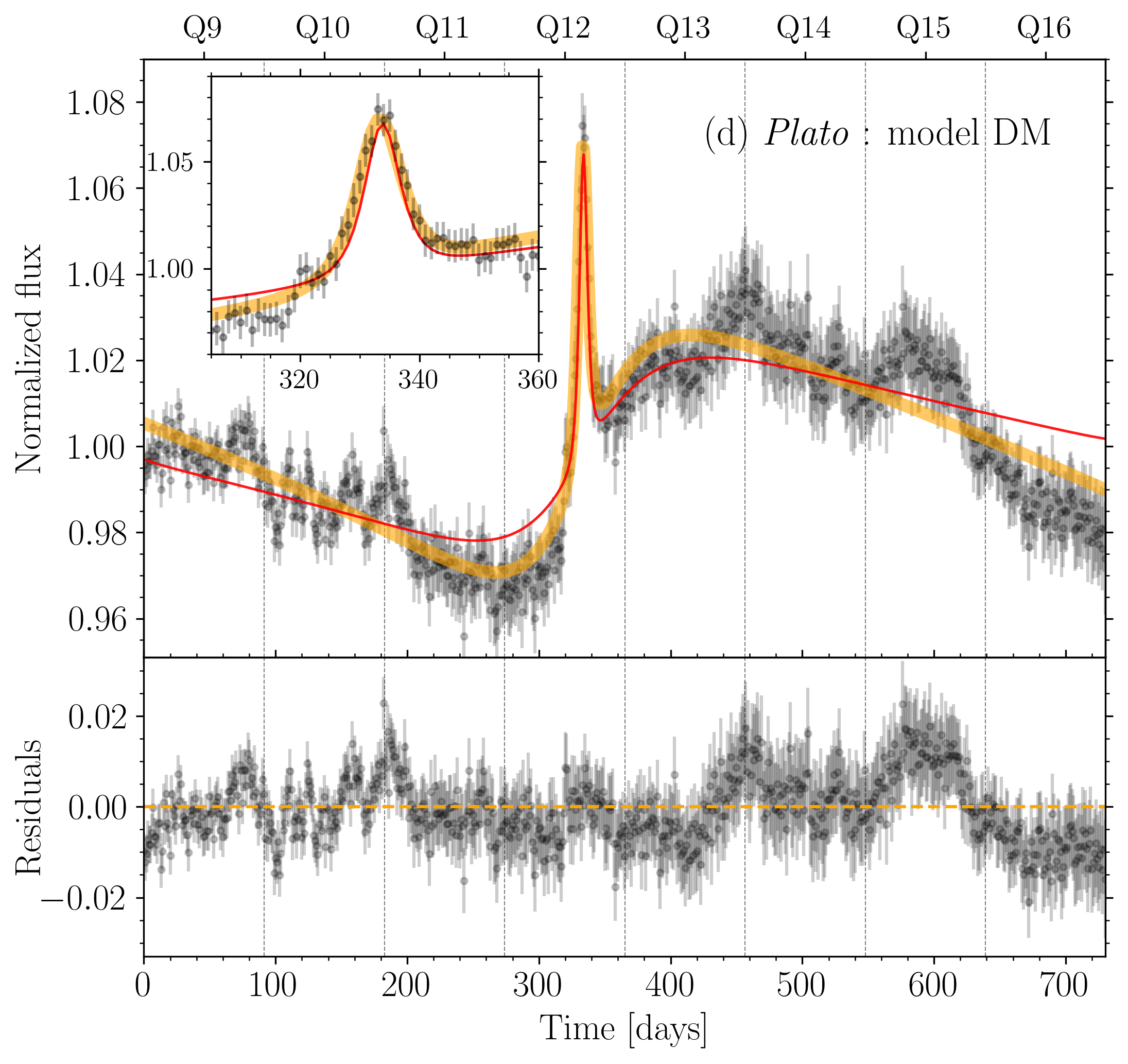}
\includegraphics[width=1.164\columnwidth]{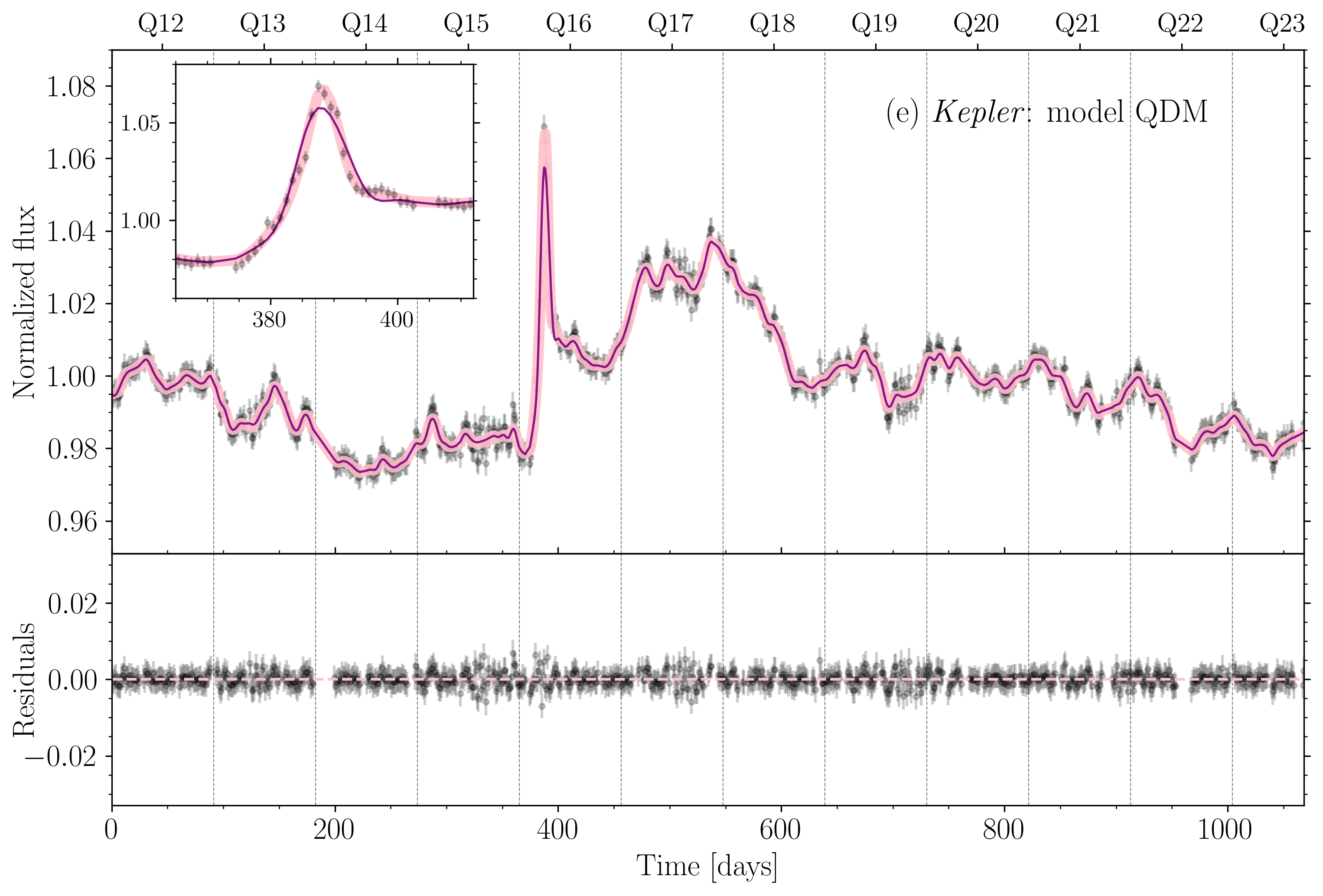}
\includegraphics[width=0.836\columnwidth]{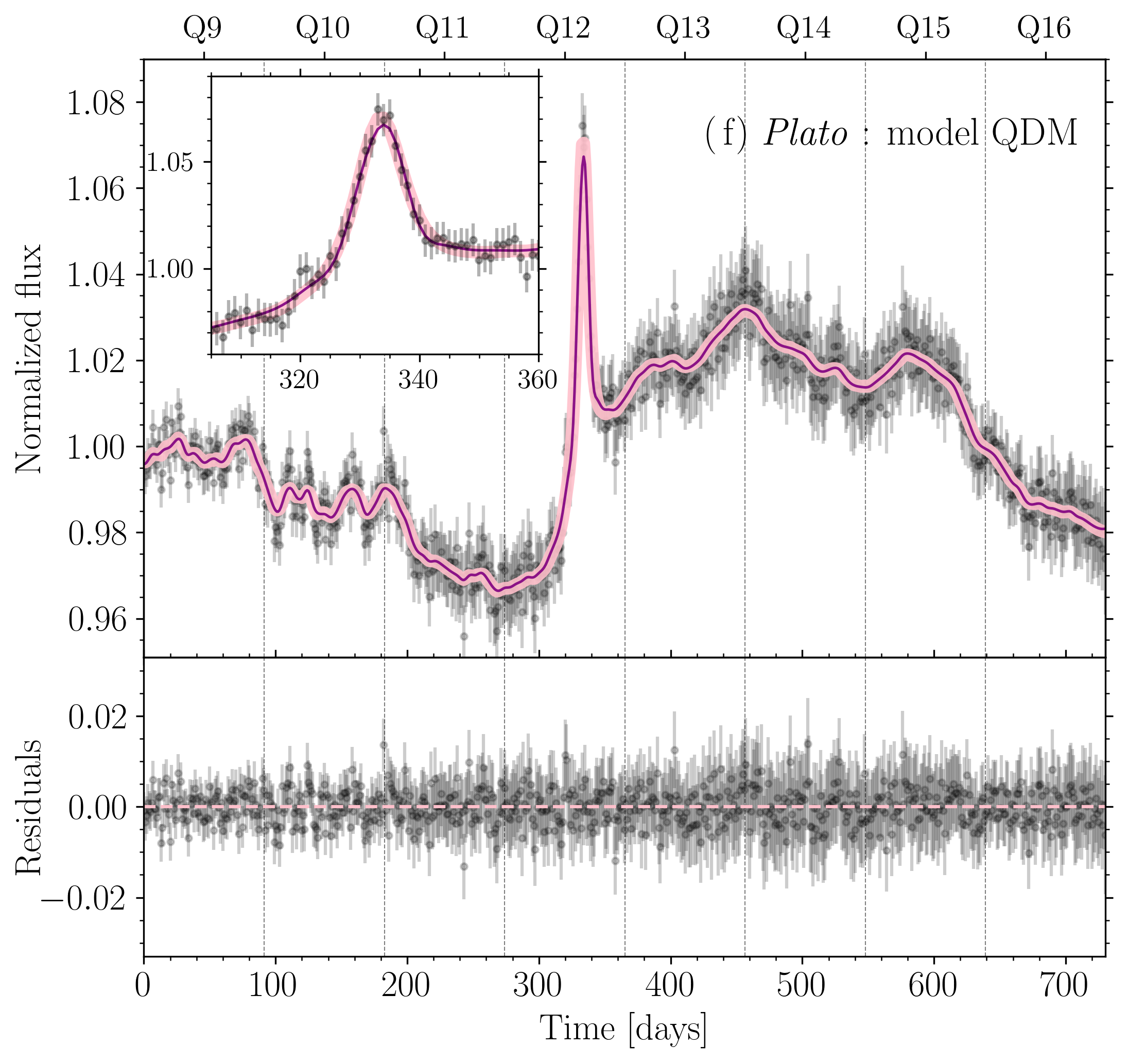}
\caption[]
{Reduced light curve of the SMBHB candidate Spikey as observed by \textit{Kepler} (left plots; from \citetalias{smith2018kepler}) and as expected by \plato{} (right plots). Both datasets have been binned to a \SI{1}{\day} sampling (black points). The upper, middle, and lower panels show a maximum likelihood fit to each dataset for the model $\mathcal{Q}$, $\mathcal{DM}$, and $\mathcal{Q\,DM}$ shown by the cyan, orange, and pink thick solid lines, respectively (corresponding to Case I). The thin solid (blue, red, and purple) lines show the corresponding model fit to the combined \kepler{}+\plato{} light curve (Case II). In each panel, a zoom-in on the SLF is shown. The bottom panels show the residuals. The vertical dotted lines show quarterly breaks every $\sim$91 days when the spacecraft realigns its solar panels towards the Sun. All subplot axes are equally scaled and thus directly comparable. Each model component $\{\mathcal{Q},\,\mathcal{D},\,\mathcal{M}\}$ for each fit of this figure is provided in Appendix~\ref{app:components}.}
\label{fig:lc_spikey_fiducial}
\end{figure*}

We highlight that our main goal is to assert whether \plato{} photometry, as an extension to the \kepler{} data (i.e. a \kepler{}+\plato{} light curve), will allow for a confident SMBHB detection. Thus, we use the model parameters of \citetalias{hu2020spikey} (see the `injected' values of Table~\ref{tab:priors}) and generate a simulated \plato{} light curve of Spikey within the aforementioned observational window of LOPN1. To validate our results we further model the observed \kepler{} dataset and simulated \plato{} dataset independently. For each of the three datasets $\{$\kepler{}, \plato{}, \kepler{}+\plato{}$\}$, our probabilistic methodology allows us to test which of the models $\{\mathcal{Q},\,\mathcal{DM},\,\mathcal{Q\,DM}\}$ provides the best fit from a Bayesian evidence perspective. Thus in total, we make nine model comparisons.

Before modelling the three datasets of Spikey, we first test the Bayesian inference performance on a synthetic $\mathcal{DM}$ model template of Spikey with added Gaussian noise. Across multiple noise levels, cadences, and durations, we find that the posterior landscape of a pure $\mathcal{DM}$ model is inherently multimodal. Particularly, a posterior bimodality in $\omega$ is omnipresent. Interestingly, one solution fits well with the injected value ($\omega_1\sim\SI{84}{\degree}$), whereas the second solution ($\omega_2\sim\SI{264}{\degree}$) violates the assumption $q\equiv M_2/M_1 \leq 1$. Hence, the latter solution essentially corresponds to a change of reference to $\omega_2\equiv\omega_1+\SI{180}{\degree}$, meaning that the primary and secondary swap places such that $L' \equiv 1 - L = L_1 / (L_2 + L_1)$. While we observe the model accuracy increase with increasing data rate, duration, and data quality, the sample size of the bimodal posteriors in $\omega$ shifts drastically with these parameters. 

Using a \SI{3}{\year} duration light curve as our fiducial model, for the above change of reference, both $\omega$-solutions agree and accurately recover $t_0$, $P$, and $e$ for all tested cases. The dominant solution generally recovers $i$ more accurately, while the less dominant solution generally recovers $\log M_1$, $\log M_2$, $\alpha$, and $L$ more accurately. Our small white-noise study also shows, as expected, that an increase in photometric precision increases the model accuracy. Unexpectedly, we find that the ability of the Bayesian inference to accurately (not precisely) model both the $\mathcal{D}$ and $\mathcal{M}$ model component simultaneously depends somewhat on the cadence. Considering instead a \SI{4}{\year} light curve duration (covering two SLFs), the recovery of $\log M_2$ and $\alpha$ is unexpectedly worse than for a \SI{2}{\year} baseline (covering one SLF), but better for recovering $\log M_1$. Overall, the above white-noise test case shows that the four parameters $\{\log M_1, \log M_2, \alpha, L\}$ are strongly correlated even in the absence of AGN variability. 

The above analysis raises the question of how \citetalias{hu2020spikey} were able to recover a smooth and monomodal posterior landscape for Spikey (cf. their Fig.~A1). It turns out that \citetalias{hu2020spikey} found a bug, which does not seem to significantly affect the inferred orbital period but similarly introduce a bimodality in $\omega$. To further test our model agreement, we perform a direct model comparison to \citetalias{hu2020spikey} by using their tailored data sampling on a DRW-subtracted light curve%
\footnote{Specifically \citetalias{hu2020spikey} use the best-fit $\mathcal{DM}$ model of the original light curve and subtract it to determine the model component $\mathcal{Q}$. This $\mathcal{Q}$ model fit is then subtracted from the original light curve, and the new light curve is binned to a \SI{9}{\day} sampling, while preserving a denser \SI{0.5}{\day} sampling around the SLF. Lastly, a final best-fit $\mathcal{DM}$ model was performed.}. %
Overall, we produces similar posterior distributions and inferred parameters within the errorbars. In conclusion, the model parameter degeneracies found as function of cadence, duration, and sampling should be explored further in future work on combining \kepler{} data with other data. For the forthcoming Bayesian analysis we use a regular sampling of \SI{1}{\day}, which was found to be a good trade-off between model accuracy and computation time.

Having investigated the inherent model degeneracies and tested the robustness of our nested sampling, we now turn to the results of the Bayesian inference of Spikey. We present in Table~\ref{tab:priors} the 50, 16, and 84 per cent quantile estimates for the modelling of the light curve of \kepler{}, \plato{}, and \kepler{}+\plato{}. Here, only the nominal \SI{2}{\year} duration \plato{} light curve was considered. As we are interested in a model comparison, for each result in Table~\ref{tab:priors}, Case I reflects an independent model fit of $\mathcal{Q}$ (i.e. best fit of $\{\tau,\,\sigma\}$) and $\mathcal{DM}$ (i.e. best fit of $\{t_0,\, P,\, i,\, e,\, \omega,\, \log M_1,\, \log M_2,\, \alpha,\, L\}$), while Case II is a combined $\mathcal{Q\,DM}$ model fit. Figure~\ref{fig:lc_spikey_fiducial} shows the outcome and Table~\ref{tab:evidences} lists the Bayesian evidence for all datasets. The $\mathcal{Q\,DM}$ model is always the preferred model with decisive evidence. To obtain further insight into these results, we supplement Fig.~\ref{fig:lc_spikey_fiducial} (showing the best fit models across the different datasets) with the model component $\{\mathcal{Q},\,\mathcal{D},\,\mathcal{M}\}$ for each model fit in Appendix~\ref{app:components}. 

\begin{table}[t!]
\caption[]{Bayesian model evidence of nested sampling for Spikey.}
\begin{center}
\begin{tabular}{ccr}
\hline\hline
Observation & Evidence & Result \\
\hline
\plato{}  & $(\ln z_{\mathcal{QDM}}-\ln z_{\mathcal{Q}})$  & $  68.36\pm0.02$ \\
\kepler{} & $(\ln z_{\mathcal{QDM}}-\ln z_{\mathcal{Q}})$  & $ 131.85\pm0.02$ \\
\kepler{}+\plato{} 
		  & $(\ln z_{\mathcal{QDM}}-\ln z_{\mathcal{Q}})$  & $ 197.76\pm0.03$ \\
\hline 
\plato{}  & $(\ln z_{\mathcal{Q}}-\ln z_{\mathcal{DM}})$   & $ 112.21\pm0.03$ \\
\kepler{} & $(\ln z_{\mathcal{Q}}-\ln z_{\mathcal{DM}})$   & $1539.57\pm0.03$ \\
\kepler{}+\plato{}  
          & $(\ln z_{\mathcal{Q}}-\ln z_{\mathcal{DM}})$   & $1898.89\pm0.04$ \\
\hline
\plato{}  & $(\ln z_{\mathcal{QDM}}-\ln z_{\mathcal{DM}})$ & $ 180.57\pm0.04$ \\
\kepler{} & $(\ln z_{\mathcal{QDM}}-\ln z_{\mathcal{DM}})$ & $1671.42\pm0.05$ \\
\kepler{}+\plato{}  
          & $(\ln z_{\mathcal{QDM}}-\ln z_{\mathcal{DM}})$ & $2096.66\pm0.06$ \\
\hline
\end{tabular}
\end{center}
\label{tab:evidences}
\end{table}

{Aside from the conclusive evidence in Table~\ref{tab:evidences}, our results also reflect that the DRW alone can explain the variation in the data to a high degree. Indeed, for the $\mathcal{Q\,DM}$ model fits, the Doppler boosting is largely absorbed by the DRW model. As reflected in Table~\ref{tab:priors}, we find an increased eccentricity ($e$), prolonged period ($P$), lower inclination ($i$), lower mass ratio ($q \equiv M_2/M1$), and a larger spectral index ($\alpha$) for the $\mathcal{Q\,DM}$ model as compared to the inferred parameters of the $\mathcal{DM}$ model. To confirm this behaviour, we further investigated the inference of model $\mathcal{Q}$ by using two deterministic models for the mean variation: the fiducial Spikey $\mathcal{DM}$ template and a Spikey $\mathcal{DM}$ template with zero eccentricity. Aligning with the above evidence, the DRW fits both dataset almost equally well, thus highlighting the flexibility of the DRW to fit AGN variability and adapt to `wrong' model input. 

In spite of this DRW-adaptability, the fact that the DRW does not fully absorb the SLF of the $\mathcal{Q}$ (and neither the $\mathcal{Q\,DM}$) model is promising for future use cases relying on Bayesian evidence for model evaluations. A similar result was also reported by \citetalias{hu2020spikey}, when comparing the $\mathcal{Q\,DM}$ model against the $\mathcal{Q}$ model. In fact, we here show that the DRW barely provides any explanatory power for the SLFs across the three datasets (see lower panels of Fig.~\ref{fig:lc_components_separate} and \ref{fig:lc_components_combined}). Since the DRW efficiently absorbs the Doppler boosting while it is an integrated component of the relativistic model $\mathcal{DM}$, it is clear that the inferred fit parameters of the $\mathcal{Q\,DM}$ model are unreliable. This suggest that parameter estimation is better done with the $\mathcal{DM}$ model alone, while the DRW model parameters are best estimated with an independent $\mathcal{Q}$ model fit on a $\mathcal{DM}$ model-subtracted light curve (the exact recipe of \citetalias{hu2020spikey}). 

Any robust Bayesian inference in the presence of a DRW signal necessitates an extended baseline or multi-epoch observational campaigns (to cover at least two SLFs). As neither the \kepler{} nor the \plato{} light curve aligns with the former requirement, care must be taken when interpreting their Bayesian inferences. As demonstrated in Table~\ref{tab:priors} for the $\mathcal{DM}$ model, \plato{}'s nominal baseline is too short for a trustworthy inference. This result is also partly shaped by an imperfect \plato{} data reduction, which causes an enhanced Doppler boosting and thus a shorter orbital period as compared to the \kepler{} data (see Fig.~\ref{fig:lc_spikey_fiducial}; thick orange line of panel d vs. c). The joined force of the \kepler{}+\plato{} dataset (with the joint fit shown in Fig.~\ref{fig:lc_spikey_fiducial}; thin red line across panel c and d) illustrates that the precision of the orbital period will increase by a factor of $\sim$100. This is a promising avenue as it will result in conclusive evidence to (de)confirm Spikey's SMBHB nature, while placing strong constraints on the orbital precession and likely an estimate of the orbital decay due to GW emission. 

We note again that the \plato{} simulation of Spikey was generated using the parameters of \citetalias{hu2020spikey}, while our re-analysis of Spikey's \kepler{} data revealed slightly different parameters. For the $\mathcal{DM}$ model, we find a shorter orbital period of \pmx{1.130}{-0.004}{+0.004} as compared to \pmx{1.144}{-0.030}{+0.029} of \citetalias{hu2020spikey}. As a result, we find some disagreements between the model precision and accuracy in the joint \kepler{}+\plato{} model fit (e.g. see again $P$ of Table~\ref{tab:priors}). Acknowledging the limitation of our re-analysis of Spikey (due to \kepler{}'s limited baseline), we finish with an interesting observation concerning $\alpha$. Compared to \citetalias{hu2020spikey}'s estimate of the spectral index $\alpha = $\,\pmx{2.09}{-0.18}{+0.29}, we find a value $\alpha = $\,\pmx{-0.3}{-0.2}{+0.4} for the $\mathcal{DM}$ model. This is more consistent with the measured value of $\alpha \sim -0.12$ from optical spectroscopy (cf. \citetalias{hu2020spikey}, footnote 2) and could indicate that Spikey is perhaps a less eccentric binary (i.e. $e\sim 0.418$) than previously reported.      

\subsection{Observations of Spikey-like objects}
\label{sec:results_observation}

While the \kepler{}+\plato{} dataset is able to determine Spikey's physical nature, it has to be seen whether \plato{} imagery can be used to reliably estimate physical properties. As Spikey is near \plato{}'s brightness limit, the answer will depend strongly on the exact performance of the baffles, optics, and detectors in flight (setting the level of light scattering, throughput, and electronic/digital noise, respectively). Meanwhile, we can gain more insight by studying the impact of the source magnitude of Spikey-like objects. Figure~\ref{fig:lc_magnitudes} shows, from top to bottom panel, a \SI{1}{\hour} cadence (black) and \SI{1}{\day} cadence (grey) light curve for the four magnitude bins $\Pb \in \{16, 17, 18, 19\}$. Overlaid on each light curve is the injected $\mathcal{Q\,DM}$ (blue) and $\mathcal{DM}$ (orange) model template. While the figure considers a potential \SI{2}{\year} mission extension of the LOPN1, it shows that we should be able to detect and model Spikey-like signatures for $\Pb\leq18$, while for $\Pb>18$, more advanced statistical techniques are most likely required. 

\begin{figure}[t!]
\center
\includegraphics[width=\columnwidth]{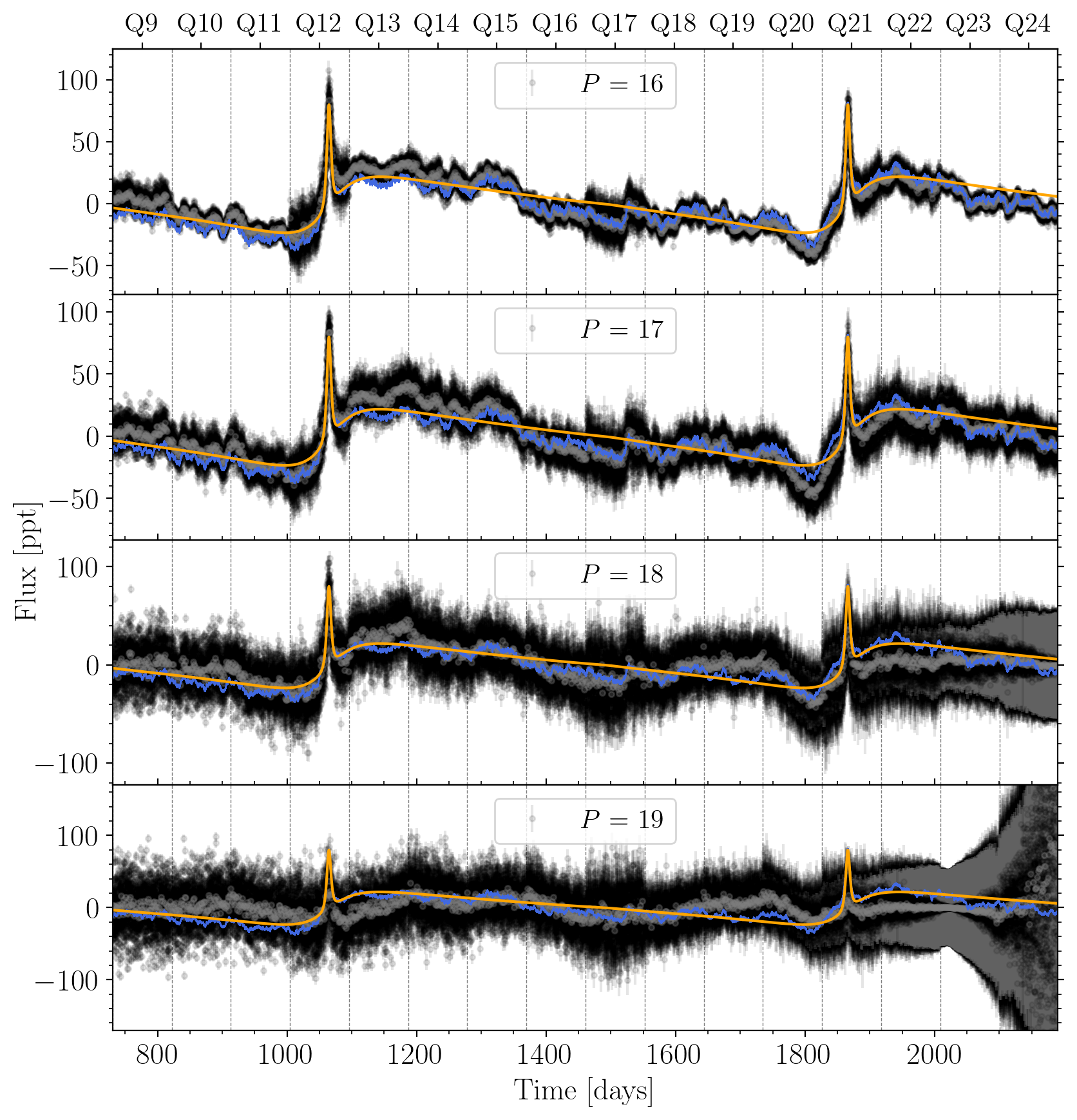}
\caption[]
{Fully reduced Spikey-like light curves binned to a \SI{1}{\hour} cadence (black points) and a \SI{1}{\day} cadence (grey points). The injected model is shown (blue line) together with the recovered $\mathcal{DM}$ maximum likelihood fit (orange line). The panels show, from top to bottom, a simulated light curve with increasing source magnitudes of 16, 17, 18, and 19, respectively.}
\label{fig:lc_magnitudes}
\end{figure}

Another important lesson from Fig.~\ref{fig:lc_magnitudes} is that \plato{}'s photometric detection limit is time dependent. Specifically, for faint sources such as AGNs, digitalisation noise increases over time. This is clearly visible as photometric features and increasing photometric errors towards the second half of the $\Pb=18$ and $\Pb=19$ light curves, where the digitalisation correction step fails due to a low flux level (in agreement with Fig.~\ref{fig:nsr_quasars}). Specifically for Spikey, for a \SI{2}{\year} LOPN1 mission extension, we find that Bayesian modelling (cf. Table~\ref{tab:priors} and Fig.~\ref{fig:lc_spikey_fiducial}) slightly improves the precision of the timings ($t_0$ and $P$), but at the expense of a worse inference for the remaining model parameters.

In preparation for a future simulation study (see Sect.~\ref{sec:discussion}), we explore the SMBHB parameter space of Spikey-like objects. We select a few key orbital configurations, such as the inclination $i \in \{\SI{84}{\degree}, [\SI{81.95}{\degree}], \SI{80}{\degree}, \SI{78}{\degree}\}$, eccentricity $e \in \{0.1, 0.3, [0.52], 0.7\}$, and argument of periapse $\omega \in \{\SI{0}{\degree}, [\SI{84.6}{\degree}], \SI{180}{\degree}, \SI{270}{\degree}\}$. The bracketed numbers reflect the fiducial model parameters of Spikey. The first three upper panels of Fig.~\ref{fig:lc_wcs} show the parameter values that result in the lowest $\mathcal{DM}$ signal detectability. It is evident from this plot that Spikey itself (if confirmed) would be a unique eccentric binary as its orbital configuration (with $\omega \sim \SI{90}{\degree}$) results in the shortest duration yet the largest secondary lens magnification possible. As we have shown with our Bayesian scheme (and \citetalias{hu2020spikey} as well), the former is vital to unambiguously assert whether Spikey's prominent flare is real or a temporal DRW excursion. Figure~\ref{fig:lc_wcs} shows that identifying less eccentric Spikey-like candidates with traditional methods (including Bayesian inference) is more challenging, while less inclined (\SI{78}{\degree}) or phase shifted orbits (\SI{270}{\degree}) would likely degrade such an object from a potential SLF candidate to yet another quasar with temporal quasi-cyclic variability. 

\begin{figure}[t!]
\center
\includegraphics[width=\columnwidth]{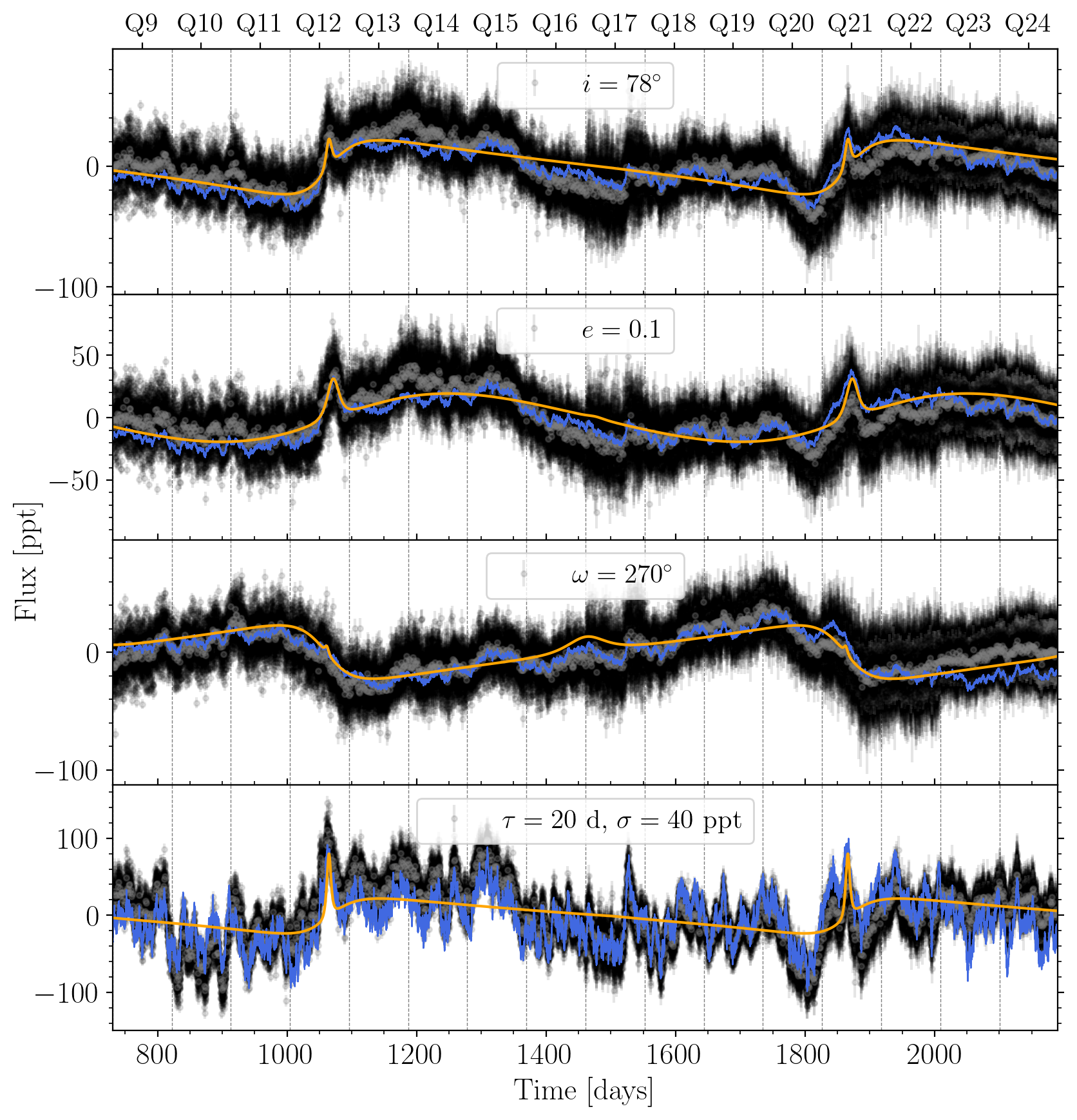}
\caption[]
{Same as Fig.~\ref{fig:lc_magnitudes} but for a Spikey-like object with $i=\SI{78}{\degree}$, $e=0.1$, $\omega=\SI{270}{\degree}$, and $(\tau, \sigma)=(\SI{10}{\day}, \SI{40}{\ppt})$ from top to bottom, respectively.} 
\label{fig:lc_wcs}
\end{figure}

\section{Discussion}\label{sec:discussion}

While Spikey is a best-case example of an SMBH photometric signature in terms of orbital phase and magnitude (and camera visibility of all 24 N-CAMs), the fact that each model component is visible in the \plato{} light curve (cf. Fig.~\ref{fig:lc_spikey_fiducial}) strongly motivates utilising \plato{} for AGN and SMBHB research. To do so, we here discuss our assumptions, how we can improve our modelling, and how \plato{} can contribute to both SMBHB searches and AGN research in general.

As a proof-of-concept study, we have implemented a physical `toy model' that ignores many aspects of gas interactions for binaries. In fact, the current picture of SMBHB evolution \citep[mainly driven by hydrodynamic simulations, e.g.][]{lai2023circumbinary, guierrez2025accretion} shows that gravitational interaction of surrounding gas forms a circumbinary accretion disc from which a low-density cavity is carved out due to gravitational perturbations asserted by the orbiting binary system. As accretion streams transfer gas across the cavity to the central binary, a circumsingle disc around either of the black holes may form \citep{dorazio2013accretion}. Thus, this study ignores a rich dynamical problem involving the circum-binary disc, while assuming that at least one of the black holes is accreting. 

We further highlight that our adopted model describes temporal flux variations of quasars with a DRW. \cite{tachibana2020deep} performed a deep learning analysis of CRTS legacy data and showed that quasar variability differs from a DRW model. While a DRW is inherently time symmetric, a temporal asymmetry is observed in optical variability. As the direction of time symmetry depends on the physical details of the region of accretion \citep{kawaguchi1998optical}, the observed asymmetry is consistent with the disc instability model \citep{takeuchi1995xray} where mass accretion happens during large-scale avalanches. As time asymmetry is predicted to be time-scale dependent, \plato{}'s high-cadence will for example allow studying AGN disc structure through their temperature profiles. Thus, in a broader context, \plato{} could be pivotal in understanding (and thus predicting) AGN variability across time scales. For SMBHB searches, this will in turns allow mitigating misinterpretation of DRW excursions for Doppler cycles or strong one-off peaks resembling SLFs. The lower panel of Fig.~\ref{fig:lc_wcs} illustrates the base of this problem for a Spikey-like light curve for when the DRW variability is comparable to the relativistic temporal variations (such that the SLF drowns in the DRW sea of peaks). While similar cases may challenge SMBHB detection, the high cadence of \plato{} will significantly help reduce such signal confusion.

Regarding the photometric detection of relativistic effects, we explored the finite-source gravitational lensing model by \citet{park2025self} in the \plato{} passband (cf.~Sect.~\ref{app:model_lensing}). As this model is computationally expensive for Bayesian inference, the point-source model was used throughout. Figure~\ref{fig:lensing_chromatic} shows that for a near edge-on orbit, the SLF amplitude difference between the two models can be up to one order of magnitude. For a system such as Spikey, the amplitude difference is much closer to unity. Moreover, assuming that the disc size is proportional to the Hill radius (i.e. outermost stable orbit), \cite{dorazio2018periodic} showed that finite-sized source lensing only becomes important for larger separation binaries. This is good news for a \plato{} monitoring programme of short-period SMBHBs.

Despite our favourable model assumptions, our results align with the high false-alarm probability of SMBHB candidates to date. We note that existing searches used Lomb-Scargle periodograms, or similar Fourier techniques, which are sensitive only to sinusoidal periodicities (which are precisely the ones that can be relatively easily mimicked by stochastic DRW variability). Furthermore, binary periodicity from both hydrodynamical modulations of the accretion and from Doppler effects for an eccentric binary are expected to be bursty/saw-tooth like. These pulse shapes would have been missed in current searches \citep{lin2026lomb, park2026self}, and will be more challenging to identify. Thus, it is unlikely that upcoming photometric surveys alone will suffice to build a firm population of confirmed SMBHBs across all potential periodicities, without follow-up observations and modelling advancements.

Future simulations should, among other things, include proper modelling of accretion disc dynamics. The open-source software \texttt{binlite} \citep{dorazio2024binlite} has already been used in similar studies to model both the circum-disc and mini-discs of the binary system \citep[e.g.][]{park2026self}. To fully exploit and quantify \plato{}'s potential for SMBHB research, we aim to perform an extensive simulation study that will cover a much larger physical and dynamical parameter of SMBHBs, including more realistic population models (e.g. using \texttt{binlite}). Herein, the idea will be to employ machine learning to train a model that can distinguish Doppler boosting and SLFs apart from spurious DRW signals and complex accretion scenarios. Looking back at Sect.~\ref{sec:catalogue_noise}, with non-linearity and digitalisation noise acting as signal suppression, machine learning will help to understand and mitigate these effects before launch. Future work will likewise allow us to study the impact of camera visibility, stellar crowding, and instrumental systematics on a global scale.

With a future simulation study planned, it should be kept in mind that self-lensing depends on the inclination and certain favourable physical conditions about accretion and disc structure. First, the chance of a SLF depends on the ratio of the angular size of the orbit to the Einstein radius, which is given by $(2a/R_s)^{-1/2}$, where $a$ is semimajor axis and $R_s$ is the binary Schwarzschild radius. Hence, if the binary separation is $\sim$100$\,R_s$, there is a few percent chance of a strong SLF, while for a modest $\sim$7\% magnification similar to Spikey, the occurrence rate can be achieved for larger binary separations. The increased probability of smaller inclination angles naturally results in SLFs of shorter duration (lower amplitude) compared to longer (larger) ones \citep{park2025self}. While this poses a challenge for upcoming time-domain facilities, it is a blessing for \plato{} as we discuss below. Secondly, not all quasars are binaries and especially not compact binaries. Considering again a binary separation of $\sim$100$\,R_s$, the lifetime of the binary system will be much shorter than the typical quasar lifetime of $\sim$\SI{e8}{\year}~\citep{Martini2004}. Thus, only a small fraction will be caught in this stage, even if all quasars are inspiraling binaries. For bright $\sim$\SI{e9}{\Msun} SMBHs with an orbital period of order a year, this fraction will be $\sim$0.1\% \citep{park2025self}. Thus, with $\sim$1100 quasars brighter than $\Pb < 17.5$ in the two LOPs, we can expect a few binaries and one SLF. While we plan to refine this rough prediction, it is clear that \plato{}'s true niche is to follow-up SMBHBs rather than discovering them.

A complementary ground-based survey to \plato{} is the LSST, which has just come online. LSST will explore the transient sky in the coming decade (in multiple passbands) and increase the sample size of photometric AGN observations by several orders of magnitude. While LSST generally will observe much fainter sources than \plato{}, its high cadence and photometric quality makes it an ideal survey for \plato{} follow-up studies. Since LSST will be conducted from the Vera C.\ Rubin observatory in Chile, \plato{} could contribute significantly to LSST's limited coverage of the northern sky while providing simultaneous observations of bright SMBHB candidates in LOPS2. 

In the era of LSST and other large scale surveys, \plato{} may particularly be useful for confirming short-duration SLF candidates. Both \cite{kelly2021gravitational} and \cite{park2026self} showed that the number of expected SLFs is a very steep function of the flare duration. Assuming a $\sim$\SI{3}{\day} cadence and requiring $\sim$10 points per flare, and therefore limiting the search to $\sim$\SI{30}{\day} or longer flares in LSST, results in a huge loss of short-period LSST candidates. For example, the decrease from a flare duration of \SI{30}{\day} to \SI{10}{\day} results in anincrease in the number of objects by more than a factor of 20 \citep[][Table~IV]{park2026self}. \plato{}'s dense data coverage allows for a confirmation of these narrow flares and a search for SMBH shadows imprinted into the flares \cite[showing up as even narrower dips in a fraction of the self-lensed sources;][]{davelaar2022self}. Similarly, from bright quasars, \plato{} could search for SLFs which were missed or very poorly sampled by larger time-domain surveys such as CRTS, ZTF, and PTF.

Another synergistic survey for \plato{} is the High-Latitude Time-Domain Survey (HLTDS) to be conducted by NASA’s Nancy Grace Roman Space Telescope \citep[\textit{Roman};][]{akeson2019roman}. Similar to \plato{}'s observing strategy, the HLTDS has a northern (ELAIS-N1) and southern (EDFS) field, tiling each field to a continuous viewing zone and thus revisiting each patch of sky every five days for two years. Each field overlap with the 12 N-CAM region of the respective LOP field (furthest away from the Galactic plane). A wide and deep field is scheduled for each HLTDS, with a central overlap for the ELAIS-N1 and a spatially offset for EDFS (to increase the coverage with ESA's \textit{Euclid} space mission). With imaging capabilities similar to the \textit{Hubble} Space Telescope (HST), \textit{Roman}'s wide (deep) field covers $\sim$\SI{10}{\deg^2} ($\sim$\SI{2}{\deg^2}). While these fields are relatively small compared to the LOPs, \textit{Roman}'s near-to-infrared wavelength coverage could allow a strong synergy with \plato{}.  
 
A strongly synergistic mission to \plato{} is the upcoming Chinese space mission called Earth 2.0 \citep[ET;][]{ge2022et}. With similar mission objectives and similar photometric precision as \plato{}, ET will be revisiting the \kepler{} field for a duration of four years starting around the time when \plato{} will have finished its nominal 2-year observation of LOPS2. ET will cover a ($\sim$4.3) smaller field than \plato{} but will be able to downlink all of its full-frame CCD images with a cadence of \SI{11}{\minute}. As it is still undecided whether \plato{} will observe LOPS2 for two, three, or four years (and potentially spend its mission extension on LOPN1), establishing an optimal synergy between \plato{} and ET would be beneficial for any research of shared targets with long periodicity present in their variability, including gravity-mode pulsators in single and binary stars \citep{ATk2024,jannsen2025mocka} as well as AGNs and SMBHBs. Linked to our present study, extending the \plato{} baseline beyond two-to-four years will dramatically help to assert the SMBHB nature of current and future candidates, including Spikey.

\section{Conclusions}\label{sec:conclusions}

We have investigated the possibility of using \plato{} to observe faint extragalactic SMBHBs. This work shows that despite the technical challenges of observing in the faintest limit of the \plato{} mission (i.e. $\mathcal{P} > 15$), it is possible to detect signatures of SMBHBs from precise high-cadence \plato{} observations. 

We used the SMBHB candidate Spikey as a first benchmark for our simulations and showed that similar Spikey-like signatures from gravitational SLFs can be detected in \plato{} observations. Accordingly, in combination with \kepler{} data, \plato{} will be able to confirm the eccentric self-lensing scenario of Spikey (and thus its nature), assuming that the SMBHB hypothesis is true. Regardless of Spikey’s true nature, the uninterrupted high-cadence and high-precision photometry delivered by \textit{Kepler} and to be delivered by \plato{} highlights that (i) an avenue of AGN variability is still to be explored at short timescales over long baselines, and (ii) Bayesian inference techniques for SMBHB research can be used as a probe for detection (and potentially confirmation). Future better modelling effort is needed if we are to account for more complex population models.

To exploit \plato{}'s future potential, we created a high-probability quasar catalogue covering the two \plato{} pointing fields. We used this catalogue to push beyond simple noise calculations and demonstrate that $\Pb \leq 17.5$ ($\Pb \leq 17$) should be enforced to retain conserved flux measurements throughout a 2-yr (4-yr) mission duration and to avoid an unbiased determination of physical system parameters. With the observing field being fixed, our catalogue contains 494 (170) quasars within the LOPS2. 

We aim to refine the current quasar catalogue further and use it with a more elaborate simulation study as a community source for future \plato{} GO applications. Particularly, a suite of simulations testing the detectability of SLFs and Doppler boosting using machine learning methods will be helpful to expose (i) the limits of \plato{} data for this type of SMBHB research, and (ii) which quasars in the future LOPs will be optimal for follow-up studies in the era of LSST. 

\section*{Data availability}

All data presented in this paper can be found on the public GitHub page  \url{https://github.com/IvS-KULeuven/plato_spikey_bayes}. This repository specifically contains Jupyter notebooks that reproduce all the Bayesian inference results presented in this paper (including all maximum-likelihood fits and and corner plots). 


\begin{acknowledgements}

This work presents results from the European Space Agency (ESA) space mission \textit{Plato}. The \textit{Plato} payload, the \textit{Plato} Ground Segment and \textit{Plato} data processing are joint developments of ESA and the \textit{Plato} mission consortium (PMC). Funding for the PMC is provided at national levels, in particular by countries participating in the \textit{Plato} Multilateral Agreement (Austria, Belgium, Czech Republic, Denmark, France, Germany, Italy, Netherlands, Portugal, Spain, Sweden, Switzerland, Norway, and United Kingdom) and institutions from Brazil. Members of the \textit{Plato} Consortium can be found at \url{https://platomission.com/}. The ESA \textit{Plato} mission website is \url{https://www.cosmos.esa.int/plato}. We thank the teams working for \textit{Plato} for all their work.
The research behind these results has received funding from the BELgian Federal Science Policy Office (BELSPO) through PRODEX grants for \textit{Plato} and {\it Gaia}.
Z.H. was supported by NASA grants 80NSSC22K0822 and 80NSSC24K0440.
We thank the authors of \citet{smith2018kepler} for sharing their reduced AGN light curves observed by \textit{Kepler}.
This project made use of the following published Python packages \texttt{Numba} , \texttt{NumPy} \citep{harris2020array}, \texttt{Pandas} \citep{mckinney2011pandas, team2022pandas}, \texttt{SciPy} \citep{virtanen2020scipy}, \texttt{Astropy} \citep{price2022astropy, price2018astropy, robitaille2013astropy}, \texttt{Matplotlib} \citep{hunter2007matplotlib}, and \texttt{corner} \citep{foremanmackey2016corner}.

\end{acknowledgements}


\bibliographystyle{aa}
\bibliography{bibliography} 

\begin{appendix}


\section{\plato{} quasar catalogue}\label{app:cat}

Using a search grid larger than each LOP, we first require that the query only returns objects with a \textit{Gaia} magnitude below $G<19$, an AGN classification \texttt{vari\_best\_class\_name = AGN}, and a relative error in (QSOC) redshift $\Delta z/z < 0.05$, as given by \textit{Gaia}'s (astrophysical) parameter database. We further refine the sample by using \textit{Gaia}'s overall quasar classification probability requiring \texttt{classprob\_dsc\_combmod\_quasar > 0.999}. This classification is derived from $\tx{G}{BP}/\tx{G}{RP}$ spectra together with several astrometric and photometric features. As in \cite{jannsen2025mocka}, we use full-frame CCD images simulated with \platosim{} to confine all objects within the field-of-view of \textit{Plato}. For each object, this allows us to monitor its N-CAMs visibility and level of contamination from nearby objects. Note that LOPs footprints we recover exceeds slightly that of \cite{nascimbeni2025plato} as we include the latest camera FOVs from the test house measurements. Specifically we allow the radial distance from the optical axis of each camera to extend to $\vartheta=\SI{19.4}{\degree}$ which corresponds to a overall transmission efficiency of 50\% on average.

To exclude sources with a spurious astrometric solution, we use the astrometric reliability diagnostic \texttt{fidelity\_v2} from \cite{rybizki2022classifier}. Compared to the typical \textit{Gaia} quality flags (e.g. \texttt{ruwe} and \texttt{visibility\_period\_used}), this method has been shown to separate `good' (1 being the best) from `bad' (0 being the worst) astrometric solutions more efficiently by an order of magnitude. We exclude sources with \texttt{fidelity\_v2 < 0.75}, which is a conservative choice adopted from \cite{elbadry2022magnetic}. As a next step, we make a proper motion cut of $\log\mu < 0.4(G-18.25) \, \text{mas} \,\, \si{\per\year}$ \citep[cf.][]{storeyfisher2024quaia}, thus omitting sources with less precise astrometry. 

Since the \textit{Gaia} AGN classification label and the quasar probabilities may be affected by spurious measurements, we further refine the number of candidates following the methodology of \cite{butler2011optimal}. The idea is to differentiate quasars from other field sources using variability metrics alone and construct a QSO variability selection diagram. A QSO variability diagram consists of two key metrics available from the \textit{Gaia} photometric pipeline, namely \texttt{non\_qso\_variability}, the QSO non-variability metric ($\tx{\chi}{False}^2$), and \texttt{qso\_variability}, the QSO variability metric ($\tx{\chi}{QSO}^2$). Figure~\ref{fig:classification_QSO} shows $\tx{\chi}{False}^2$ vs. $\tx{\chi}{QSO}^2$ for the combined quasar candidates in the two LOPs before (black asterisks) and after (blue triangles) the cuts we perform. Also shown are the subset of quasars with a detected host galaxy (orange circles) and the subset of quasars located at a redshift $z>3$ (pink squares). 

\begin{figure}[h!]
\center
\includegraphics[width=\columnwidth]{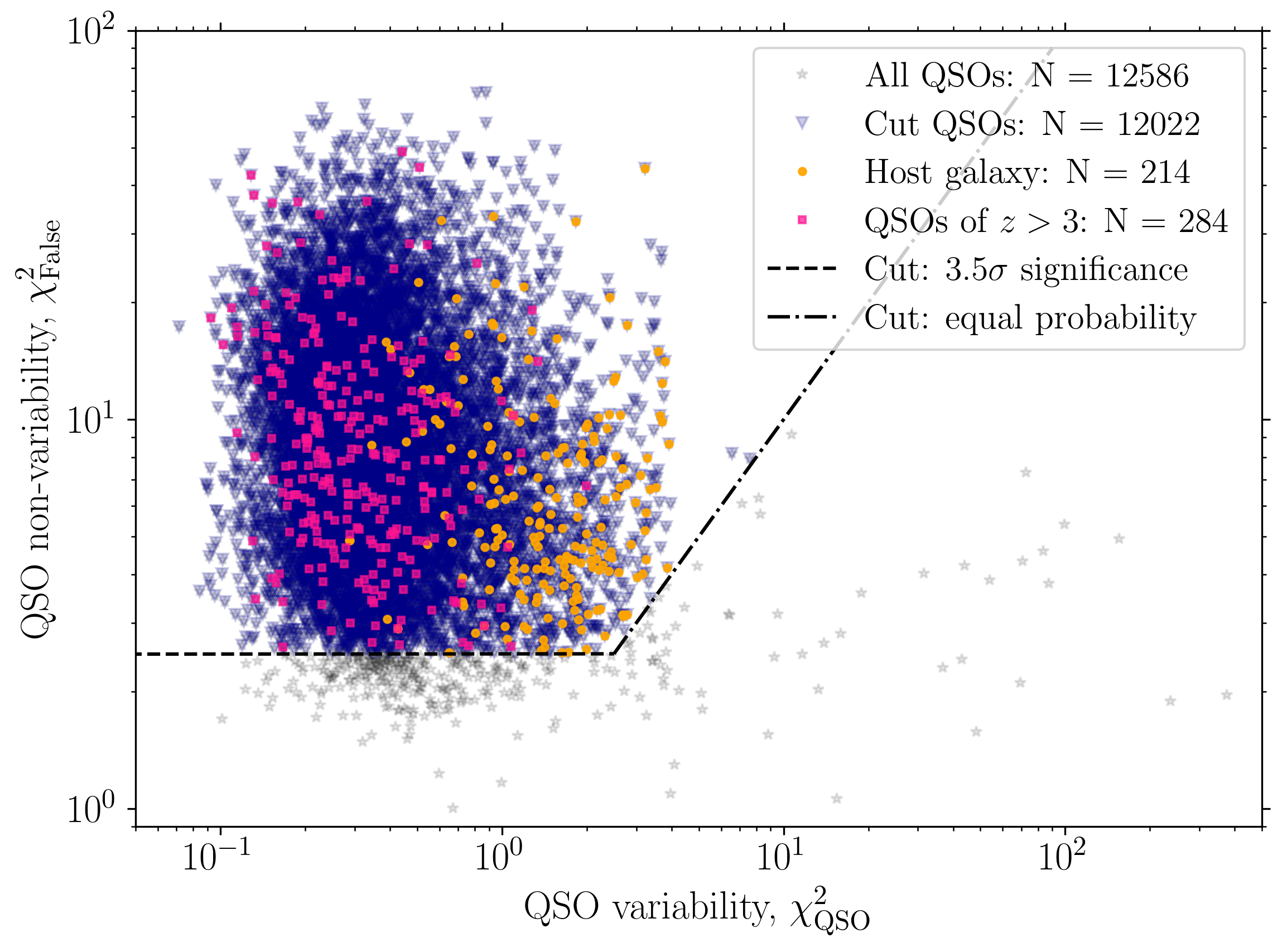}
\caption[]
{QSO variability selection diagram showing the QSO non-variability metric ($\tx{\chi}{False}^2$) versus the QSO variability metric ($\tx{\chi}{QSO}^2$) as defined by \cite{butler2011optimal}. The values of the two metrics are from the \textit{Gaia} DR3 parameter database. The black asterisks and the blue triangles show the combined QSO candidates from the LOPS2 and LOPN1 before and after the applied cuts, respectively. The orange circles show the subset with a visible host galaxy and the pink squares show all QSOs with a redshift larger than $z>3$. The black dashed-dotted line represents a cut based on equal probability ($\tx{\chi}{QSO}^2 = \tx{\chi}{False}^2$) between the two hypotheses. The black dashed horizontal line represents a $\geq 3.5 \sigma$ significance cut against false positives for $\tx{\chi}{False}^2 > 2.5$.} 
\label{fig:classification_QSO}
\end{figure}

A location in Fig.~\ref{fig:classification_QSO} represents a QSO classification $\chi^2$-likelihood. For example, the dashed-dotted black line (defined by $\tx{\chi}{QSO}^2 = \tx{\chi}{False}^2$) represents a line of equal odds in favour of the hypothesis that the object is a quasar (above the line) compared to not being a quasar (below the line). The dashed black line is a significance cut of $\geq 3.5 \sigma$ against false positives. Highly probably QSO candidates are confined in the upper left corner (above the dashed and dashed-dotted lines) of Fig.~\ref{fig:classification_QSO}. Since the short-timescale variability empirically increases with increasing intrinsic quasar brightness \citep[cf.][Fig.~3b]{butler2011optimal}, it is expected that quasars with a visual host galaxy counterpart statistically have a smaller $\tx{\chi}{QSO}^2$ compared to the bulk of the quasars due to their low redshift, which is what we observe.

As a last step to increase the sample purity, we cross-match our sample with the all-sky \textit{Gaia}-unWISE spectroscopic quasar catalogue (Quaia) of \cite{storeyfisher2024quaia}. Quaia effectively accounts for extinction due to its integrated ground-based spectroscopic observations. Figure\,\ref{fig:redshift_colour_correlation} shows the Gaia colour against the Quia redshift for the quasars in our \plato{} catalogue.

\begin{figure}[h!]
\center
\includegraphics[width=\columnwidth]{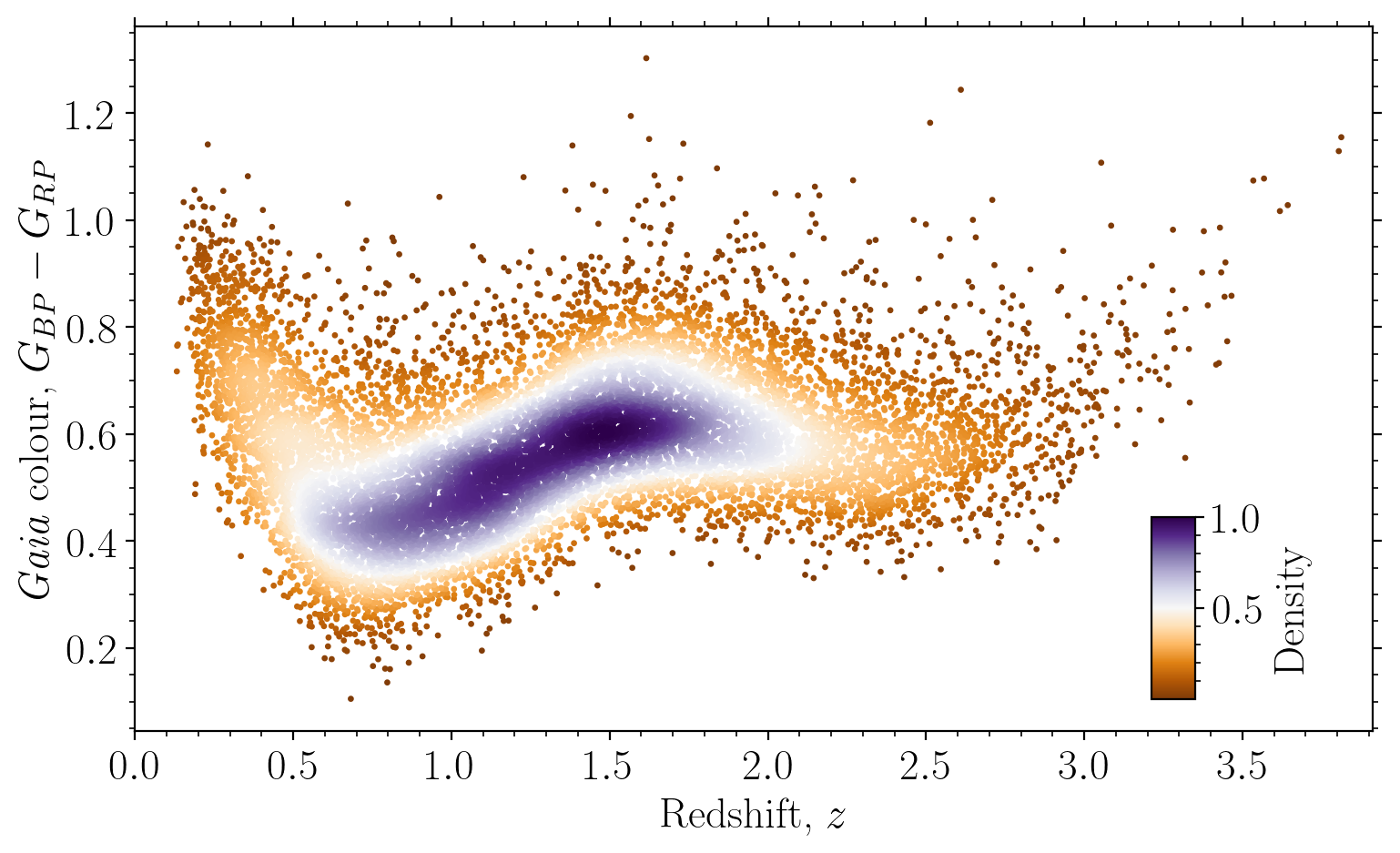}
\caption[]
{\textit{Gaia} colour, $\tx{G}{BP}-\tx{G}{RP}$, plotted as a function of the source's redshift, $z$. The sources are colour coded according to their
normalised Gaussian density.} 
\label{fig:redshift_colour_correlation}
\end{figure}


\section{Model components}\label{app:model}

\subsection{Orbital dynamics}

We model the binary system following Keplerian orbits with eccentricity $e$. To simplify the notation in the following we define the total mass of the two-body system as $M \equiv M_1 + M_2$ (with $M_1$ ($M_2$) being the most (least) massive body) and the mass ratio as $q \equiv M_2/M_1 \leq 1$. We follow \cite{murray2010keplerian} to define the orbital elements in three dimensions, showing that the $(x, y, z)$ cartesian coordinates in the orbital plane of the observer are
\begin{align}\label{eq:xyz}
x &= r \, (\cos\Omega \cos(\omega+f) - \sin\Omega \sin(\omega+f)\cos i) \,, \\
y &= r \, (\sin\Omega \cos(\omega+f) + \cos\Omega \sin(\omega+f)\cos i) \,, \nonumber \\
z &= r \sin(\omega+f)\sin i \,, \nonumber
\end{align}
with $i$ the orbital binary inclination relative to the LOS (such that $i = \SI{90}{\degree}$ corresponds to an edge-on orbit), $\omega$ is the argument of periapse, and $\Omega$ is the longitude of ascending node (for which we anchor at $\Omega=\pi/2$). The radial separation of two bodies at a given instant in time is   
\begin{equation}\label{eq:r}
r = a \, (1 - e \cos f_e) \,,
\end{equation}
with $a$ being the semi-major axis of the elliptical orbit determined by Kepler's third law
\begin{equation}\label{eq:a}
a = \paren{\frac{G M P^2}{4 \pi^2}}^{1/3} \,,
\end{equation}
with $G$ being the gravitational constant, and $P$ the orbital period in the binary rest frame. The orbital period in the observer frame of reference changes with redshift $z$, following $T = P (1+z)$. To calculate the Keplerian motion we use three angular anomalies: the mean anomaly $f_m$, the eccentric anomaly $f_e$, and the true anomaly $f$. Using a time of ephemeris $t_0$, per time step $t$ we can compute the mean anomaly as 
\begin{equation}
f_m = 2 \pi \frac{(t-t_0)}{T} \,.
\end{equation}
With $f_m$ we use the Newton--Raphson method to calculate the eccentric anomaly $f_e$ from Kepler's equation: $f_m = f_e - e \sin f_e$. Lastly, with $f_e$ we determine the orbital motion in $(x,y,z)$ of Eq.~\eqref{eq:xyz} using the true anomaly
\begin{equation}\label{eq:f}
f = 2 \tan^{-1}\sqrt{\frac{1+e}{1-e}} \, \tan\paren{\frac{f_e}{2}} \,.
\end{equation}

\subsection{Doppler boosting}\label{app:model_doppler}

Relativistic Doppler boosting of an accretion disc surrounding either of the SMBH binary components can induce periodic variability. A boosted signal observed at frequency $\nu$ is related to the rest-frame frequency $\nu_0$ by the expression $\nu = \mathcal{D} \, \nu_0$. Assuming an intrinsic power-law spectrum for the flux of a stationary source $F_{\nu_0}$, \cite{dorazio2015relativistic} showed that since the number of photons is $\propto F_{\nu}/\nu^3$, the observed flux due to Doppler boosting can be written as $F_{\nu} = \mathcal{D}^{3-\alpha} \, F_{\nu_0}$. The relativistic Doppler factor is thus defined as
\begin{equation}
\mathcal{D}_i^{3-\alpha} = \parenf{\gamma \paren{1 - \frac{v_{r,i}}{c}}^{(3 - \alpha)}}^{-1} \,,
\end{equation}
where the subscript $i$ denotes the i$^{th}$ binary component, $\gamma \equiv (1-v^2/c^2)^{-1/2}$ the Lorentz factor, which in turn depends on the (squared) velocity defined by \citep[e.g.][]{murray2010keplerian}
\begin{equation}
v_i^2 = \frac{G M}{c^2} \paren{\frac{M_i}{M}}^2 \paren{\frac{2}{r}-\frac{1}{a}} \,.
\end{equation}
The radial velocity, being the projection of the velocity vector onto the LOS, is given by
\begin{equation}\label{eq:v_r}
v_{r,i} = v_z + K_i \parenf{\cos(\omega + f) + e \cos \omega} \,,
\end{equation}
with $v_z$ being the proper motion of the barycentre. Defining the mass indices $i(j=2)=1$ and $i(j=1)=2$, the RV semi-amplitude for each component can be written as
\begin{equation}\label{eq:K}
K_i = \frac{2\pi}{P}\frac{M_j}{M}\frac{a \sin I}{\sqrt{1-e^2}} \,.
\end{equation}

\subsection{Gravitational lensing}\label{app:model_lensing}

\subsubsection*{Point source model}

We start by addressing the gravitational lensing of a point source. Defining a time of ephemeris $t_0$ as the mid-time point when the primary lenses the secondary, the magnification is given by \citep{paczynski1986gravitational}
\begin{equation} \label{eq:point_source}
\tx{\mathcal{M}}{PS}(u) = \frac{u^2 + 2}{u\sqrt{u^2 + 4}} \,,
\end{equation}
where $u = \text{Re}\{u_1, \, u_2\} = \delta/\theta_E$ is the projected separation between the primary and secondary black hole in units of angular Einstein radii
\begin{equation}
\theta_E \equiv \sqrt{\frac{4 G M_i}{c^2 \tx{D}{rel}}} \,,
\end{equation}
with $\tx{M}{l}$ being the mass of the lens, $\tx{D}{rel}^{-1} \equiv \tx{D}{l}^{-1}-\tx{D}{s}^{-1}$ is the time-dependent distance between between the lens and source along the LOS, with $\tx{D}{l}$ and $\tx{D}{s}$ being the distance to the lens and source, respectively. The angular separation between the lens and source on the sky in cartesian coordinates
\begin{equation}
\delta = \sqrt{(x_1-x_2)^2 + (y_1-y_2)^2} \,,
\end{equation}
where $(x_1, x_2, y_1, y_2)$ is found from Eq.~\eqref{eq:xyz}. To save computation time, we follow the software implementation of \texttt{binlite} \citep{dorazio2024binlite} and set the mass of the lens to $M_1$ and switch for $M_2$ whenever $z_1 < 0$ (as we defined the $z$-axis pointing towards the observer). This also implies a switch in the time-dependent coordinates of $\tx{D}{l} = -z_1$ and $\tx{D}{s} = -z_2$ whenever $M_1$ and $M_2$ is the lens, respectively.

\subsubsection*{Finite source model}

According to \citet[][Fig.~2]{dorazio2018periodic}, for binaries of total mass $\lesssim 10^8 M_\odot$ and orbital periods of a few years, the angular size of the accretion disc becomes comparable to the Einstein radius. As a result, the point source approximation does not hold, and it is necessary to incorporate the size of the source's accretion disc. We adopt the model designed by \cite{dorazio2018periodic} to account for finite source effects. This model is based on the thin-disc, multi-colour accretion disc emission model from \cite{shakura1973black}, which extends from the innermost stable circular orbit (ISCO), $r_{\rm ISCO} = 6 G \tx{M}{s}/c^2$, to the tidal truncation radius of the source, $r_{\rm tidal}=0.27 q^{0.3} a$ \citep[cf.][]{roedig2014observational}. The temperature profile of the disc is given by
\begin{equation}\label{T(r) function}
\sigma T^4(r)=\frac{3G\,\tx{M}{s}\,\dot{\tx{M}{s}}}{8\pi r^3} \left[1-\left(\frac{r_{\rm ISCO}}{r}\right)^{1/2}\right] \,,
\end{equation} 
where $r$ is the radial distance from the central SMBH and $\dot{\tx{M}{s}}$ is the accretion rate of the lensed SMBH, for which an Eddington accretion rate is assumed. For $r<\tx{r}{ISCO}$ and $r>\tx{r}{tidal}$, we set $T(r)=0$. The resulting flux from the disc is given by
\begin{equation}\label{eq:black-body}
F_\nu (r) = \pi B_\nu [T(r)] \,,
\end{equation}
where $B_\nu$ is the Planck function. We calculate the lensing magnification by evaluating the expression
\begin{equation}\label{eq:finite_source}
\mathcal{M}_{\rm FS}(u', v', \nu) = \frac{\int_0^{2\pi}\int_0^{\infty}F_\nu(u',v')\mathcal{M}_{\rm PS}(u')u'du'dv'}{\int_0^{2\pi}\int_0^{\infty}F_\nu(u',v')u'du'dv'} \,,
\end{equation}
where $\mathcal{M}_{\rm PS}(u')$ is given in Eq.~\eqref{eq:point_source}, evaluated in the lens-centred polar coordinates $(u,v)$
\begin{equation}
r_* = r_E \sqrt{u_0^2 + u^2 - 2 u_0 u \cos(v-v_0)} \,,
\end{equation}
\begin{equation}
r = r_*\sqrt{\cos^2\theta + \frac{\sin^2\theta}{\cos^2(\pi/2-J)}} \,,
\end{equation}
\begin{equation}
\sin \theta = \frac{u\sin v - u_0 \sin v_0}{\sqrt{(u\sin v - u_0 \sin v_0)^2+(u\cos v - u_0 \cos v_0)^2}} \,.
\end{equation}
Here $(u_0, v_0)$ is the position of the secondary, where $u_0$ is in units of Einstein radii, while $v_0$ is given in radians. $J$ is the inclination of the source disc relative to the LOS. Compared to point-source solution, here the complex quantity of each $u$ binary component is defined as
\begin{equation} \label{eq:u}
u_i = \frac{a}{R_{E,i}} \sqrt{\cos \phi_i^2 + \sin I^2 \sin \phi_i^2} \,.
\end{equation}
with the Einstein radius defined as
\begin{equation}
R_{E,i} = \sqrt{2 \, R_{S,i} \, a \cos I \, \sin \phi_i} \,, 
\end{equation}
and the Schwarzschild radius is defined
\begin{equation}
R_{S,i} = \frac{2 G M_i}{c^2} \,.
\end{equation}
The orbital phase of each SMBH component is given by $\phi_i = 2 \pi (t - t_0) / P$ (thus $\phi_1 \equiv \phi_2 \pm \pi$).

\begin{figure}[t!]
\center
\includegraphics[width=\columnwidth]{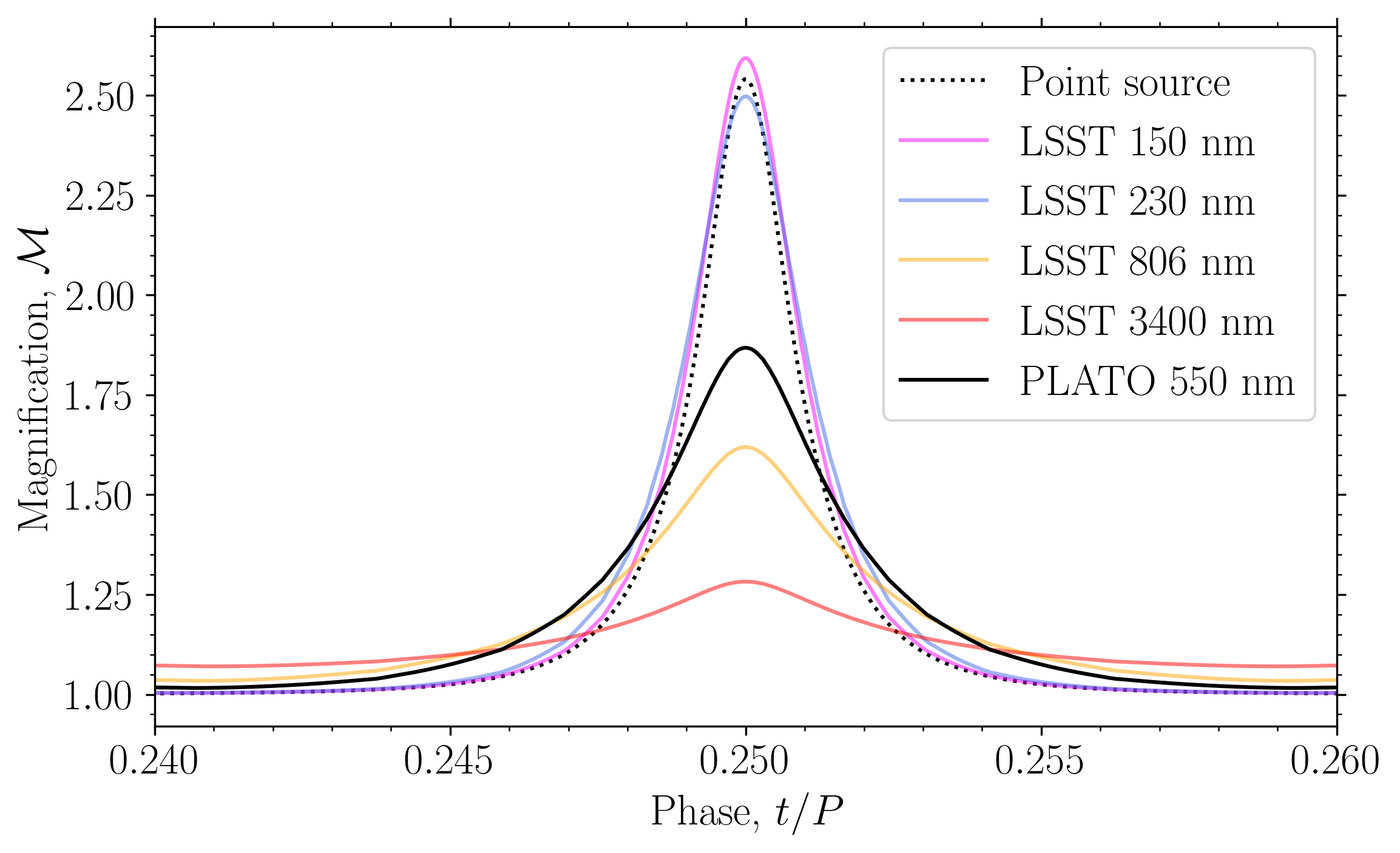}
\caption[]
{Illustration of the gravitational lensing event of the finite-source accretion disc observed at $(I, J)=(\SI{0.28}{\degree}, \SI{45}{\degree})$. Together with multi-passband observations expected for future colour filters of LSST, the plot shows the magnification of a point source (black dotted line) and specifically for a finite source observed in the central wavelength of the \textit{Plato} passband (black solid line).} 
\label{fig:lensing_chromatic}
\end{figure}

Figure~\ref{fig:lensing_chromatic} is a zoom-in on a synthetic light curve showing the magnification using the finite source model description. Apart from the results for a few colour filters of the LSST survey (coloured solid lines), the figure shows the magnification of a point source (cf. Eq.~\eqref{eq:point_source}; black dotted line) and the magnification of a finite-source accretion disc as observed by the \textit{Plato} passband (cf. Eq.~\eqref{eq:finite_source}; black solid line). We generated these models using a disc inclination of $J=\pi/4$. As SLFs are generally more likely for small $I$, lensing signatures from discs aligned relative to the orbit are almost achromatic. In contrast, misaligned configurations show an increasing magnification in shorter wavelength (cf. Fig.~\ref{fig:lensing_chromatic}) as more of the inner and hotter disc parts are revealed. It has been demonstrated that the magnification $\mathcal{M}_{\rm \nu}^{\rm FS}$ does not vary much according to $J$ \citep[cf.][Fig.~4 and 5]{dorazio2018periodic}. However, this assumption depends on whether or not obscuration from a dusty torus takes place, which is less well-known for SMBH binary systems \citep{kelly2021gravitational}.

\subsection{Quasar variability}\label{app:model_quasar}

Intrinsic QSO variability in the optical domain is typically modelled as a damped random walk \citep[DRW;][]{kozlowski2010quantifying, mcleod2010modeling}. DRW variability follows a correlated random Gaussian signal that can be described by a characteristic (damping) timescale $\tau$ and a long-term standard deviation of variability $\sigma$. We model DRW in the time domain following \cite{mcleod2010modeling}. To readily explore the regimes of the DRW model, we exploit the power spectral density (PSD) of the DRW given by
\begin{equation}\label{eq:v_r}
\text{PSD}(\nu) = \frac{\tau^2 \, \sigma^2}{1 + (2\pi \, \nu \, \tau)^2} \,.
\end{equation}
This formalism has a characteristic break frequency $(2\pi \, \tau)^{-1}$ dividing two regimes: for $\nu > (2\pi \, \tau)^{-1}$, the model corresponds to a red signal model with $\text{PSD} \propto \nu^{-2}$, and for $\nu < (2\pi \, \tau)^{-1}$, the PSD is independent of $\nu$, thus corresponding to a white signal model.

\subsection{Synthetic light curve of Spikey}\label{app:model_quasar}

Figure.~\ref{fig:model_spikey} shows our regenerated synthetic light curve of Spikey. This plot was generated using the best-fit model parameters of \citetalias[][Table~1]{hu2020spikey} and the model description in Eq.~\eqref{eq:QDM}.

\begin{figure}[h!]
\center
\includegraphics[width=\columnwidth]{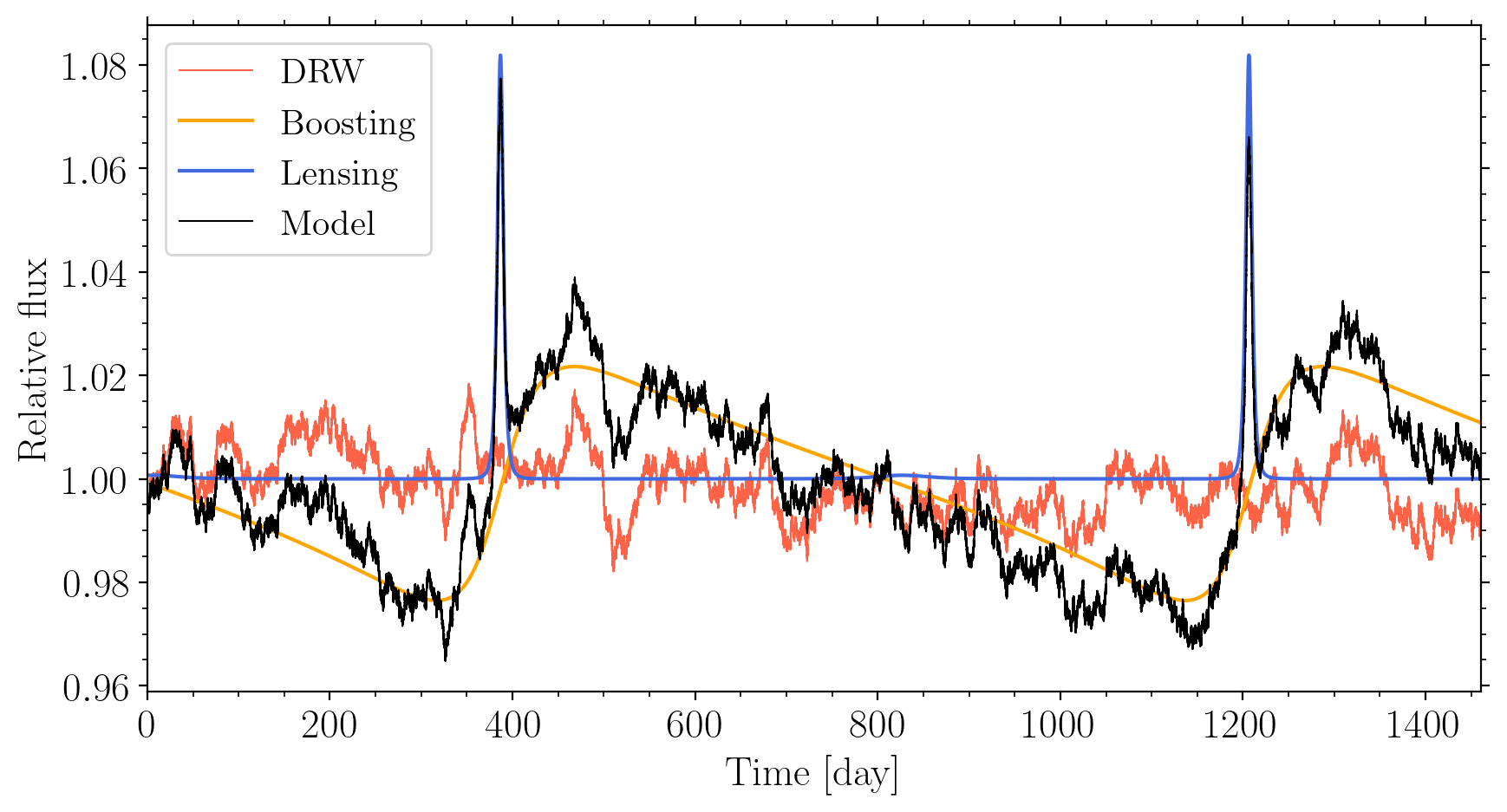}
\caption[]
{Synthetic light curve imitating the \textit{Kepler} observation of Spikey. The components of DRW (red line), Doppler beaming (orange line), and gravitational self-lensing (blue line), together with the combined model (black line), are shown.} 
\label{fig:model_spikey}
\end{figure}


\section{Posterior distributions of Spikey}\label{app:posterior}

We here present the posterior distributions for the modelling of the \kepler{}+\plato{} light curve. Figure~\ref{fig:corner_combined_Q_and_DM} shows the corner plot of the $\mathcal{Q}$ (top right) and $\mathcal{DM}$ (bottom left) model fit, while Fig.~\ref{fig:corner_combined_QDM} shows the corner plot of the $\mathcal{Q\,DM}$ model fit. The posterior distributions (and model fits) presented in this paper are freely available on our GitHub project (see Data Availability section).

\begin{figure*}[b!]
\centering
\settototalheight{\dimen0}{\includegraphics[width=2\columnwidth]{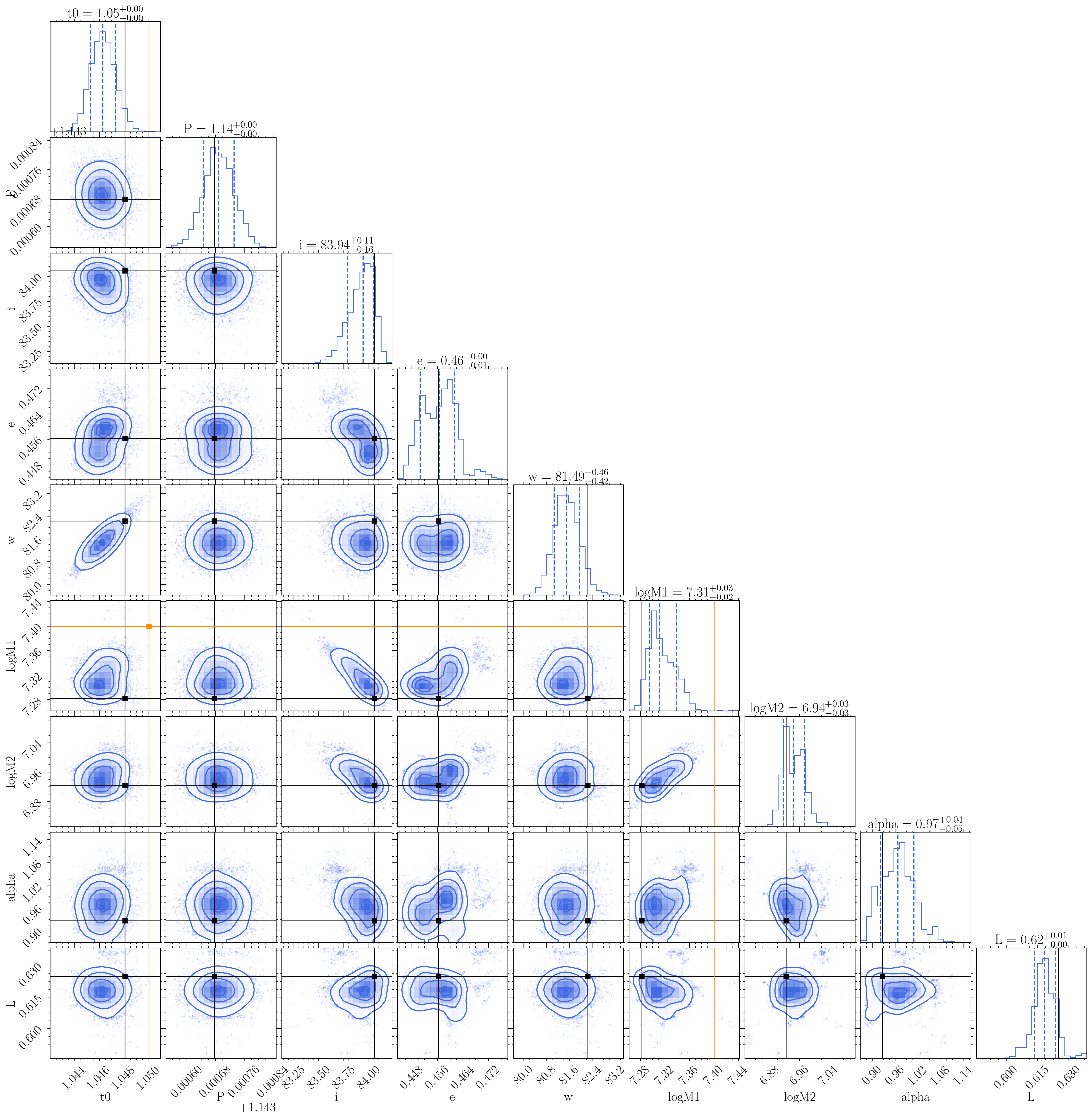}}%
\includegraphics[width=2\columnwidth]{corner_spikey_combined_DM}%
\llap{\raisebox{\dimen0-7cm}{%
\includegraphics[height=7cm]{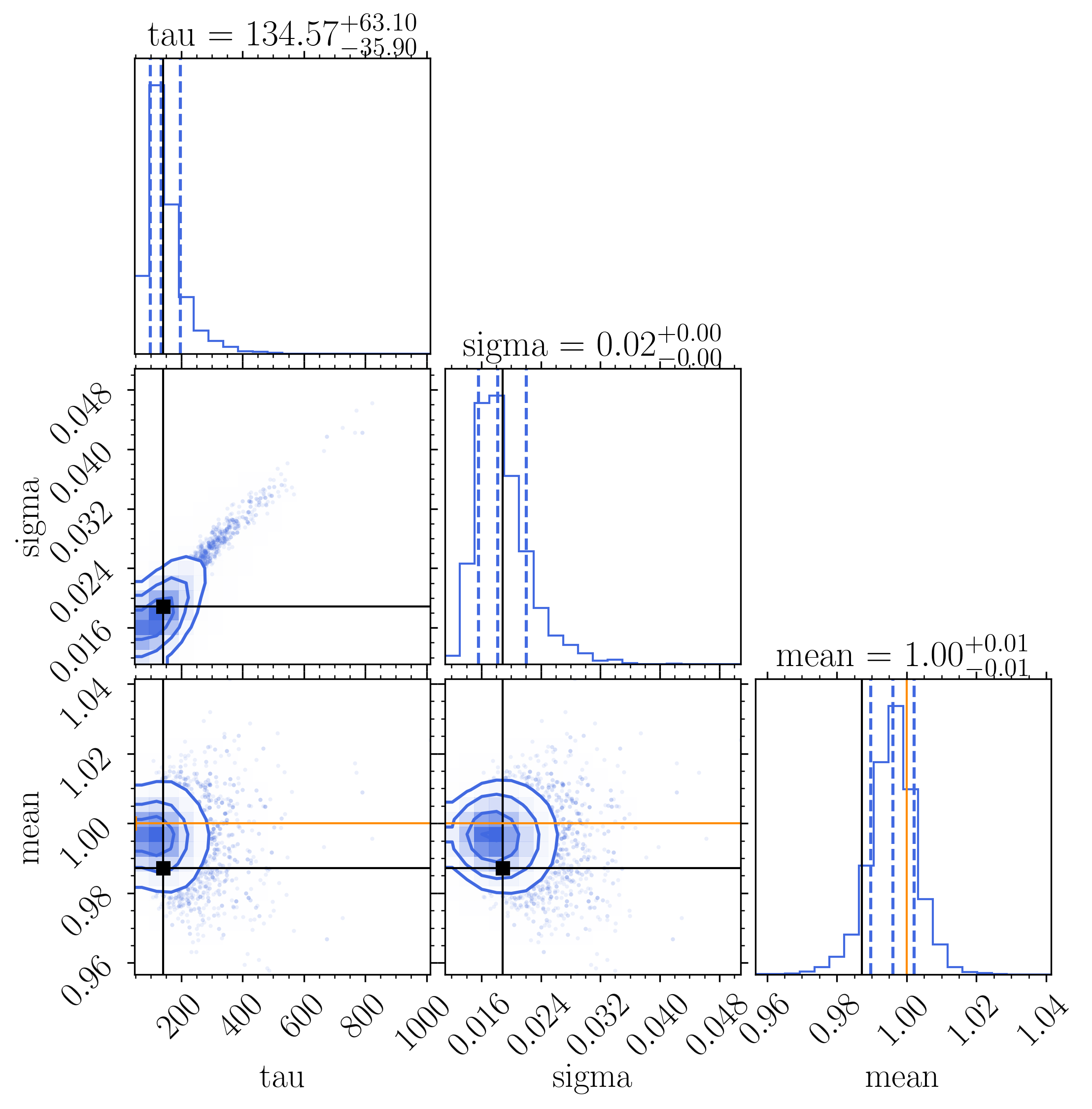}%
}}
\caption{Corner plot of the $\mathcal{Q}$ (top right) and $\mathcal{DM}$ (bottom left) model fit. The posterior distributions (blue scatter points) are shown along with the 16, 50, and 84 per cent quantiles (outer to inner blue contour lines, respectively), together with the maximum likelihood values (black lines) and the injected values from \citetalias{hu2020spikey} (orange lines). The upper diagonal of subpanels show the 2D projected distributions with the mean posterior value reported above each subpanel.}
\label{fig:corner_combined_Q_and_DM}
\end{figure*} 

\begin{figure*}[h!]
\center
\includegraphics[width=2\columnwidth]{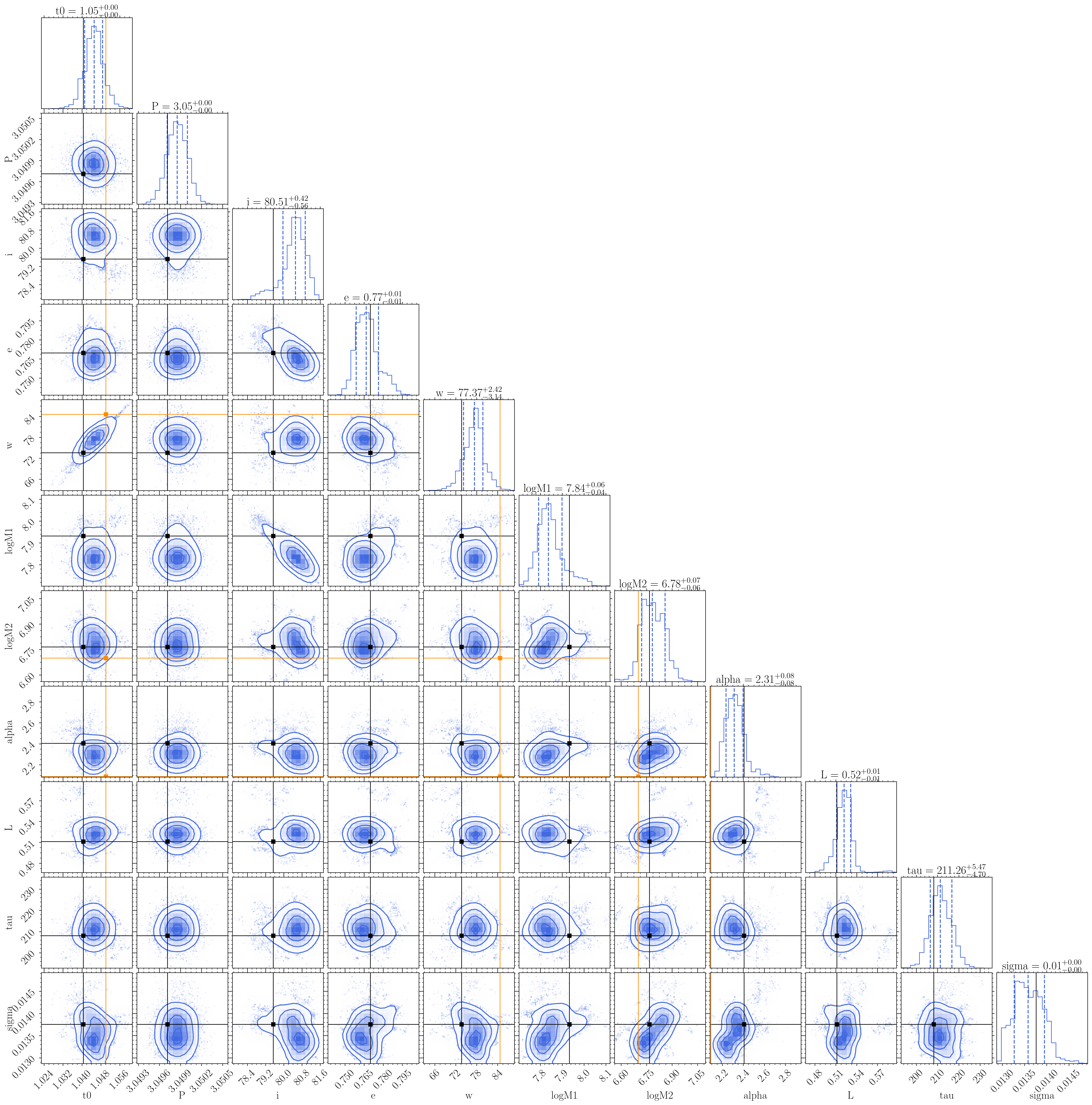}
\caption[]
{Corner plot for the $\mathcal{Q\,DM}$ model of the \kepler{}+\plato{} light curve. Same description as Fig.~\ref{fig:corner_combined_Q_and_DM}.} 
\label{fig:corner_combined_QDM}
\end{figure*}


\section{Model fit components of Spikey}\label{app:components}

We provide two figures showing the individual model components of our modelling to Spikey from Sect.~\ref{sec:results_spikey}. Figure~\ref{fig:lc_components_separate} shows the modelling of the \kepler{}-only and \plato{}-only, and Fig.~\ref{fig:lc_components_combined} shows the modelling of the combined \kepler{}+\plato{} dataset. In each figure, the top, middle, and bottom panels show the $\mathcal{Q}$, $\mathcal{DM}$, and $\mathcal{Q\,DM}$ model fit, respectively. In every plot, each model component is shown with a solid line: $\mathcal{Q}$ (red), $\mathcal{D}$ (orange), and $\mathcal{M}$ (blue), together with the combined model $\mathcal{Q\,DM}$ (pink).

\begin{figure*}[t!]
\center
\includegraphics[width=1.164\columnwidth]{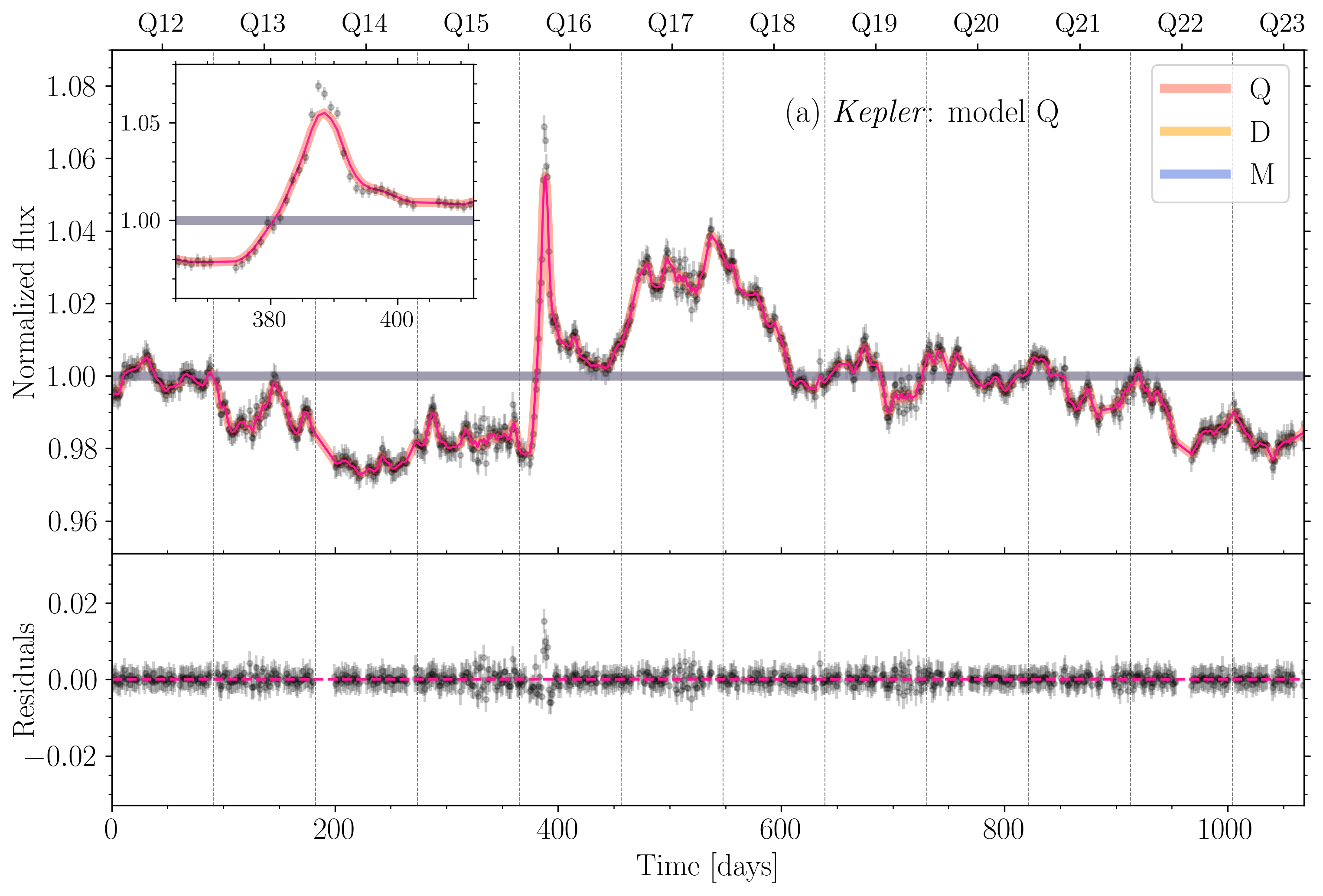}
\includegraphics[width=0.836\columnwidth]{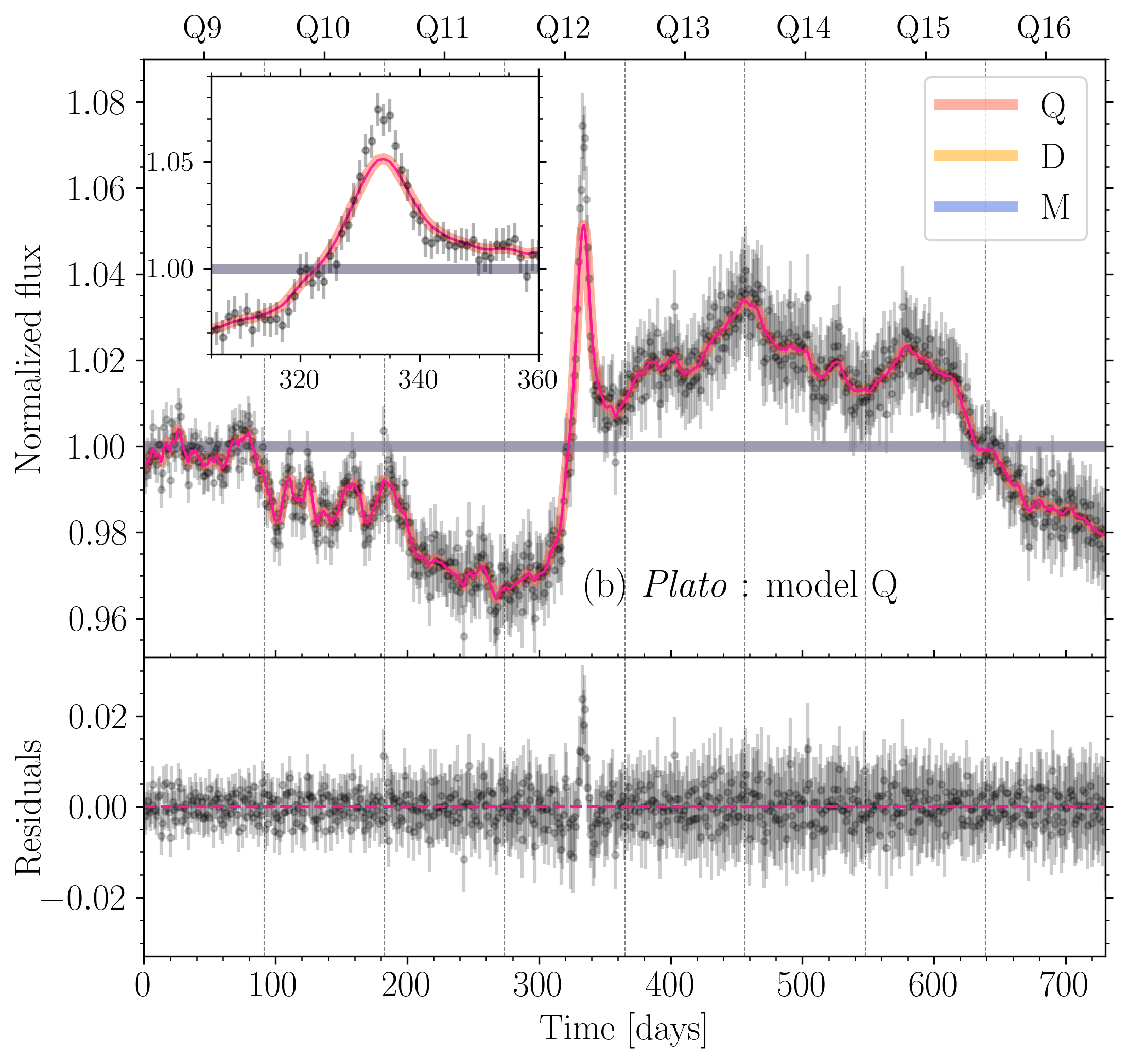}
\includegraphics[width=1.164\columnwidth]{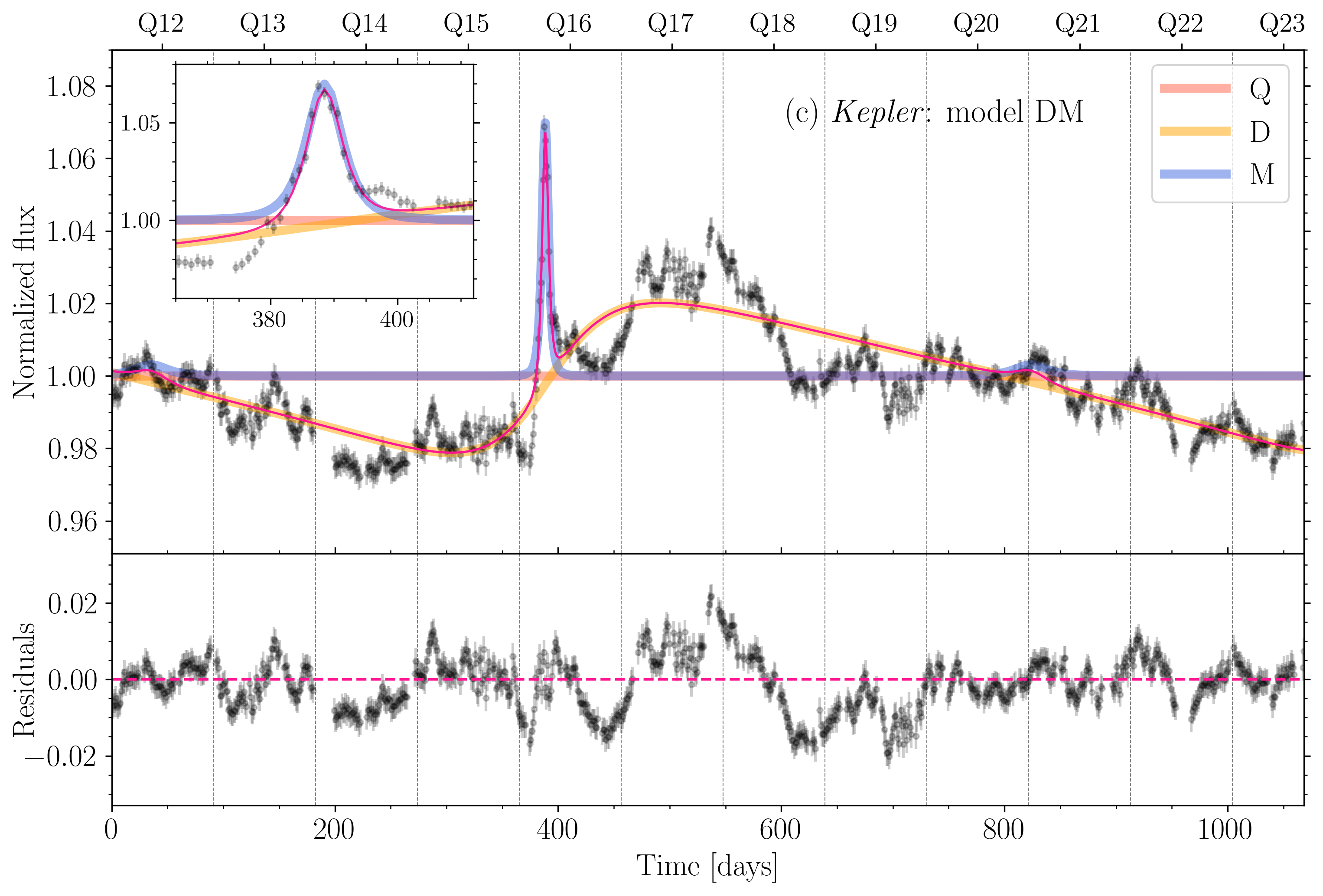}
\includegraphics[width=0.836\columnwidth]{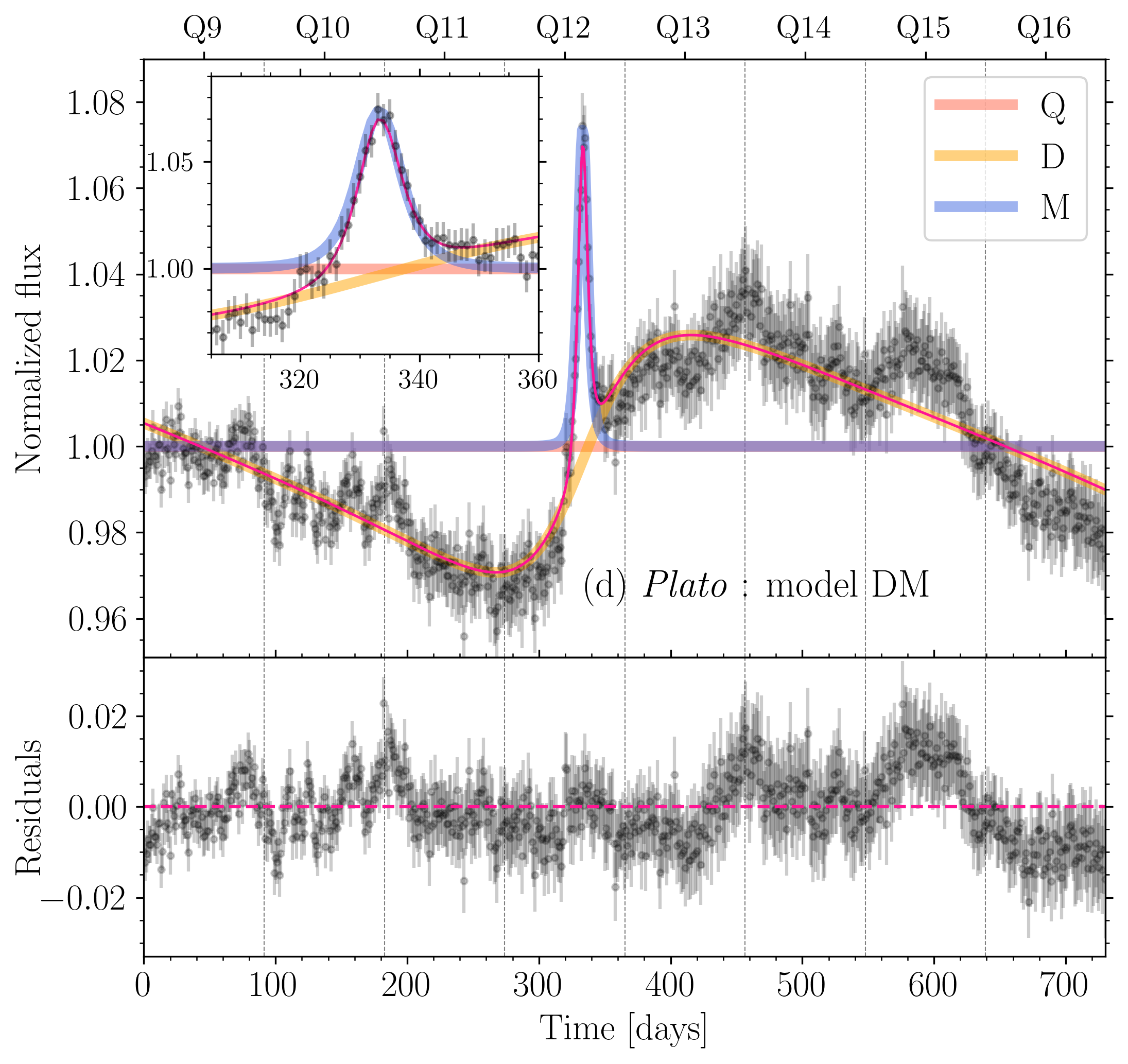}
\includegraphics[width=1.164\columnwidth]{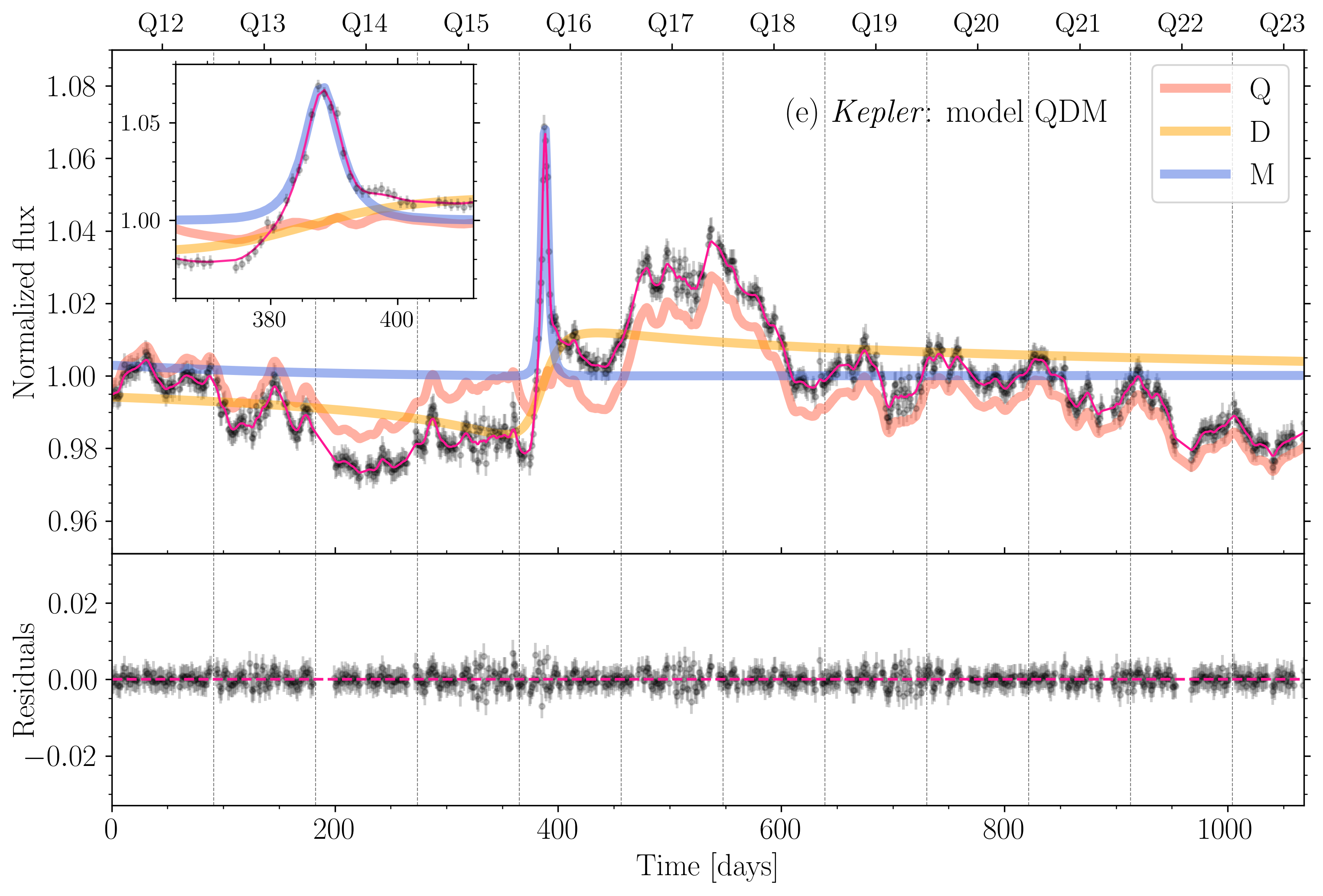}
\includegraphics[width=0.836\columnwidth]{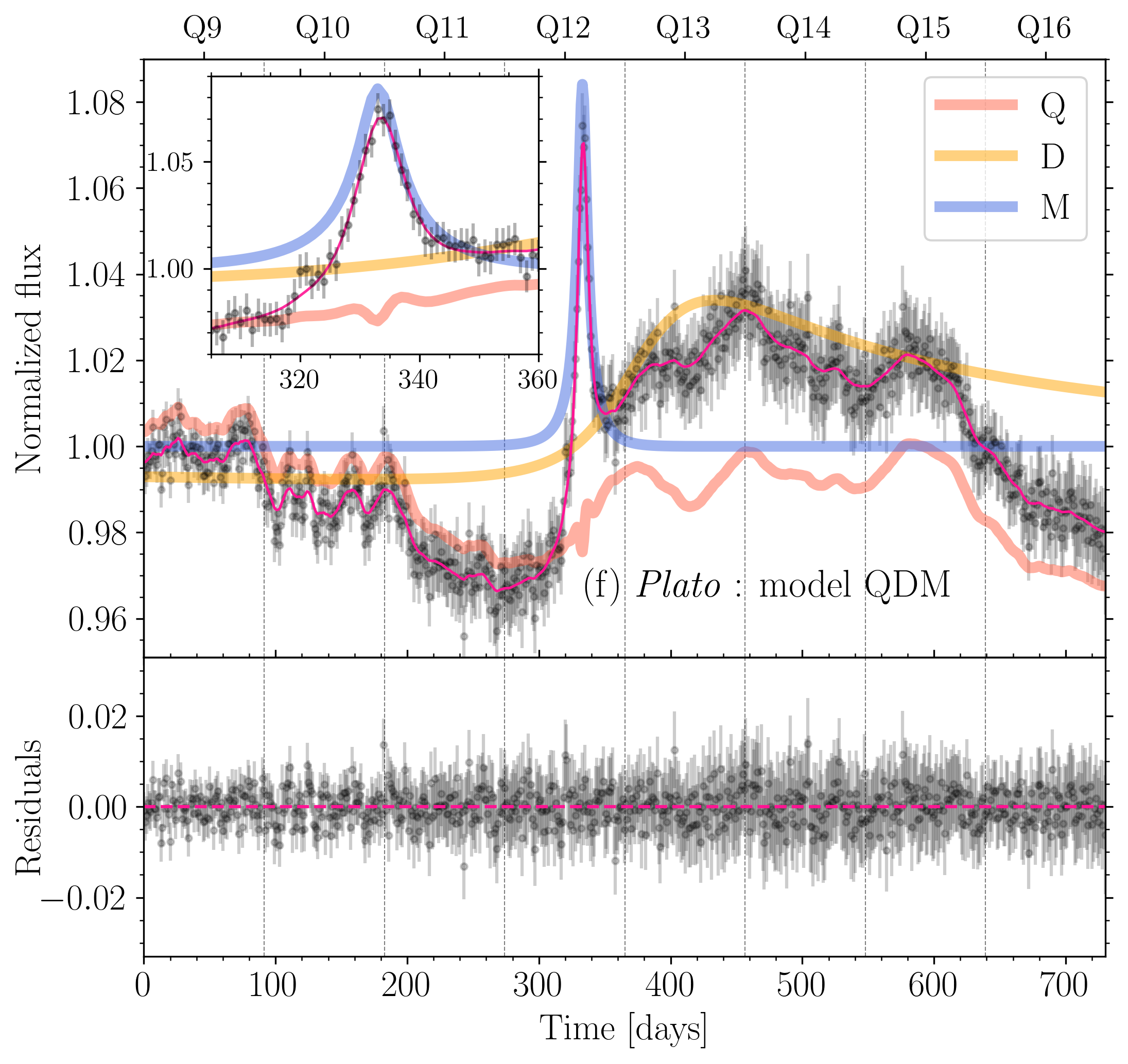}
\caption[]
{Model components of the fit to the \kepler{}-only (left panels) and \plato{}-only (right panels) dataset. Shown is the fit of $\mathcal{Q}$ (panel a and b), $\mathcal{DM}$ (panel c and d) and $\mathcal{Q\,DM}$ (panel e and f). Each model component is clearly shown with a solid line for $\mathcal{Q}$ (red), $\mathcal{D}$ (orange), $\mathcal{M}$ (blue), and the combined model $\mathcal{(Q)\,DM}$ (pink). The lower subpanel of each plot shows the residuals to the combined model $\mathcal{(Q)\,DM}$.}
\label{fig:lc_components_separate}
\end{figure*}

\begin{figure*}[t!]
\center
\includegraphics[width=1.164\columnwidth]{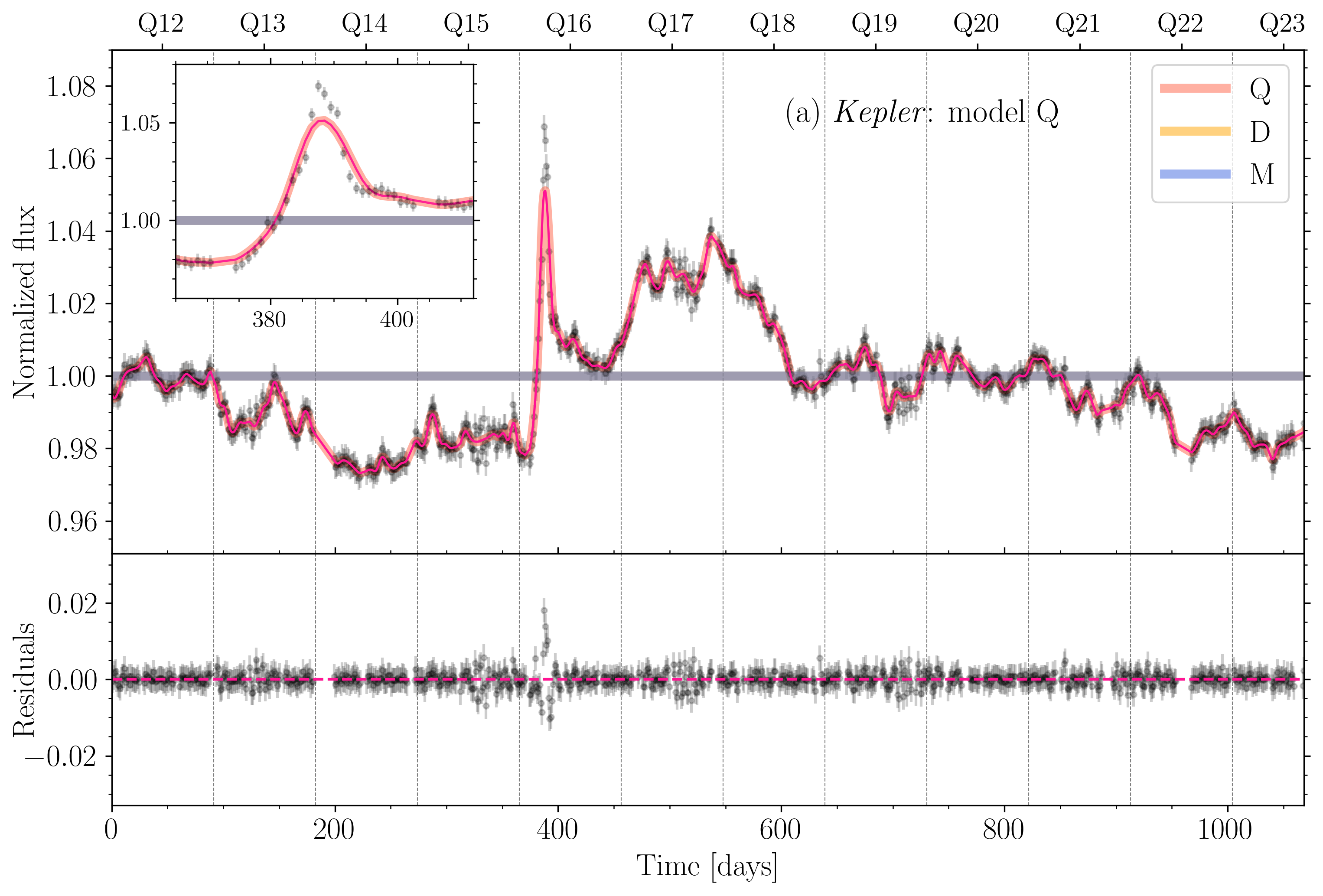}
\includegraphics[width=0.836\columnwidth]{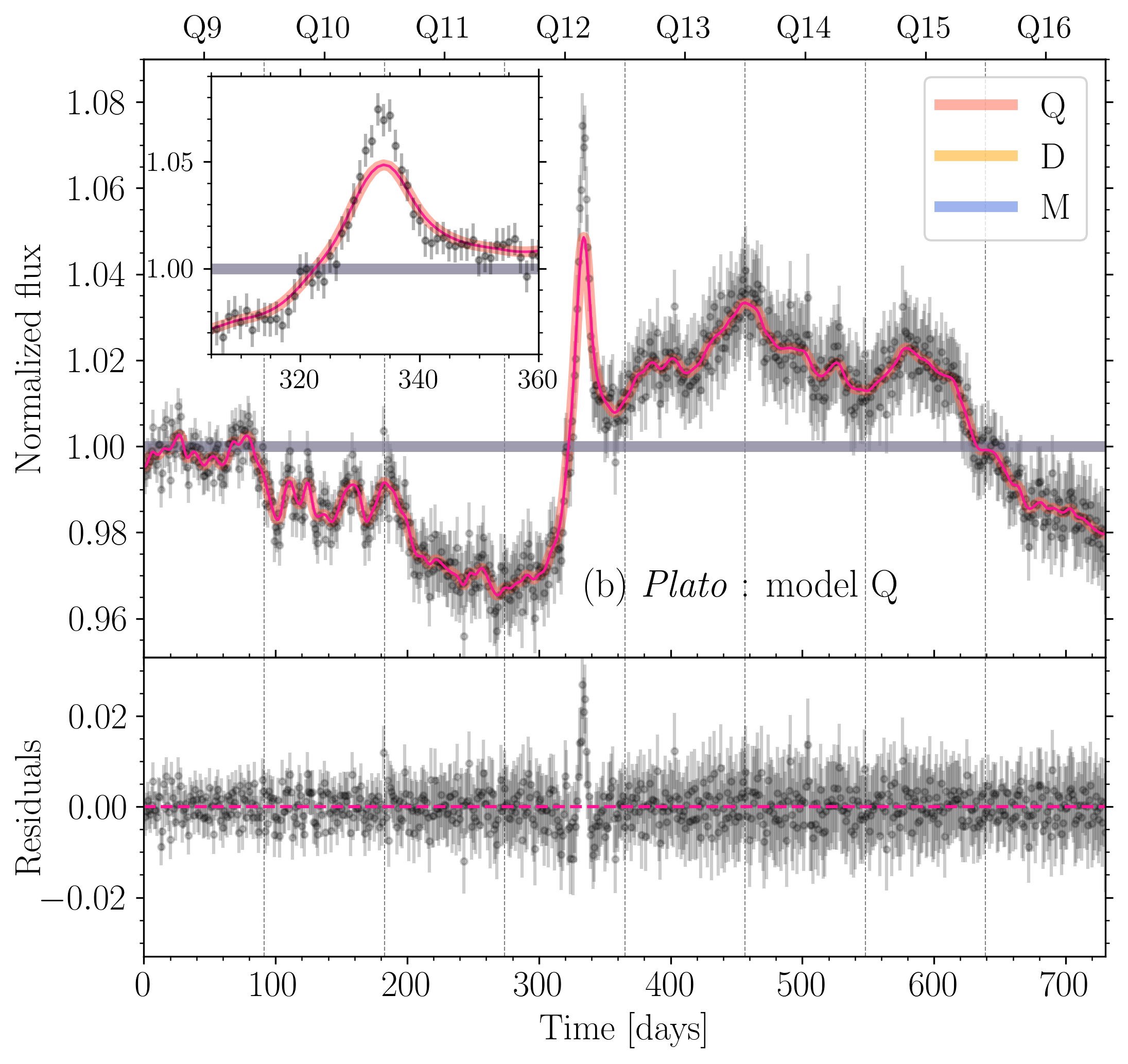}
\includegraphics[width=1.164\columnwidth]{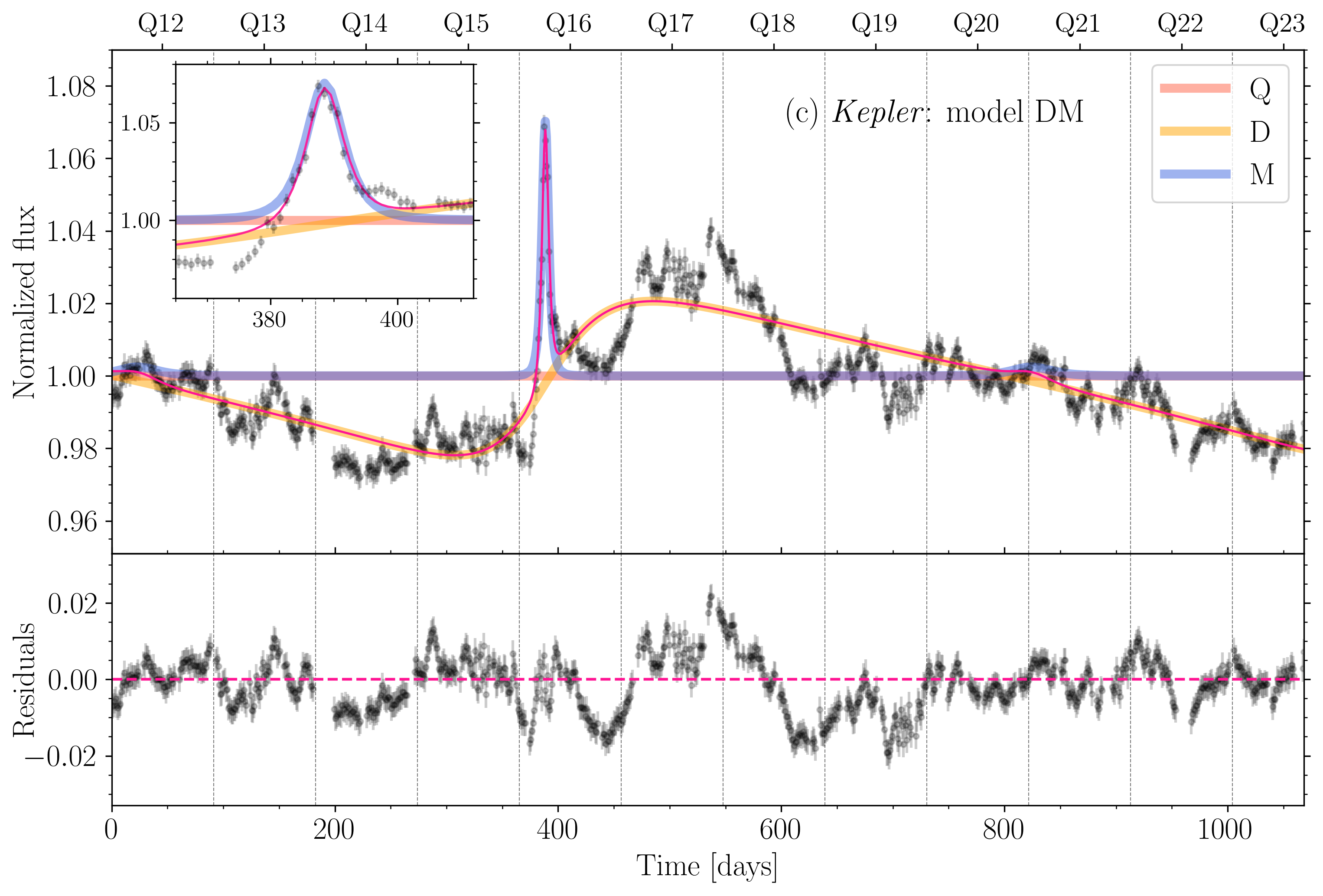}
\includegraphics[width=0.836\columnwidth]{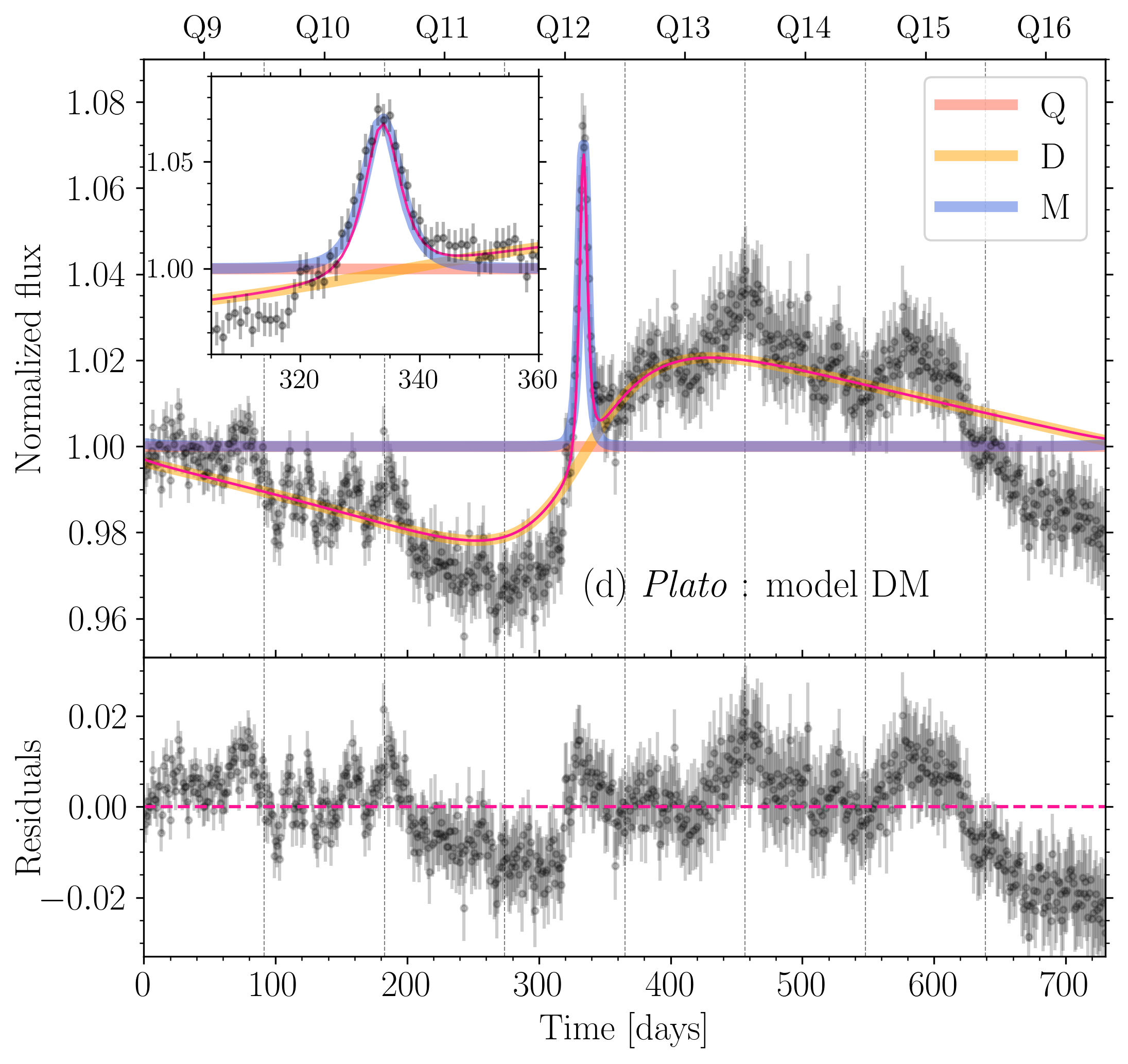}
\includegraphics[width=1.164\columnwidth]{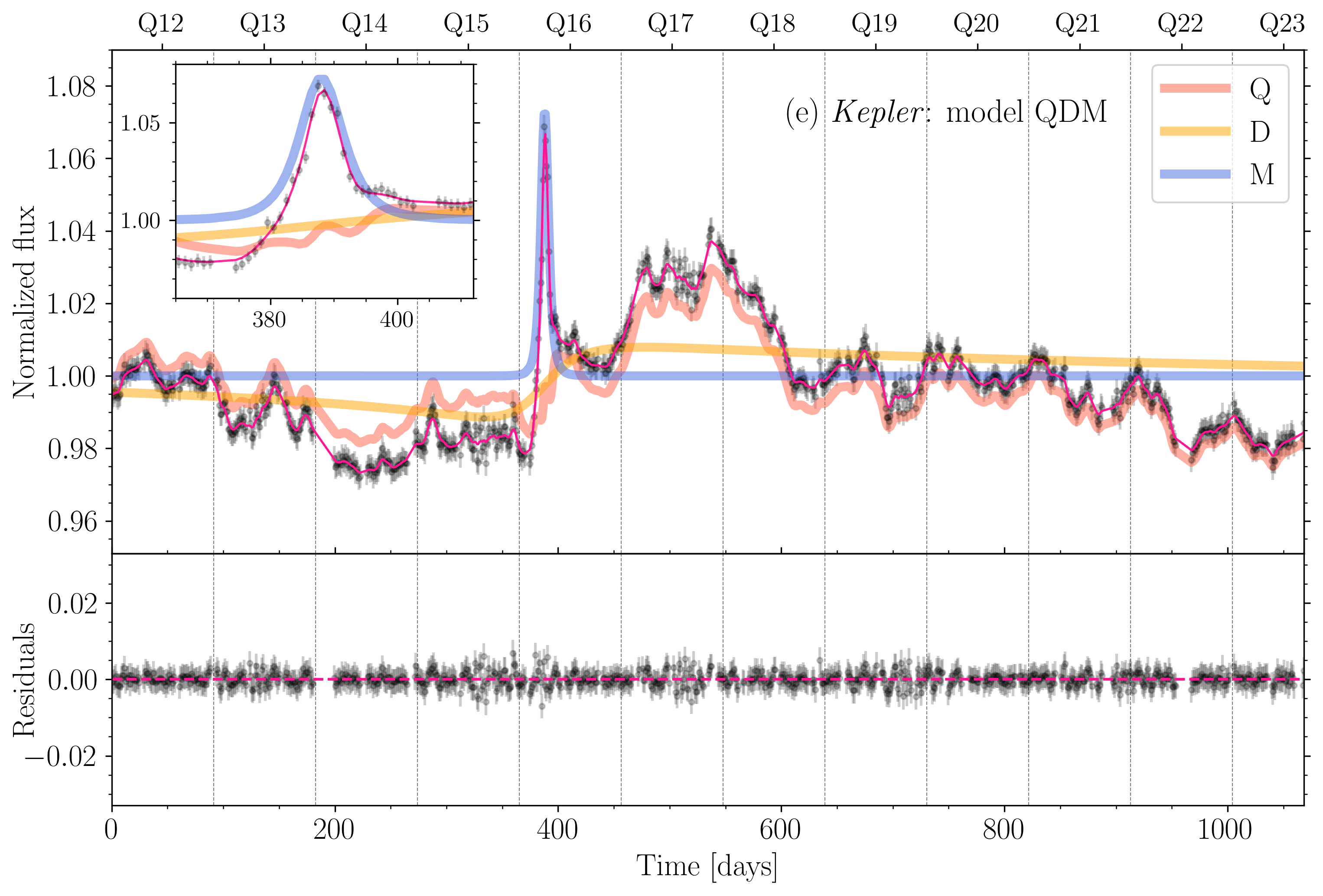}
\includegraphics[width=0.836\columnwidth]{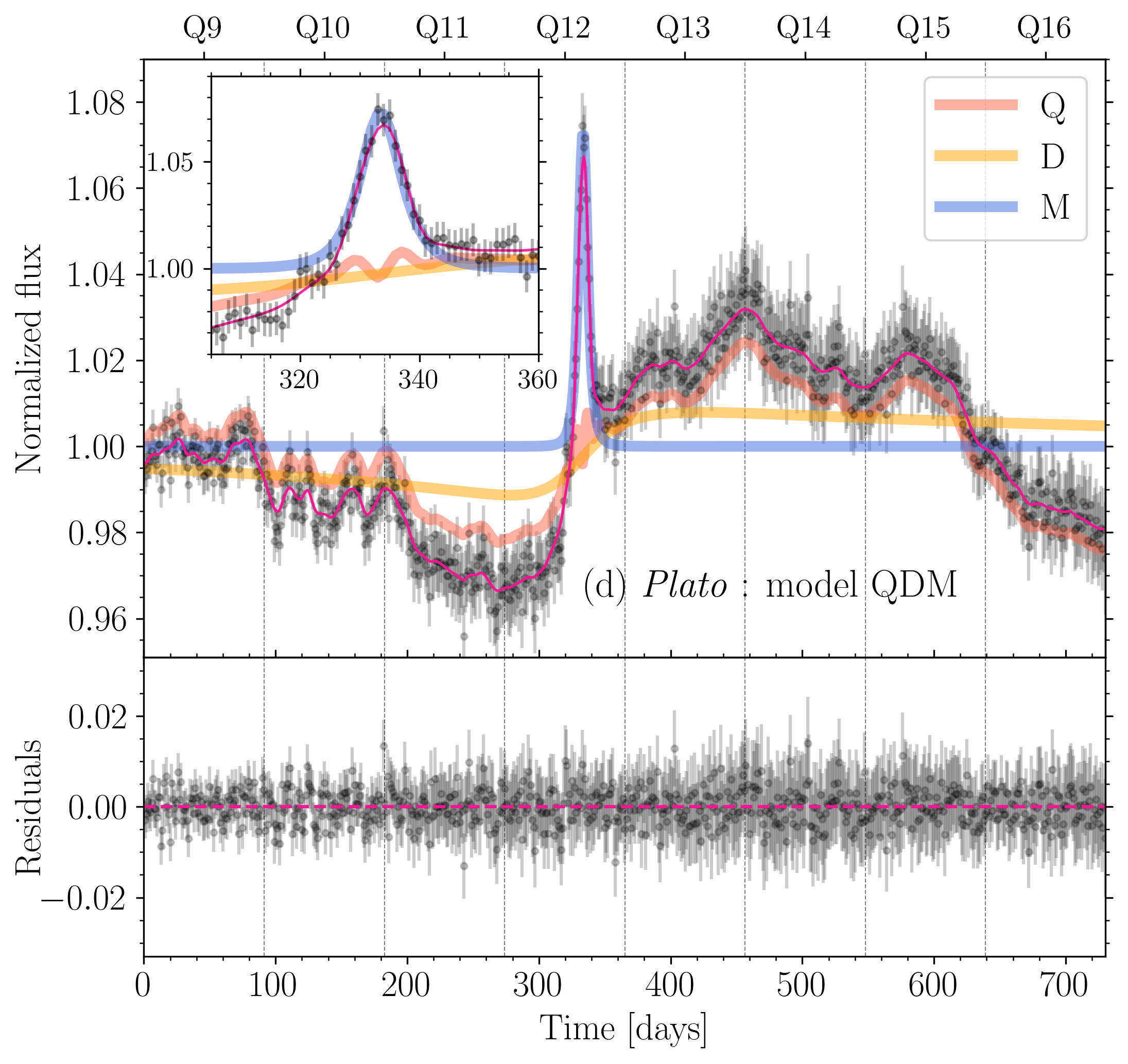}
\caption[]
{Model components of the fit to the combined \kepler{}+\plato{} dataset. Same description as Fig.~\ref{fig:lc_components_separate}.}
\label{fig:lc_components_combined}
\end{figure*}

\end{appendix}

\end{document}